\pdfoutput=1
\documentclass[floatfix,reprint,aps,prc,showpacs,showkeys,eqsecnum,amsmath,amsfonts,amssymb,twocolumn,groupedaddress,superscriptaddress,nofootinbib,noeprint]{revtex4-2}

\usepackage{graphicx}    

\usepackage{dcolumn}     
\usepackage{bm}          
\usepackage{hyperref} 
\usepackage{float} 

\usepackage{amsmath} 
\usepackage{amssymb} 
\usepackage{amsfonts} 
\usepackage{latexsym}
\usepackage{enumitem} 

\usepackage{booktabs} 

\usepackage{cleveref}

\usepackage[english]{babel}
\usepackage[expansion=all,babel]{microtype}

\usepackage[usenames]{xcolor}

\usepackage{mathtools}

\usepackage{multirow}

\usepackage{longtable}
\usepackage{rotating}

\newcommand{\sat}[1]{{#1}_{\mathrm{sat}}}
\newcommand{\sym}[1]{{#1}_{\mathrm{sym}}}
\newcommand{\eff}[1]{{#1}_{\mathrm{eff}}}

\usepackage{txfonts}

\newcommand{\be}{\begin{equation}}
\newcommand{\ee}{\end{equation}}
\newcommand{\ba}{\begin{eqnarray}}
\newcommand{\ea}{\end{eqnarray}}
   
\newcommand{\Msun}{\mathrm{M}_\odot}

\makeatletter
	\newcommand{\vast}{\bBigg@{2.85}}
\makeatother

\begin{document}

\preprint{Grant\#}

\author{Mikhail V. Beznogov}
\email{mikhail.beznogov@nipne.ro}
\affiliation{National Institute for Physics and Nuclear Engineering (IFIN-HH), RO-077125 Bucharest, Romania}

\title{Bayesian Inference of the Dense Matter Equation of State built upon Covariant Density Functionals}
\author{Adriana R. Raduta}
\email{araduta@nipne.ro}
\affiliation{National Institute for Physics and Nuclear Engineering (IFIN-HH), RO-077125 Bucharest, Romania}

\date{\today}

\begin{abstract}
A modified version of the density dependent covariant density functional model proposed in [T. Malik, M. Ferreira, B. K. Agrawal and C. Provid\^encia, ApJ 930, 17 (2022)] is employed in a Bayesian analysis to determine the equation of state (EOS) of dense matter with nucleonic degrees of freedom. Various constraints from nuclear physics and microscopic calculations of pure neutron matter (PNM) along with a lower bound on the maximum mass of neutron stars (NSs) are imposed on the EOS models to investigate the effectiveness of progressive incorporation of the constraints, their compatibility as well as correlations among parameters of nuclear matter and properties of NSs. Our results include the different roles played by pressure and energy per particle of PNM in constraining the isovector behavior of nuclear matter; tension with the values of Dirac effective mass extracted from spin-orbit splitting; correlations between the radius of the canonical mass NS and second- and third-order coefficients in the Taylor expansion of energy per particle as a function of density; correlation between the central pressure of the maximum mass configuration and Dirac effective mass of the nucleon at saturation. For some of our models the tail of the NS maximum mass reaches $2.7\,\Msun$, which means that the secondary object in GW190814 could have been a NS.
\end{abstract}

\pacs{}        
\keywords{Dense matter EOS, neutron stars} 

\maketitle

\section{Introduction}
\label{sec:Intro}

The knowledge of the equation of state (EOS) of dense matter is a key ingredient needed to understand the structure of neutron stars (NSs). Calculations of many NSs global parameters, including those which are observationally accessible, e.g., maximum mass, radius, moment of inertia, tidal deformability, requires exclusively the relation between pressure and energy density. As such, schematic models, e.g., piecewise polytropes, spectral parametrizations and parametrizations of the speed of sound, have allowed to explore the relation between EOS and NS properties, the most famous example being the relation between EOS stiffness and NS maximum mass. At variance with this, modeling of thermal evolution, damping of oscillation modes and pulsar glitches requires also information on matter constituents and, thus, a model of nuclear matter (NM).

A NS EOS can in principle be built for any model of nuclear effective interaction. In doing so one implicitly assumes that the functional relation deduced from nuclear data --- which probe a narrow density domain around nuclear saturation density $n_{\mathrm{sat}} \approx 0.16 ~ \mathrm{fm}^{-3} \approx 2.7 \times 10^{14} ~ \mathrm{g/cm^3}$ and essentially symmetric matter, i.e., with an equal number of protons and neutrons --- and, in some cases, also microscopic calculations of neutron matter remains valid up to several times the value of $n_{\mathrm{sat}}$ and for arbitrary values of proton fraction $Y_\mathrm{p}$. The latter quantity can be determined, for any value of matter density, from the $\beta$-equilibrium condition which links particles' chemical potentials. As such, a large number of models have been proposed, based on different techniques to treat the many body problem of strongly interacting particles; for a review see Refs.~\cite{Oertel_RMP_2017,Burgio_PPNP_2021}. While their predictions are in fair agreement over baryon number density range $n_{\mathrm{B}} \lesssim n_{\mathrm{sat}}$, the results diverge as the density increases, which reflects the poor knowledge of the nuclear interactions. The dispersion among model predictions was nevertheless evocative for the correlations between NSs properties and the isovector behavior of the EOSs, an example in this sense being offered by the relation between internal pressure of matter at intermediate densities ($1.5 n_{\mathrm{sat}} \lesssim n_{\mathrm{B}} \lesssim 3 n_{\mathrm{sat}}$), slope of the symmetry energy and radii of canonical mass NSs \cite{Lattimer_PhysRep_2007,Fortin_PRC_2016}. On the other hand, comparison of the EOS models built within various frameworks revealed a certain model dependence of the results and a certain rigidity in exploring the associated parameter spaces. A common issue with non-relativistic approaches is causality violation at high densities.


Recent progress in multi-messenger astrophysics has provided valuable new constraints on the behavior of dense neutron rich matter. As such, the measurement, since 2010, of several massive pulsars with masses around 2~$\Msun$ \cite{Demorest_Nature_2010,Antoniadis2013,Arzoumanian_ApJSS_2018,Cromartie2020,Fonseca_2021} fueled intensive studies on the EOS stiffness, particularly relevant in connection with the onset of non-nucleonic degrees of freedom \cite{Sedrakian_PPNP_2022}. Measurement of the combined tidal deformability of the two NS in the GW170817 event~\cite{Abbott_PRL119_161101,Abbott_ApJ2017ApJ_L12}, with an estimated total mass $M_\mathrm{T}=2.73^{+0.04}_{-0.01}\,\Msun$ and a mass ratio $0.73 \leq q=M_2/M_1 \leq 1$, supplied constraints strong enough to rule out a certain number of realistic, i.e., nuclear physics motivated, EOS based on the behavior at intermediate densities. Finally, X-ray observations and analysis of rotating hot spot patterns by NICER made possible the determination, with an uncertainty of the order of a few km, of the radii of the canonic mass pulsar PSR J0030+0451 \cite{Miller_2019,Riley_2019} and the massive millisecond pulsar J0740+6620 \cite{Miller_may2021,Riley_may2021}. The huge potential of these observational data is best exploited by statistical analyses, which allow for systematic explorations of the broad model spaces. Two recent examples are provided by Bayesian analyses in Refs. \cite{Miller_may2021,Raaijmakers_may2021}. Their common features include usage of parametrized EOS models, e.g., piece-wise polytropes and models of the speed of sound, and the three categories of NS constraints listed above, i.e., a $\approx 2\,\Msun$ lower limit on the maximum NS mass, tidal deformabilities from GW170817 and GW190425~\cite{Abbott_ApJL_2020} and the two joint mass and radius estimates by NICER. Ref. \cite{Raaijmakers_may2021} additionally posed constraints from chiral effective field theory ($\chi$EFT) calculations of neutron matter and AT2017gfo, the kilonova associated with GW170817. Common conclusions include a certain sensitivity of posterior distributions to the EOS model, e.g., the higher end of the maximum mass distribution; significant sensitivity of the results to included data sets; reduction of uncertainties in the posterior distributions upon progressive incorporation of various data. The latter aspect results in a reduction of uncertainties in estimated NS radii by a factor of 2 (for canonic mass NS) to 3 (for massive NS), the higher convergence at low densities being explained by the dominance of low density probes.

Statistical analyses of cold catalyzed matter with astrophysics constraints have been performed also using phenomenological models of baryonic interaction. Different techniques, such as Bayesian, frequentist or machine learning, have been employed. Some of the models exclusively assume nucleonic degrees of freedom~\cite{Lim_EPJA_2019,Zhang_ApJ_2019,Ferreira_PRD_2020,Guven_PRC_2020,Ferreira_PRD_2021,Ferreira_JCAP_2021,Ghosh_EPJA_2022,Malik_ApJ_2022,Patra_PRD_2022,Malik_RMF_NL_2023,Papakonstantinou_2023}, while hyperons~\cite{Malik_PRD_2022,Ghosh_FASS_2022} or a hadron to quark phase transition~\cite{Annala_PRL_2018,Annala_Nature_2020,Blaschke_EPJA_2021,Huth_PRC_2021,Li_ApJ_2021} are envisaged in others. The partial loss of flexibility, with respect to parametric models, is largely compensated by the physical underpinning; inclusion of saturation properties of NM and, in some cases, data from heavy ion collisions; the potential to provide insights into NSs particle composition; possibility to investigate NM properties in environments other than those probed by NSs and terrestrial nuclear experiments. In the following we shall elaborate on models with nucleonic degrees of freedom, relevant for the purpose of our present work. Most of the studies performed so far employ non-relativistic approaches and adopt  for the energy density functional different solutions: Taylor expansion with respect to deviations from saturation and isospin symmetry \cite{Zhang_ApJ_2019,Ferreira_PRD_2020,Patra_PRD_2022}; semiagnostic metamodelling \cite{Guven_PRC_2020}; a $\chi$EFT inspired expansion in terms of neutrons and protons Fermi momenta \cite{Lim_EPJA_2019}. A technical issue that has to be checked is whether causality is respected at least up to the density corresponding to the central value of the most massive configuration. Very recently generalizations of covariant density functionals (CDF), very successful in describing NS matter~\cite{Dutra_RMF_PRC_2014,Fortin_PRC_2016}, have been proposed \cite{Ghosh_EPJA_2022,Malik_ApJ_2022,Malik_RMF_NL_2023}. In Ref.~\cite{Ghosh_EPJA_2022} the authors make use of a model with non-linear couplings and implement a frequentist approach; Ref.~\cite{Malik_ApJ_2022} employs a density dependent (DD) parametrization and performs a Bayesian analysis. Bayesian analyses of CDF EOS are undertaken also in Ref.~\cite{Malik_RMF_NL_2023}, but this time starting from models with non-linear couplings. The conclusions reached by the authors of Ref.~\cite{Malik_ApJ_2022} include the following: for the assumed functional of nucleon interactions the maximum NS mass can be as large as $2.5\,\Msun$; in agreement with the predictions of standard DD CDFs like DD2~\cite{DD2} and DDME2~\cite{Lalazissis_PRC_2005}, nucleonic direct URCA does not operate in stable stars; maximum value of the speed of sound is about $\sqrt{2/3}c$; at 90\% credible interval (CI) the pressure as a function of number density in pure neutron matter (PNM) and in NS matter agrees with corresponding ``bands'' obtained by $\chi$EFT calculations of Hebeler at al.~\cite{Hebeler_ApJ_2013}, used as one of the constraints, and with the pressure distribution conditioned on GW170817 from Ref.~\cite{Abbott_PRL_121}, respectively. The first consensus testifies the ability of DD CDF models to comply with the low-density EOS of PNM constrained by ab initio models. The second one demonstrates the compatibility between the low density behavior, fine tuned on $\chi$EFT, the $2\,\Msun$ lower limit, imposed in Ref.~\cite{Malik_ApJ_2022} but disregarded in \cite{Abbott_PRL_121}, and the tidal deformability in GW170817. Regarding the agreement of the joint probability distribution of the NSs masses and radii of Ref.~\cite{Malik_ApJ_2022} with astrophysics data, the following observations are in order. The 50\% credible region (CR) of marginalized posteriors for the mass and radius obtained in the GW170817 analysis of each NS in the binary lie within the 90\% CR domain of the posterior $M-R$ distribution of Ref.~\cite{Abbott_PRL_121}. The latter also agrees with constraints from the analysis of the millisecond pulsar PSR J0030+0451 NICER X-ray data~\cite{Miller_2019,Riley_2019} and data corresponding to PSR J0740+6620~\cite{Miller_may2021}.

The aim of our work is twofold. The first is to investigate the effectiveness of progressive incorporation of various constraints available from \emph{ab initio} calculations, e.g., pressure and energy per particle of PNM, and nuclear experiments, e.g., effective mass, on determining NS EOS. The second is to search for correlations between the various parameters of the model, properties of NM and properties of NSs. The study is performed within a modified version of the model proposed by Malik et al.~\cite{Malik_ApJ_2022}. The analysis is performed by means of Bayesian inference. Posterior distributions are built upon various sets of constraints. The paper is organized as follows. The DD CDF model is presented in Sect.~\ref{ssec:Model}; constraints are discussed in Sect.~\ref{ssec:Constraints} and Markov Chain Monte Carlo (MCMC) procedure we employ is described in Sect.~\ref{ssec:MCMC}. The results are presented in Sect.~\ref{sec:Results} and conclusions in Sect.~\ref{sec:Conclusion}.

\section{Setup}
\label{sec:Setup}

Homogeneous matter in the core of a NS is treated within a modified version of the DD CDF model proposed in Ref.~\cite{Malik_ApJ_2022}, which we discuss below. In the leptonic sector both electrons and muons are accounted for. For the outer and inner crusts we adopt the models by Haensel, Zdunik and Dobaczewski~\cite{HDZ_1989} and Negele and Vautherin~\cite{NV_1973}, respectively. Crust and core EOSs are smoothly matched at the arbitrary value of $n_\mathrm{B} = 0.08~\rm{fm}^{-3}$, which means that EOS dependence of the crust-core transition density is disregarded. Alteration of intermediate mass NS radii by up to a few percents \cite{Fortin_PRC_2016} will nonetheless be innocuous for our study as no radius-related constraints are imposed in our statistical procedure.

\subsection{The model}
\label{ssec:Model}

CDF or, equivalently, relativistic mean field (RMF) models have been successfully used to describe infinite NM, hot and cold stellar matter and atomic nuclei. They rely on the Walecka model, which treats particles as fundamental degrees of freedom that interact with each other through the exchange of scalar and vector mesons. In the simplest, but still realistic, case three mesons exist:
the scalar-isoscalar $(\sigma)$, responsible for long-range attraction;
the vector-isoscalar $(\omega)$, responsible for short-range repulsion;
the vector-isovector $(\rho)$, which distinguishes between neutrons and protons and introduces the isospin symmetry.
In spite of the quantum field formulation, CDF models are phenomenological as the strength functions are fine tuned such as accord with experimental data is obtained.

In this work we adopt the model proposed by Malik et al.~\cite{Malik_ApJ_2022}; for a detailed description, the interested reader is referred to the original paper. Similarly to standard DD CDF models like DD2 \cite{DD2} and DDME2 \cite{Lalazissis_PRC_2005}, this model assumes that nucleon-meson couplings are density dependent and expresses them in terms of the couplings values at saturation, $\Gamma_{M,0}$,
which introduces the generic notation
\begin{equation}
  \Gamma_M(n)=\Gamma_{M,0} h_M(x),
 \label{eq:hM}  
\end{equation}
where $M=\sigma,\omega,\rho$; $x=n/n_{\mathrm{sat}}$ is the normalized particle density and $h_M(x)$ accounts for the density variation of the strength functions. For the isovector field the form of refs.~\cite{DD2,Lalazissis_PRC_2005} is assumed,
\begin{equation}
  h_{\rho}(n)=\exp\left[-a_{\rho}\left( x-1\right)\right],
  \label{eq:hiv}
\end{equation}
while for the isoscalar fields $M=\sigma,\omega$ a simplified dependence is postulated,
\begin{equation}
  h_M(n)=\exp\left[-\left(x^{a_M}-1 \right) \right].
  \label{eq:his}
\end{equation}
This simplification is meant to limit the number of parameters of the model and, thus, ease the survey~\cite{Malik_ApJ_2022}. Overall, as follows from Eqs.~\eqref{eq:hM},\eqref{eq:hiv} and \eqref{eq:his}, we have six parameters that \emph{completely} determine the model: $\{\Gamma_{\sigma,0}, \Gamma_{\omega,0}, \Gamma_{\rho,0}, a_{\sigma},a_{\omega},a_{\rho}\}$. We will call them (model) input parameters.

Given that the saturation point is determined by the balance between attraction, mediated by $\sigma$, and repulsion, mediated by $\omega$, there is a strong interdependence between $n_{\mathrm{sat}}$, which enters eq. (\ref{eq:hM}), on the one hand and $\{\Gamma_{\sigma}, \Gamma_{\omega}, a_{\sigma}, a_{\omega} \}$ on the other hand. We account for this situation by self-consistently computing $n_{\mathrm{sat}}$ for each set of the model parameters. This does not, nonetheless, guarantee that the saturation point exists in a density domain of arbitrarily-chosen size and even less that it corresponds to the actual saturation density of symmetric nuclear matter (SNM). If it does not, the sampler rejects such a set of input parameters, see details in Sect.~\ref{ssec:MCMC}. In Ref.~\cite{Malik_ApJ_2022} another strategy was employed~\footnote{T. Malik, C. Provid\^encia, private discussion.}. It consists in treating $n_{\mathrm{sat}}$ as a free parameter and then checking whether it complies (within a tolerance) with the definition of the saturation point, mathematically given by $P(n_{\mathrm{sat}})=0$, where $P$ stands for pressure. The latter condition was implemented in the likelihood, where the constraint on $P(\sat{n})$ was treated  as a Gaussian with zero mean and small standard deviation (SD). We have explicitly checked that small discrepancies between the results of our work and those of Ref.~\cite{Malik_ApJ_2022} are exclusively due to this feature.

\subsection{Constraints}
\label{ssec:Constraints}

In this section we discuss the various types of constraints posed on our EOS models; their domains of values; the different sets of constraints implemented in each of the six cases we considered in this work. 

\renewcommand{\arraystretch}{1.05}
\setlength{\tabcolsep}{4.0pt}
\begin{table}
	\caption{Constraints posed on EOS models. For each quantity, we provide the median value and the standard deviation. For $M_{\mathrm{G}}^*$, we specify the threshold value. Constraints from SNM and NS are identical to those in Ref.~\cite{Malik_ApJ_2022}; $\left(E/A\right)_i$ and $P_i$, with $i=1,2,3$, stand for energy per particle and pressure of PNM at the densities of 0.08, 0.12 and 0.16 $\mathrm{fm}^{-3}$; standard deviations of $\left(E/A\right)_i$ and $P_i$ in PNM have been extracted from $\chi$EFT calculations of Ref.~\cite{Hebeler_ApJ_2013}. $\eff{m}$ corresponds to the Dirac effective mass of the nucleon at $\sat{n}$; $m_\mathrm{N}$ represents the nucleon mass, for which the value of 939.0 MeV is considered here.}
	\begin{tabular}{lcccc}
		\toprule
		\toprule
    	Quantity           & Units                 & Value   & Std. deviation & Ref. \\
    	\midrule
    	$n_{\mathrm{sat}}$ & $\mathrm{fm}^{-3}$    & 0.153 	 & 0.005 & \cite{Typel_NPA_1999}     \\
    	$E_{\mathrm{sat}}$ & MeV 			   	   & $-16.1$ & 0.2 	 & \cite{Dutra_RMF_PRC_2014} \\
    	$K_{\mathrm{sat}}$ & MeV 			       & 230 	 & 40 	 & \cite{Todd_PRL_2005,Shlomo_EPJA_2006} \\ 
    	$J_{\mathrm{sym}}$ & MeV 			   	   & 32.5    & 1.8   & \cite{Essick_PRC_2021}  \\
    	$\eff{m}$          & $m_\mathrm{N}$        & 0.55    & 0.05  & \cite{Typel_PRC_2005} \\
    	$\left(E/A\right)_1$ 	  		   & MeV 		 		   &  9.50   & 0.52  & \cite{Hebeler_ApJ_2013} \\
    	$\left(E/A\right)_2$ 	  		   & MeV 			   	   & 12.68   & 1.20  & \cite{Hebeler_ApJ_2013} \\
    	$\left(E/A\right)_3$ 	  		   & MeV 			   	   & 16.31   & 2.13  & \cite{Hebeler_ApJ_2013} \\
    	$P_1$ 	  		   & MeV/$\mathrm{fm}^{3}$ & 0.509   & 0.093 & \cite{Hebeler_ApJ_2013} \\
    	$P_2$ 	  		   & MeV/$\mathrm{fm}^{3}$ & 1.238   & 0.302 & \cite{Hebeler_ApJ_2013} \\
    	$P_3$ 	 		   & MeV/$\mathrm{fm}^{3}$ & 2.482   & 0.687 & \cite{Hebeler_ApJ_2013} \\
    	$M_{\mathrm{G}}^*$ & $\Msun$ 			   & $>2.0$  &  ---  & \cite{Fonseca_2021}     \\
    	\bottomrule
    	\bottomrule
	\end{tabular}
  \label{tab:constraints}
\end{table}
\setlength{\tabcolsep}{2.0pt}
\renewcommand{\arraystretch}{1.0}

{\em Nuclear matter.} 
EOS of cold NM links the energy per nucleon, $E/A$, to neutron ($n_\mathrm{n}$) and ($n_\mathrm{p}$) proton number densities. Experimental data collected in terrestrial laboratories probe only a limited domain around the saturation density, defined as the minimum of $E(n_{\mathrm{B}})/A$, which is equivalent to $P=\left[n_{\mathrm{B}}^2 \partial \left(E/A \right)/\partial n_{\mathrm{B}}\right]|_{n_{\mathrm{B}}=n_{\mathrm{sat}}}=0$, and isospin asymmetry, $\delta=\left(n_\mathrm{n}-n_\mathrm{p}\right)/n_{\mathrm{B}}=1-2Y_\mathrm{p} \approx 0$. Here $Y_\mathrm{p}=n_\mathrm{p}/n_{\mathrm{B}}$ denotes the proton fraction and $n_{\mathrm{B}}=n_\mathrm{n}+n_\mathrm{p}$. As such, $E(n_{\mathrm{B}},\delta)/A$ is customarily split into an isoscalar and an isovector term,
\begin{equation}
  E(n_{\mathrm{B}},\delta)/A=E_0(n_{\mathrm{B}},0)+\delta^2 E_{\mathrm{sym}}(n_{\mathrm{B}},0),
  \label{eq:split} 
\end{equation}
that are further Taylor expanded with respect to deviation from saturation $\mathcal{X}=\left(n_{\mathrm{B}}-n_{\mathrm{sat}}\right)/3n_{\mathrm{sat}}$,
\begin{align}
  E_0(n_{\mathrm{B}},0) &=& \sum_{i=0,1,2,...} \frac1{i!} X_{\mathrm{sat}}^{(i)} \mathcal{X}^i, 
  \label{eq:Taylor_is}\\
  E_{\mathrm{sym}}(n_{\mathrm{B}},0) &=& \sum_{j=0,1,2,...} \frac1{j!} X_{\mathrm{sym}}^{(j)} \mathcal{X}^j,
  \label{eq:Taylor_iv}
\end{align}
with $X_{\mathrm{sat}}^{(i)}=\left(\partial^i E_0(n_{\mathrm{B}},0)/\partial \mathcal{X}^{(i)}\right)|_{n_{\mathrm{B}}=n_{\mathrm{sat}}}$ and $X_{\mathrm{sym}}^{(j)}=\left(\partial^j E_{\mathrm{sym}}(n_{\mathrm{B}},0) / \partial \mathcal{X}^{(j)}\right)|_{n_{\mathrm{B}}=n_{\mathrm{sat}}}$.

Out of the different coefficients $X_{\mathrm{sat}}^{(i)}$ and $X_{\mathrm{sym}}^{(j)}$ only the low order ones are known accurately \cite{Margueron_PRC_2018}. They are the energy per nucleon of symmetric saturated matter $E_{\mathrm{sat}}=X_{\mathrm{sat}}^{(0)}$, the compressibility modulus $K_{\mathrm{sat}}=X_{\mathrm{sat}}^{(2)}$ and symmetry energy at saturation $J_{\mathrm{sym}}=X_{\mathrm{sym}}^{(0)}$.

Similarly to Ref.~\cite{Malik_ApJ_2022} we shall constrain the EOS models on $n_{\mathrm{sat}}$, $E_{\mathrm{sat}}$, $K_{\mathrm{sat}}$ and  $J_{\mathrm{sym}}$ only. The same domains as in Ref.~\cite{Malik_ApJ_2022} will be used. They are specified in Table~\ref{tab:constraints}.

\emph{Pure neutron matter} (PNM).
A well established microscopic approach to describe nuclear forces at low energies is $\chi$EFT with nucleon and pion degrees of freedom, where the hierarchy of two (NN), three (3N) and weaker higher-body forces is explained and theoretical uncertainties are estimated. The structure of 3N interactions makes PNM particularly suitable for $\chi$EFT studies. In this work estimates of the energy per particle and pressure at the densities of 0.08, 0.12, 0.16~${\rm fm}^{-3}$, calculated by Hebeler et al.~\cite{Hebeler_ApJ_2013}, are used as constraints. For the median values and SDs, see Table~\ref{tab:constraints}. The uncertainty band results mainly from variations of the couplings involved in three-nucleon interactions. Calculations with constraints on the pressure and a double compared with Ref.~\cite{Hebeler_ApJ_2013} SDs were performed only for comparison with Ref.~\cite{Malik_ApJ_2022}. With one exception, the rest of the calculations use the original SDs both for energy per particle and pressure. We nevertheless note that artificial increase of SDs can be justified on account of differences between the results of $\chi$EFT calculations performed by various groups, including those considered here; see Fig.~6 of Ref.~\cite{Raaijmakers_may2021} for the behavior of $P(n_\mathrm{B})$ up to $\approx 1.1\,n_{\mathrm{sat}}$. Finally, in one of our calculations constraints from PNM will be completely disregarded. Comparison with results of other calculations will disclose the constraining effectiveness of these data.

Notice that implementation of constraints from $\chi$EFT in Bayesian inferences of dense matter EOS in literature does not always imply that conditions are imposed on both $P$ and $E/A$. For instance, only constraints on pressure are implemented in Ref.~\cite{Malik_ApJ_2022}, while Ref.~\cite{Hebeler_ApJ_2013} implements constraints from both $P$ and $E/A$.

{\em Effective Mass.} In RMF models the single particle energy spectrum is defined by scalar and vector self-energies. The scalar self-energy enters the definition of the Dirac effective mass $\eff{m}=m-\Sigma_S$. It is defined as the sum over all scalar fields of the products between the mean-field value of the meson field and the corresponding coupling to nucleons. In the simple case considered here, where only the scalar-isoscalar $\sigma$ meson is accounted for, it reduces to $\Sigma_S=\Gamma_{\sigma} \sigma$. The tight correlation between $\Sigma_S$ and the energy splitting between single-nucleon levels of identical orbital angular momentum but different total angular momentum allows to determine the Dirac effective mass in SNM at saturation. For most DD CDF the value extracted this way is $\eff{m} \approx 0.55\,m_\mathrm{N}$ \cite{Typel_PRC_2005}. We nevertheless note that this value leads to Landau effective masses in tension with those extracted from experimental data on level densities~\cite{Blaizot_NPA_1981} and giant resonances~\cite{Bonasera_PRC_2018}. According to Refs.~\cite{Typel_PRC_2005,Typel_EPJA_2020} reconciliation between energy spacing between spin-orbit partner states in finite nuclei and level density requires more sophisticated couplings, e.g., tensor forces, which are beyond the scope of present study. Constraints on $\eff{m}$ will be imposed in order to check on the one hand the compatibility between the posterior distributions built upon the DD CDF model introduced in Sect.~\ref{ssec:Model} and spectroscopic data. On the other hand, this will allow us to test the effectiveness of constraints on $\eff{m}$. The SD is arbitrarily taken equal to $0.05\,m_\mathrm{N}$. 

{\em Neutron Stars.} Only the condition on the lower bound of maximum NS mass $M_{\mathrm{G}}^*$ is implemented. In most cases the conservative value of $2\,\Msun$ is assumed. It roughly corresponds to the central value of PSR J0348+0432 mass at 68.3\% CI \cite{Antoniadis2013} and the lower limit of MSP J0740+6620 mass with 68.3\% CI \cite{Fonseca_2021}. In one situation a slightly larger value ($2.2\,\Msun$) is used. This will allow us to investigate the sensitivity to the possible future discovery of NS of higher mass and the potency of this constraint.

Table~\ref{tab:runs} lists the different choices done in regards to the constraints from $\chi$EFT calculations of PNM and lower limit of $M_{\mathrm{G}}^*$. The constraints on $\sat{n}$, $\sat{E}$, $\sat{K}$, $\sym{J}$ are the same for all runs, the considered values being specified in Table~\ref{tab:constraints}. Most of the results discussed in Sect.~\ref{sec:Results} correspond to run~5 that makes up our fiducial model.

\renewcommand{\arraystretch}{1.05}
\setlength{\tabcolsep}{6.0pt}
\begin{table}
	\caption{Sets of constraints, other than those on $\sat{n}$, $\sat{E}$, $\sat{K}$, $\sym{J}$, considered in our runs. $n \times X_\mathrm{PNM}$, with $X=P, (E/A)$, stays for the input from $\chi$EFT calculations of PNM at next-to-next-to-next-to leading order ($\mathrm{N^3LO}$) whose SDs have been augmented by a factor of $n$. Run~1 is equivalent to the situation considered in Ref.~\cite{Malik_ApJ_2022}. Run~5 constitutes our fiducial case.}
	\begin{tabular}{lcccc}
		\toprule
		\toprule
    	Run            &    $P_\mathrm{PNM}$        & $\left(E/A\right)_\mathrm{PNM}$ & $m_\mathrm{eff}$  &   $M_{\mathrm{G}}^*$ \\
    	\midrule
    	0            &       ---                  &     ---                     &  ---       &    2.0            \\
        1            & $2 \times \mathrm{N^3LO}$  &     ---                     &  ---       &       2.0            \\
        2            & $1 \times \mathrm{N^3LO}$  &     ---                     &  ---       &       2.0            \\
        3            & $1 \times \mathrm{N^3LO}$  &     ---                     &  ---       &       2.2            \\
        4            &     ---                    & $1 \times \mathrm{N^3LO}$   &  ---       &       2.0            \\
        5            & $1 \times \mathrm{N^3LO}$  & $1 \times \mathrm{N^3LO}$   &  ---       &       2.0            \\
        6            & $1 \times \mathrm{N^3LO}$  &     ---                     & \checkmark &       2.0            \\
  	\bottomrule
    \bottomrule
	\end{tabular}
  \label{tab:runs}
\end{table}
\setlength{\tabcolsep}{2.0pt}
\renewcommand{\arraystretch}{1.0}

\subsection{Markov Chain Monte Carlo}
\label{ssec:MCMC}

\renewcommand{\arraystretch}{1.05}
\setlength{\tabcolsep}{10.0pt}
\begin{table}
	\caption{Domains of values for the effective interaction parameters (i.e., input parameters), taken from Ref.~\cite{Malik_ApJ_2022}.}
	\begin{tabular}{cccc}
		\toprule
		\toprule
    	Parameter          & Units & Min.   & Max. \\
    	\midrule
  $\Gamma_{\sigma,0}$ & MeV & 7.5 & 13.5 \\
  $\Gamma_{\omega,0}$ & MeV & 8.5 & 14.5 \\
  $\Gamma_{\rho,0}$   & MeV & 2.5 & 8.0 \\
  $a_{\sigma}$        &  -- & 0.0 & 0.30 \\
  $a_{\omega}$        &  -- & 0.0 & 0.30 \\
  $a_{\rho}$          &  -- & 0.0 & 1.30 \\
    	\bottomrule
    	\bottomrule
	\end{tabular}
  \label{tab:inputdomains}
\end{table}
\setlength{\tabcolsep}{2.0pt}
\renewcommand{\arraystretch}{1.0}

The input parameters of the model, $\mathbf{\Theta}$, are the six parameters of the effective interaction introduced in Sect.~\ref{ssec:Model}, i.e., $\{\Gamma_{\sigma,0}, \Gamma_{\omega,0},\Gamma_{\rho,0}, a_{\sigma}, a_{\omega}, a_{\rho} \}$. Their domains of values, identical to those used in Ref.~\cite{Malik_ApJ_2022}, are given in Table~\ref{tab:inputdomains}. Their prior distributions are considered uniform (``uninformative''), as previously done in Ref.~\cite{Malik_ApJ_2022}.

The posterior distributions of different quantities, including the input parameters, are subjected to various sets of  constraints discussed in Sect.~\ref{ssec:Constraints}. With the exception of the lower limit of $M_{\mathrm{G}}^*$, which can be regarded as an ``experimental'' information, all these constraints involve sophisticated theoretical modeling. As such, a series of questions arise. The first one is to what extent it is reasonable to assume that these quantities are uncorrelated. The standard procedure to extract the values of $\sat{n}$ and $\sat{E}$ assumes either processing of experimental charge radii and nuclear binding energies within the liquid drop model or, alternatively, statistical analysis of empirical nuclear energy-density functionals calibrated to the above mentioned data. Simultaneous derivation, within a model, of two quantities from a data set is sufficient to expect correlations among their values. The confirmation is provided by the values obtained within different categories of models in the literature, for a compilation see Table~I in Ref.~\cite{Margueron_PRC_2018}.

The situation when the values of state variables are obtained from a function at different values of its argument, as it is the case of $\{\left(E/A\right)_i\}$ and $\{P_i\}$ with $i=1,...,3$ in PNM, is another situation where correlations are expected. This issue was discussed at length in Ref.~\cite{Somasundaram_2021} (in Supplemental materials), where a method to account for such correlations has been proposed. It consists in supplementing the likelihood (see below) with a term proportional to the inverse of the covariance matrix. Its applicability assumes access to the data sets from which the variances of the presumably correlated data have been calculated. This is obviously not the case of the data in Ref.~\cite{Hebeler_ApJ_2013}, where only uncertainty ``bands'' are provided. As such, we shall assume here that the various constraints on energy per baryon and/or pressure are independent. More detailed investigations on such correlations, that we anticipate to be stronger between neighboring points, and their consequences will be published elsewhere. 

Another question regards the interpretation of uncertainties in Table~\ref{tab:constraints} for which two situations are envisaged. The first one corresponds to $\sat{n}$, $\sat{E}$, $\sat{K}$ and $\sym{J}$ whose uncertainties can be interpreted as SDs of a normally distributed variables. The reason is that these quantities are commonly extracted from statistical analysis of distributions provided by collections of phenomenological models, for a review see Ref.~\cite{Margueron_PRC_2018}. The second situation corresponds to the predictions of models that, ideally, account for all uncertainties. One could claim that quantities $X_i$ in PNM calculated by $\chi$EFT should fall into this category. If this is the case, compliance with the uncertainty ``band'' $[X_\mathrm{min}, X_\mathrm{max}]$ should be enforced, which means that all models that provide for at least one of these quantities values outside the above domain should be rejected. In this work we shall not adopt this strategy.
In default of more detailed data and for the sake of simplicity in this work all constraints will be treated on the same footing. To be more specific, we assume that no correlation exists among the various constraints and that all data are normally distributed. Though not explicitly stated, the same working hypotheses were employed in Ref.~\cite{Malik_ApJ_2022}. 

\looseness=-1
Given the model parameters $\mathbf{\Theta}$ and the constraints $\mathbf{D}$, the likelihood is given by:
\begin{align}
	\mathcal{L}_q \left(\mathbf{\Theta} | \mathbf{D} \right) = \prod_{i=1}^{N_q} \frac{1}{\sqrt{2 \pi}\sigma_i} \exp{\left[ -\frac{1}{2} \left( \frac{d_i - \xi_i(\mathbf{\Theta}) }{\sigma_i} \right)^2 \right]},
	\label{eq:Likelihood}
\end{align}
where $q$ labels the ``run'' under consideration; the index $i$ runs over all the constraints other than the one corresponding to the lower limit of the maximum mass; the number of these constraints is $N_q$; $d_i$ and $\sigma_i$ denote the constraint and its SD reported in Table~\ref{tab:constraints}; $\xi_i(\mathbf{\Theta})$ corresponds to the value the model defined by the parameter set $\mathbf{\Theta}$ provides for the quantity $i$. For the implementation of the $M_{\mathrm{G}}^*$-related constraint and the solution adopted when $\sat{n}$ falls out of the density interval where it is sought for, see below.

For sampling from the posterior distribution, as it is the case here, normalization factors are not relevant. This means that the overall normalization factor in the Bayes theorem (a.k.a evidence or Bayes factor) as well as the normalization in the likelihood function \eqref{eq:Likelihood} can be safely disregarded. Out of the different available techniques for Bayesian inference and sampling from the posterior distributions, we have chosen the \texttt{emcee} package~\footnote{The documentation is available at \url{https://emcee.readthedocs.io} and the github page \url{https://github.com/dfm/emcee}} \cite{emcee,emcee3}. \texttt{Emcee} is a pure-Python implementation of affine invariant MCMC \cite{Goodman_2010}.

Similar to other samplers, \texttt{emcee} works with the logarithm of the likelihood. This feature reduces the likelihood, Eq.~\eqref{eq:Likelihood}, to the following non-normalized log-likelihood:
\begin{align}
	\log \mathcal{L}_q \propto  -\frac{1}{2} \sum_{i=1}^{N_q}  \left(\frac{d_i - \xi_i(\mathbf{\Theta}) } {\sigma_i} \right)^2 = -\chi_q^2.
	\label{eq:Chi2}
\end{align}
The constraint on the lower limit of the maximum mass, not accounted for in the above summations, is treated as a ``hard wall''. More precisely, if -- for the given set of model parameters -- the value of $M_\mathrm{G}^*$ exceeds the threshold value of $2\,\Msun ~(2.2\,\Msun)$, the value of $\chi^2$ [Eq.~\ref{eq:Chi2}] is not modified. Otherwise, $\chi^2$ is set to $10^6$. Given that the typical values of $\chi^2$ are $\approx 1$ and the maximum ones are around 15, setting $\chi^2$ to $10^6$ makes the sampler reject the trial set from the Markov chain. This is equivalent to an impenetrable ``wall''. The same strategy is adopted for the situation when $\sat{n}$, self-consistently calculated (by fixed point iterations) for every set $\mathbf{\Theta}=\{\Gamma_{\sigma,0}, \Gamma_{\omega,0}, \Gamma_{\rho,0}, a_{\sigma},a_{\omega},a_{\rho}\}$, falls out of the (arbitrary) domain $0.01 \leq n_\mathrm{B} \leq 0.3~{\rm fm^{-3}}$. Note that in Ref.~\cite{Malik_ApJ_2022} the saturation is enforced by supplementing the likelihood function \ref{eq:Likelihood} by the factor $1/\sqrt{2 \pi 0.01^2} \exp \left[ - \left( P\left(\mathbf{\Theta'}\right)/0.01\right)^2 / 2\right]$, where $P$ stays for the pressure of SNM calculated at $\sat{n}$, treated as a free parameter. Here $\mathbf{\Theta'}=\{ \mathbf{\Theta}, \sat{n}\}$. By favoring parameter sets whose $\sat{n}$ complies with the saturation definition, this extra constraint naturally selects consistent EOS models. This procedure nevertheless leads to posteriors which differ from those produced by our procedure. More details will become obvious in Sect.~\ref{sec:Results}.

For the success of the MCMC method it is crucial to be sure that the sampler has actually converged to the posterior distribution and that one has enough independent samples of the posterior. We have performed several different tests of the convergence. First of all, we have calculated the integrated autocorrelation time $\tau$ of the Markov chain for each of the six input parameters. In the following we will refer to the longest autocorrelation time among those six parameters. \texttt{Emcee} documentation recommends that the length of the chain $L$ should be at least $50 \tau$. We have run the chains up to  $\approx 200 \tau$. Taking into account that we have used 400 walkers, that means that we have  $\approx 80\,000$ \emph{independent} samples from the posterior; in total we have 1\,000\,000 samples for each run. Apart from that, we made sure to keep the acceptance fraction within the recommended range ($0.2-0.5$) and we have disregarded the first $\approx 10 \tau$ steps of the chain as a burn-in (documentation recommends to disregard at least $(2-3) \tau$ steps as a burn-in).

As the second important test of convergence we performed ``bootstrap'' tests of stability of the posterior distributions. To this aim, the histograms of the model parameters for consecutive slices of 200 chain steps (which is roughly 20 autocorrelation times since $\tau \approx 10$) have been plotted and the 10\%, 50\% and 90\% quantiles have been computed. The values of the corresponding quantiles varied by less than 1\% between different slices of the chain and these variations were random; there were no systematic drifts. Visual inspection has also demonstrated no signs of any drift of the histograms. 

Taking into account that there were no indications of multimodal distributions and some additional independent comparisons (see below), we consider that the above-mentioned tests provide sufficient evidence of proper convergence of the sampler. We have not performed Gelman-Rubin test \cite{Gelman_1992} because, as argued in \texttt{emcee} documentation, it is designed for independent runs of the same model, and walkers in the same run are not independent.

Still, to additionally validate the results, posteriors corresponding to the fiducial run (5 in Table~\ref{tab:runs}) have been calculated also using a dynamic nested sampling method \cite{Skilling_2004,Skilling_2006,Higson_2019}, implemented in the Python package \texttt{dynesty} \footnote{The documentation is available at \url{https://dynesty.readthedocs.io} and the github page is \url{https://github.com/joshspeagle/dynesty}} \cite{dynesty,dynesty2}. Since we were not interested in computing the evidence, we have performed dynamic nested sampling optimized for obtaining posterior distributions. Nested sampling, unlike MCMC (without tempered chains), can ``naturally'' deal with multimodal distributions provided that there are enough live points. \texttt{Dynesty} documentation recommends to start with at least 250 live points and add at least another 50 for each supposed mode of a multimodal distribution. We employed 3\,000 initial live points with no limit on adding additional batches of live points should the sampler require it. 

After $\approx 750\,000$ likelihood evaluations, we ended up with $\approx 33\,000$ posterior samples.
The results were very close to the results of our MCMC run. The relative difference in the quantiles of the model parameters did not exceed 1\% and was typically even less (we tested 5\%, 50\% and 95\% quantiles). This difference is comparable with the differences that we observed in the ``bootstrap'' test. Thus, we can conclude that 
(i) our MCMC results are reliable, the sampler has definitely converged
and
(ii) the statistical uncertainty of the computed quantiles does not exceed 1\%, which is much less than the widths of the posterior distributions (for the latter see Sect.~\ref{sec:Results}).  

Note that during the calculations no constraints were imposed to ensure the thermodynamic stability of NS EOS (which translates into $dP/dn_\mathrm{B}>0$ for any $n_\mathrm{B}$), but rather this was checked afterwards by analyzing the posterior. It came out that: i) out of $\approx 100\,000$ independent posterior samples from run~5 there was not a single sample that presented such instabilities and ii) out of the $\approx 95\,000$ independent samples from run~0 there was only one that showed a tiny  instability over a narrow density domain around $\sat{n}/2$. Given that runs~5 and 0 are our most and least constrained runs, respectively, we conclude that NS EOS models built in this work do not present mechanical instabilities.


\section{Results}
\label{sec:Results}

%
\begin{figure}
	\includegraphics[]{"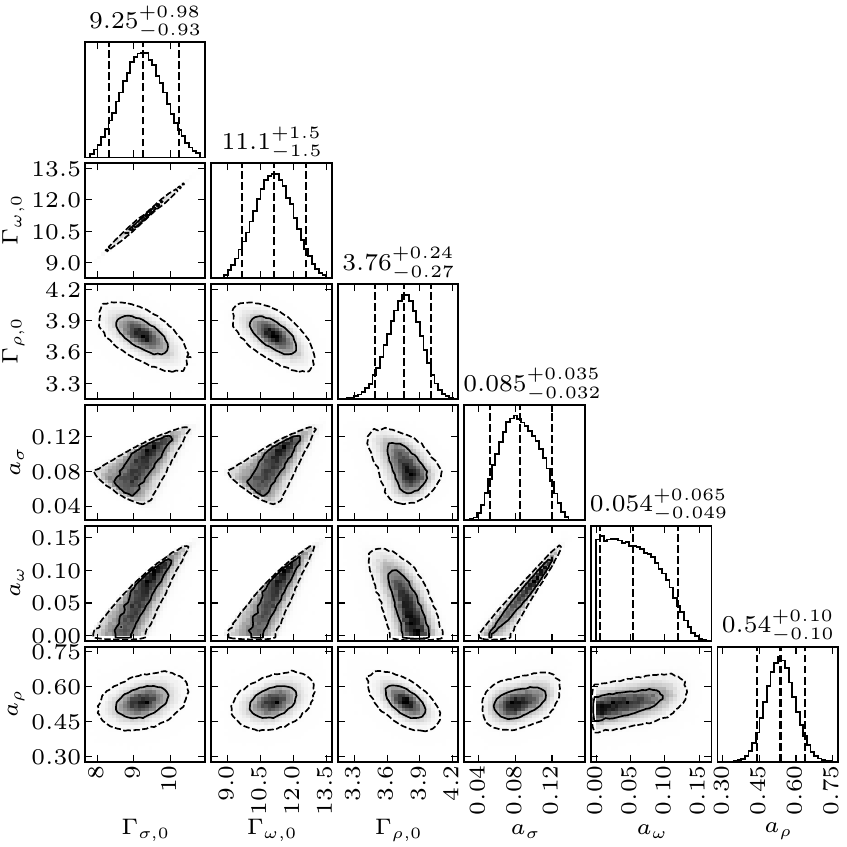"}
	\caption{1D and 2D marginalized posterior PDFs of the input parameters of our DD CDF model. $\Gamma_{M,0}$ are expressed in MeV; $a_M$ are dimensionless ($M=\sigma$, $\omega$, $\rho$). Dashed vertical lines in 1D histograms correspond to 5\%, 50\% and 95\% quantiles; thus, the values on the top of 1D histograms denote the median and its 90\% CI. The solid and the dashed contours on 2D histograms show 50\% and 90\% CR, respectively. Results correspond to run~5 in Table~\ref{tab:runs}, our fiducial model.}
	\label{Fig:BasePar_Cor_1xP_1xE}
\end{figure}

In this section we shall first analyze the posterior probability distributions for our fiducial model (5 in Table~\ref{tab:runs}). Correlations among input parameters; parameters of the isoscalar and isovector channels of the nuclear EOS; selected properties of NSs will be discussed. Then, modifications brought by the various sets of constraints will be highlighted by confronting marginalized probability density functions (PDFs) corresponding to runs~0, 3, 5 and 6 in Table~\ref{tab:runs}. Marginalized PDFs of both input parameters and calculated parameters of NM, PNM and NS will be considered. Finally, correlations among all these quantities will be sought for. Two alternative methods will be used: 2D histograms and Kendall rank correlation coefficients.

\subsection{The fiducial case}
\label{ssec:case5}

We choose run~5 in Table~\ref{tab:runs} as our fiducial model. To be more explicit, the posteriors are conditioned on constraints on $\sat{n}$, $\sat{E}$, $\sat{K}$, $\sym{J}$; pressure ($P$) and energy per particle ($E/A$) in PNM at $n_\mathrm {B}=0.08,~0.12,~0.16~\mathrm{fm}^{-3}$ as calculated within the $\chi$EFT in Ref.~\cite{Hebeler_ApJ_2013}; $M_{\mathrm{G}}^* \geq 2 \Msun$. For the values of these quantities, see Table~\ref{tab:constraints}.

The corner plot of the input parameters of the model is provided in Fig.~\ref{Fig:BasePar_Cor_1xP_1xE}. This figure shows a strong correlation among the strengths of $\sigma$ and $\omega$ mesons at saturation and a somewhat weaker correlation among the parameters which regulate their density dependence. The strong correlation among $\Gamma_{\sigma,0}$ and $\Gamma_{\omega,0}$ is attributable to the strong constraints imposed on $\sat{n}$ and $\sat{E}$; the weaker correlation among $a_{\sigma}$ and $a_{\omega}$ is due to the weaker constraint imposed on $\sat{K}$. Weak correlations exist also among the other sets of parameters that govern the isoscalar channel as well as between $\left(\Gamma_{\sigma,0},\Gamma_{\rho,0}\right)$ and $\left(\Gamma_{\omega,0},\Gamma_{\rho,0}\right)$.  We note that, with the exception of $a_{\sigma}$ and $a_{\omega}$, the PDFs of the input parameters are (roughly) Gaussian. The qualitative features of 2D posterior PDFs in this figure resemble those of Fig.~1 in Ref.~\cite{Malik_ApJ_2022}. With the exception of the 1D PDFs of  $a_{\sigma}$ and $a_{\omega}$, the same holds for the other 1D posteriors. Differences with respect to Ref.~\cite{Malik_ApJ_2022} are nevertheless obvious in the values of the quantiles. They exclusively stem from the different handling of the saturation point, for details see the discussion in Sect.~\ref{ssec:MCMC}. 

\begin{figure}
	\includegraphics[]{"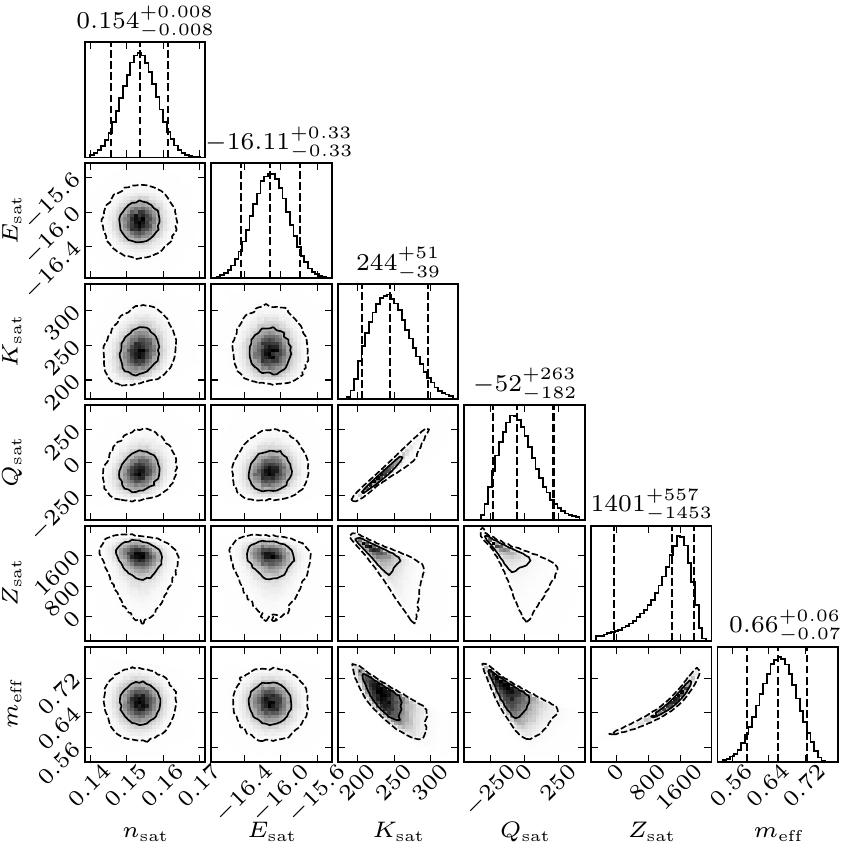"}
	\caption{1D and 2D marginalized posterior PDFs of the parameters describing the isoscalar behavior of the NM EOS; $\sat{E}$, $\sat{K}$, $\sat{Q}$ are expressed in MeV; $\sat{n}$ is expressed in $\mathrm{fm}^{-3}$. Information on the Dirac effective mass of the nucleon in symmetric saturated NM, expressed in $m_\mathrm{N}$, is also provided. Dashed vertical lines in 1D histograms correspond to 5\%, 50\% and 95\% quantiles; thus, the values on top of 1D histograms denote the median and its 90\% CI. The solid and the dashed contours on 2D histograms show 50\% and 90\% CR, respectively. Results correspond to run~5 in Table~\ref{tab:runs}, our fiducial model.}
	\label{Fig:DerPar_sat_Cor_1xP_1xE}
\end{figure}
\begin{figure}
	\includegraphics[]{"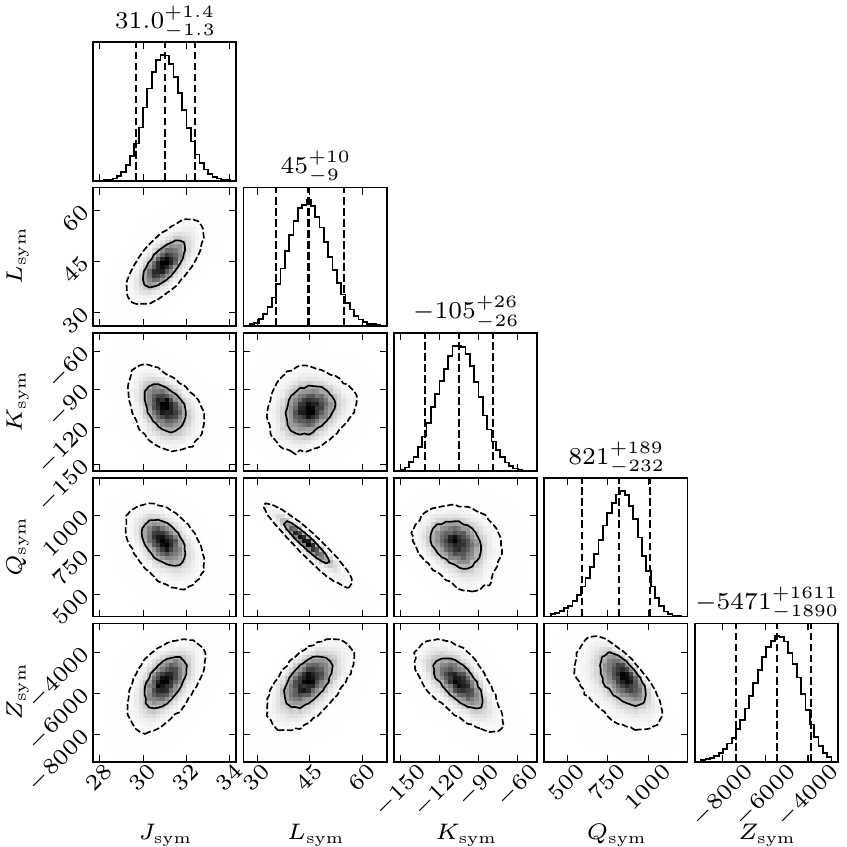"}
	\caption{1D and 2D marginalized posterior PDFs of the parameters describing the isovector behavior of the NM EOS. All quantities are expressed in MeV. Dashed vertical lines in 1D histograms correspond to 5\%, 50\% and 95\% quantiles; thus, the values on top of 1D histograms denote the median and its 90\% CI. The solid and the dashed contours on 2D histograms show 50\% and 90\% CR, respectively. Results correspond to run~5 in Table~\ref{tab:runs}, our fiducial model.}
	\label{Fig:DerPar_sym_Cor_1xP_1xE}
\end{figure}

We now proceed and analyze the posteriors of NM parameters. Also considered is the Dirac effective mass of the nucleon in symmetric saturated matter. As no correlations among isoscalar and isovector channels were identified, the properties of these channels will be discussed separately. 

Fig.~\ref{Fig:DerPar_sat_Cor_1xP_1xE} demonstrates the corner plot of the isoscalar parameters supplemented with information regarding $\eff{m}$. Only two strong correlations exist. The first one holds between the coefficients of second- and third-order terms in Eq.~\eqref{eq:Taylor_is}, $\sat{K}$ and $\sat{Q}$, and exists also in Ref.~\cite{Malik_ApJ_2022}. Linear correlations among these two quantities have been previously observed in the case of nonrelativistic Skyrme and Gogny interactions and explained by the presence of a single density-dependent term in the nuclear force ~\cite{Khan_PRC_2013}. Posterior distributions of a non-relativistic agnostic model conditioned on various sets of nuclear and astrophysical data and $\chi$EFT calculations also manifest a linear correlation between $\sat{K}$ and $\sat{Q}$ ~\cite{Guven_PRC_2020}, thus suggesting a more universal feature. The second correlation links $\sat{Z}$ and $\eff{m}$. Supposing that it is not an artefact of this model, it might help constraining the high-order parameter $\sat{Z}$. 

Other comments are as follows. The median value of $\eff{m}$ is significantly higher than in standard DD CDF models, as DD2 \cite{DD2}and DDME2 \cite{Lalazissis_PRC_2005}. As previously noted in relation with Fig.~\ref{Fig:BasePar_Cor_1xP_1xE} and for the same reason, some differences exist between our results and those in Ref.~\cite{Malik_ApJ_2022}. For example, we have obtained $\approx 13$~MeV higher median value of $\sat{K}$ than the one in Ref.~\cite{Malik_ApJ_2022}, which is related to our higher values of $M_\mathrm{G}^*$. 

The corner plot of the isovector parameters is provided in Fig.~\ref{Fig:DerPar_sym_Cor_1xP_1xE}. The only notable correlation one can see is the negative correlation between $\sym{L}$ and $\sym{Q}$. This result qualitatively agrees with that in Ref.~\cite{Malik_ApJ_2022} but quantitative differences are too important to be attributable to $\sat{n}$-related technicalities. The explanation for this situation resides in that -- with constraints on both pressure and energy per particle in PNM -- our fiducial model has more constraints in the isovector channel than the one used in Ref.~\cite{Malik_ApJ_2022}. Indeed, the latter corresponds to run~1 in Table~\ref{tab:runs}. Consequences of the number of constraints and/or SDs of the target parameters of PNM also explain i) why correlations among $\sym{L}$ and $\sym{Z}$, observed in Ref.~\cite{Malik_ApJ_2022}, are not present in Fig.~\ref{Fig:DerPar_sym_Cor_1xP_1xE} and ii) why all CI we report in Table~\ref{tab:Posteriors_1} for run~5 are by a factor of 2 (for low-order parameters) to 3 (for high-order parameters) smaller than those reported for run~1 in Table~\ref{tab:Posteriors_0}, which corresponds to the situation considered in Ref.~\cite{Malik_ApJ_2022}. For a more detailed discussion, see Sect.~\ref{ssec:runs}.

\begin{figure*}
	\includegraphics[]{"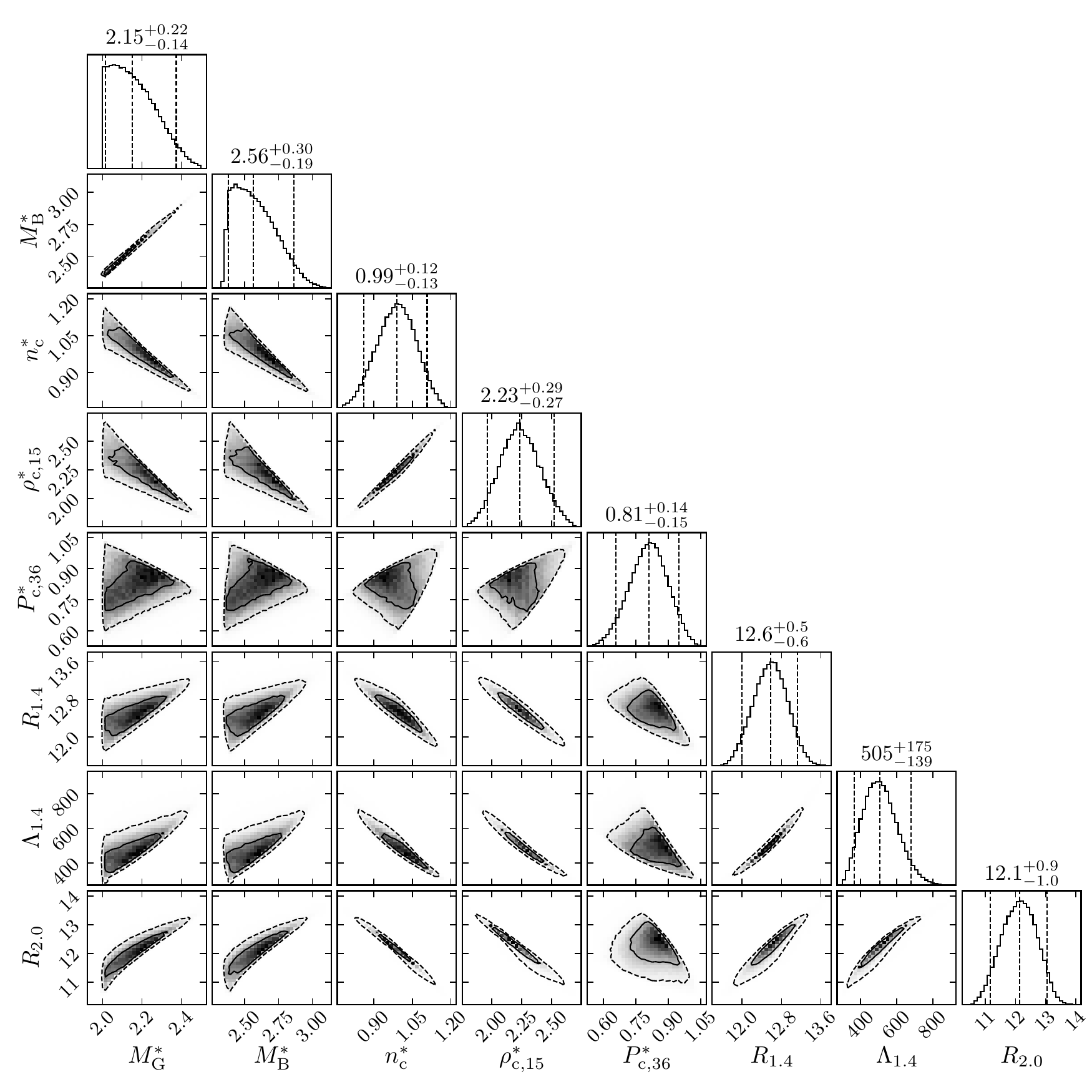"}
	\caption{1D and 2D marginalized posterior PDFs of selected properties of NS: maximum gravitational and baryonic masses (expressed in $\Msun$), central nucleon number density (in $\mathrm{1/fm^3}$), energy density (in $10^{15}\mathrm{g/cm^3}$) and pressure (in $10^{36}\mathrm{dyn/cm^2}$) of the maximum mass configuration; radius (in km) and tidal deformability (dimensionless) of the canonic mass NS; radius (in km) of the $2\,\Msun$ NS. Dashed vertical lines in 1D histograms correspond to 5\%, 50\% and 95\% quantiles; thus, the values on top of 1D histograms denote the median and its 90\% CI. The solid and the dashed contours on 2D histograms show 50\% and 90\% CR, respectively. Results correspond to run~5 in Table~\ref{tab:runs}, our fiducial model.}
	\label{Fig:DerPar_NS_Cor_1xP_1xE}
\end{figure*}

The corner plot of NS-related quantities is provided in Fig.~\ref{Fig:DerPar_NS_Cor_1xP_1xE}. Several correlations are present. Those among $M_{\mathrm{G}}^*$ and $M_{\mathrm{B}}^*$ or $n_{\mathrm{c}}^*$ and $\rho_{\mathrm{c},15}^*$ are trivial in the sense that they hold among quantities that, modulo the EOS, are equivalent. The correlation between $R_{1.4}$ and $\Lambda_{1.4}$ is equally easy to understand given the analytic expression of $\Lambda=2/3 k_2 \left( R/M\right)^5$, with $k_2$ representing the second gravito-electric Love number, which, similarly to the radius, is EOS-dependent. More interesting are the anti-correlations $n_{\mathrm{c}}^* - M_{\mathrm{G}}^*$,  $n_{\mathrm{c}}^* - R_{1.4}$ and $n_{\mathrm{c}}^* - R_{2.0}$. All of them can be explained considering that high $M_{\mathrm{G}}^*$ values require the EOS to be stiff, which assumes a certain incompressibility of the matter. The latter feature prevents large values of central density to be achieved as well as NSs in mass range $1.4 \leq M/\Msun \leq 2$ to have small radii. We note that a related anti-correlation, $n_{\mathrm{c}}^* - R^*$, was signalled in CDF models with non-linear couplings~\cite{Malik_RMF_NL_2023} as well as in models based on parametrizations of the speed of sound~\cite{Rezzolla_2022}; here $R^*$ represents the radius of the most massive configuration. Rather strong and positive correlations hold also between $R_{2.0}$ on the one hand and $R_{1.4}$ and $M_{\mathrm{G}}^*$ on the other hand. The remaining correlations $n_{\mathrm{c}}^* - M_{\mathrm{B}}^*$, $\rho_{\mathrm{c}}^* - M_{\mathrm{G}}^*$,  $\rho_{\mathrm{c}}^* - M_{\mathrm{B}}^*$, $n_{\mathrm{c}}^* - \Lambda_{1.4}$, $\rho_{\mathrm{c}}^* - \Lambda_{1.4}$, $\rho_{\mathrm{c}}^* - R_{1.4}$, $R_{2.0}-\rho_{\mathrm{c}}^*$ $R_{2.0}-\Lambda_{1.4}$ and $R_{2.0}-M_{\mathrm{B}}^*$ are natural consequences of the correlations explained before. Quantitative estimates of these correlations will be provided in Sect.~\ref{ssec:runs}, where their sensitivity to the constraints from PNM will be also tested.

\subsection{Alternative constraints from PNM and constraints on $\eff{m}$}
\label{ssec:runs}

%
\begin{figure*}
	\includegraphics[]{"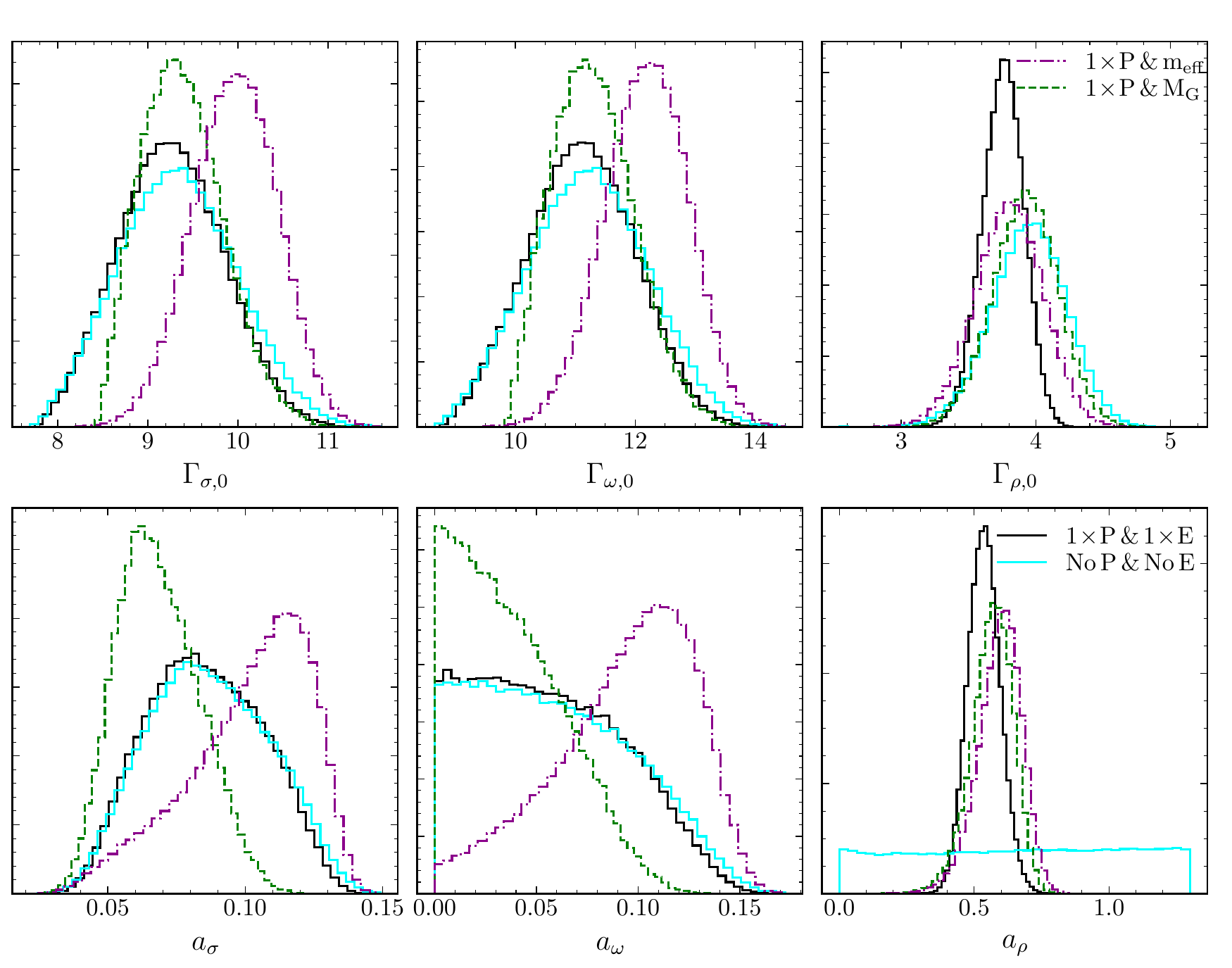"}
	\caption{Marginalized posteriors of the input parameters of our DD CDF model for runs~0, 3, 5 and 6. $\Gamma_{M,0}$ are expressed in MeV; $a_M$ are dimensionless ($M=\sigma$, $\omega$, $\rho$).}
	\label{Fig:BasePar_All_Hist}
\end{figure*}

\begin{figure*}
	\includegraphics[]{"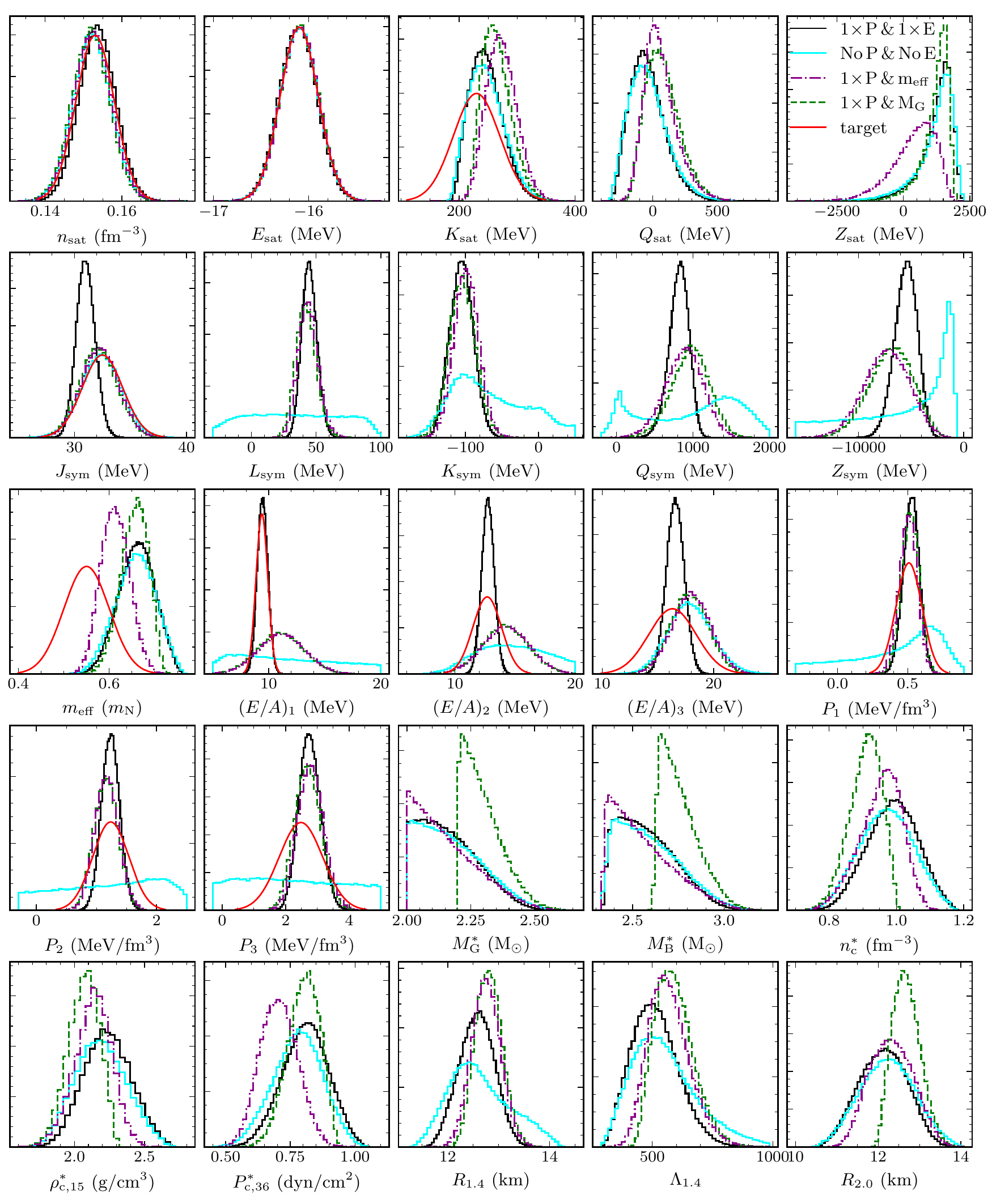"}
	\caption{Marginalized posteriors of NM parameters, energy per particle and pressure in PNM, $\eff{m}$ and selected properties of NSs. Considered are the runs~0, 3, 5 and 6. Target distributions for constrained parameters are also shown following Table~\ref{tab:constraints}. Note that not all of the constraints are used for each run, see Table~\ref{tab:runs} and main text for details.}
	\label{Fig:DerPar_All_Hist}
\end{figure*}

In this section we shall investigate the sensitivity of the posterior distributions of all of the above addressed quantities to the constraints from PNM. The effectiveness of the constraints on $\eff{m}$ will be also discussed. To this aim, the posterior PDFs of the input parameters corresponding to runs~0, 3, 5 and 6 are confronted in Fig.~\ref{Fig:BasePar_All_Hist}. Corresponding posterior distributions for the parameters of NM, PNM and NSs are depicted in Fig.~\ref{Fig:DerPar_All_Hist}. The median values as well as 68\% and 90\% CI of the seven runs mentioned in Table~\ref{tab:runs} are listed in Tables~\ref{tab:Posteriors_0} and \ref{tab:Posteriors_1} for all input and output quantities.

Fig.~\ref{Fig:BasePar_All_Hist} shows that PNM-related conditions only impact the quantities that regulate the isovector channel. Indeed, the posteriors of $\Gamma_{\sigma,0}$, $\Gamma_{\omega,0}$, $a_{\sigma}$, $a_{\omega}$ are almost identical for the least constrained run~0 and the most constrained run~5. The situation of posteriors of $\Gamma_{\rho,0}$ and $a_{\rho}$ is different. The lack of constraints from PNM leads, in what regards $\Gamma_{\rho,0}$, to a PDF for run~0 slightly shifted and slightly wider with respect to the one of run~5. The relative stability of the PDF of $\Gamma_{\rho,0}$, which accounts for the behavior of the isovector channel around $\sat{n}$, is due to the constraint on $\sym{J}$, identical in all our runs.
	
At variance with these, constraints from PNM strongly influence the PDF of $a_{\rho}$, which governs the behavior of the symmetry energy away from saturation. Lack of constraints from PNM is reflected in a flat distribution for run~0, which means no preference for a particular $\sym{E}(n_\mathrm{B})$ dependence, whereas a large number of constraints entails a strongly peaked distribution for run~5. Comparison between the curves corresponding to runs~2 and 4 (not illustrated here) shows that the constraints on $E/A$ are not equivalent with those on its derivative. PDF of each of these intermediately constrained runs are wider than the PDF of run~5. For more, see Tables \ref{tab:Posteriors_0} and \ref{tab:Posteriors_1}.

Constraints on $\eff{m}$ translate into constraints on $\Gamma_{\sigma,0}$, see the discussion in Sect.~\ref{ssec:Constraints}, and, due to the correlation between $\Gamma_{\sigma,0}$ and $\Gamma_{\omega,0}$, also into constraints on $\Gamma_{\omega,0}$. The curves corresponding to run~6 confirm this scenario. They also show that the PDFs of the parameters that govern the density dependence of the $\sigma$ and $\omega$ mesonic fields are modified as well. The latter modification is due to the constraint on $\sat{K}$. To a small extent the constraints on $\eff{m}$ also impact the PDFs of the two parameters in the isovector sector, though this can not be judged from Fig.~\ref{Fig:BasePar_All_Hist} as run~2 is absent from it. Constraints on the lower bound of the maximum NS masses impact the PDFs of the four parameters that deal with the isoscalar channel; $a_{\sigma}$, $a_{\omega}$, which account for the density dependence, are the most affected. It is straightforward to see that the PDFs of run~3 are much narrower than their counterparts corresponding to run~5, due to more stringent constraints.

Fig.~\ref{Fig:DerPar_All_Hist} confronts the PDFs of the calculated properties of NM, PNM and NS corresponding to various constraints. Target distributions of the constrained parameters are also shown (see Table~\ref{tab:constraints}). Note that not all of the constraints are used for each run (see Table~\ref{tab:runs}). It comes out that the PDFs of $\sat{E}$ are almost identical for all runs and are in excellent agreement with their target distributions. To a lesser extent this is also the case of PDFs of $\sat{n}$. As easy to anticipate and for the same reasons discussed above, the parameters of the isoscalar and isovector sectors are almost exclusively sensitive to constraints imposed to the respective channel. Run~0 produces for all $\sym{X}$ parameters, with the exception of $\sym{J}$, very wide PDFs. We particularly note that non-negligible numbers of models have negative values of $\sym{L}$ and/or positive values of $\sym{K}$, which is something that does not occur in any other run; $\sym{Q}$ has a bi-modal distribution. Wide PDFs are obtained also for $(E/A)_i$ with $i=1,2$ and $P_j$ with $j=1,2,3$. The explanation for the relatively narrow distribution of $(E/A)_3$ coincides with the one we provided above for the PDF of $\Gamma_{\rho,0}$ and is related to the constraint imposed on $\sym{J}$. In $70\%$ to $90\%$ of the cases negative values of $\sym{L}$ result in negative values of $P_j$, which is unphysical. Here the lower limit corresponds to $P_3$ and the upper limit to $P_1$.
In what regards NS observables, the most important differences between runs~0 and 5 are obtained for radii and tidal deformabilities of intermediate mass NSs. 

The small SD of $E_1$ explains in part the effectiveness of the constraints from the energy per particle of PNM in the case of run~5. We note that the effect of conditions on $(E/A)_{i=1,2,3}$ on $\sym{X}$ is opposite in what regards low- and high-order parameters. More precisely, run~5 is characterized by the lowest values of $\sym{J}$ and the highest values of $\sym{Z}$. The same is the case of run~4, not illustrated here.
The constraint on $\eff{m}$ impacts all $\sat{X}$ parameters as well as the PDFs of NSs quantities; little or no effect is seen on $\sym{X}$, $\left(E/A\right)_i$ and $P_i$. The target distribution of $\sym{J}$ is reached only when the isovector channel is controlled via pressure of PNM, that is in runs~6, 3 and 2 (not shown); the target distributions of $\sat{K}$ and $\eff{m}$ are never met. After incorporation of the constraint on Dirac effective mass the posterior probability distribution of $m_{\mathrm{eff}}$ shifts toward lower values and gets narrower though it still remains very different from the target distribution. The inability of the present parametrization to fit $m_{\mathrm{eff}}$ values extracted from experimental data should be attributed to its excessive simplicity. This deficiency could, in principle, be cured if extra terms accounting for tensor couplings are included in the functional. This is, however, beyond the scope of present work. We finally note that the constraints on $\eff{m}$ alter the posterior of maximum gravitational and baryonic masses in the sense that the values closer to the threshold are favored over larger values. 
Larger values for the lower bound on $M_{\mathrm{G}}^*$ shift the PDFs of $\sat{K}$, $\sat{Q}$ and $\sat{Z}$ to upper values; all these PDFs also become somewhat narrower. PDFs of $\sym{X}$ for runs~3 and 6 are similar, which means that these quantities are not affected. Median values of the PDFs corresponding to $R_{1.4}$, $\Lambda_{1.4}$, $R_{2.0}$ are larger for run~3 than for run~5, while the opposite is the case of $n_{\mathrm{c}}^*$ and $\rho_{\mathrm{c}}^*$.

In runs~6 and 1 (not shown), the tail of the maximum gravitational mass of NS extends up to about $2.71\,\Msun$\, \footnote{In other runs the tail of $M_\mathrm{G}^*$ distribution goes up to $2.65 - 2.70\,\Msun$.}, meaning that the secondary object in GW190814 with an estimated mass of $2.6\,\Msun$ \cite{GW190814} could have been a NS.

\renewcommand{\arraystretch}{1.7}
\setlength{\tabcolsep}{5.5pt}
\begin{table*}
	\caption{Parameters of marginalized posterior distributions. Data correspond to runs~0, 1, 2, and 3. Medians (Med.), 68\% and 90\% CI are shown. The median values and the values of the CI are rounded to 3 and 2 significant digits, respectively; \emph{all} trailing zeros after the decimal point are removed. The units are as follows: $\Gamma_{M,0}$, $X_{\mathrm{sat}}^{(i)}$, $X_{\mathrm{sym}}^{(j)}$ and $(E/A)_k$ are in MeV; $a_M$ and $\Lambda_{1.4}$ are adimensional; $m_{\mathrm{eff}}$ is in $m_\mathrm{N}$; $P_k$ is in MeV$/$fm$^3$; $n_{\mathrm{sat}}$ and $n_{\mathrm{c}}^*$ are in 1$/$fm$^3$; $M_{\mathrm G}^*$ and $M_{\mathrm{B}}^*$ are in $\Msun$; $R_{1.4}$ and $R_{2.0}$ are in km; $\rho_{\mathrm{c};15}^*$ is in $10^{15}$ g$/$cm$^3$; $P_{\mathrm{c};36}^*$ is in $10^{36}$ dyn$/$cm$^2$.}
	\begin{tabular}{lcccccccccccc}
		\toprule
		\toprule
		\multirow{2}{*}{Par.}  & \multicolumn{3}{c}{run 0}                                & \multicolumn{3}{c}{run 1}                                & \multicolumn{3}{c}{run 2}                                & \multicolumn{3}{c}{run 3}                              \\
		\cmidrule(lr){2-4}
		\cmidrule(lr){5-7}
		\cmidrule(lr){8-10}
		\cmidrule(lr){11-13}
		& Med.      & 68\% CI               & 90\% CI                & Med.      & 68\% CI               & 90\% CI                & Med.     & 68\% CI                & 90\% CI                & Med.     & 68\% CI                & 90\% CI              \\
		\midrule
		$\Gamma_{\sigma,0}$    & $9.32$   & $^{+0.67}_{-0.64}$     & $^{+1.1}_{-1}$         & $9.33$   & $^{+0.65}_{-0.62}$     & $^{+1.1}_{-0.99}$      & $9.33$   & $^{+0.63}_{-0.6}$      & $^{+1}_{-0.97}$        & $9.36$   & $^{+0.49}_{-0.43}$     & $^{+0.83}_{-0.64}$   \\
		$\Gamma_{\omega,0}$    & $11.2$   & $^{+1}_{-1}$           & $^{+1.6}_{-1.6}$       & $11.2$   & $^{+0.98}_{-0.96}$     & $^{+1.6}_{-1.5}$       & $11.2$   & $^{+0.94}_{-0.93}$     & $^{+1.5}_{-1.5}$       & $11.3$   & $^{+0.74}_{-0.66}$     & $^{+1.2}_{-0.99}$    \\
		$\Gamma_{\rho,0}$      & $3.95$   & $^{+0.27}_{-0.28}$     & $^{+0.44}_{-0.48}$     & $3.91$   & $^{+0.26}_{-0.28}$     & $^{+0.43}_{-0.47}$     & $3.89$   & $^{+0.24}_{-0.25}$     & $^{+0.4}_{-0.43}$      & $3.92$   & $^{+0.24}_{-0.24}$     & $^{+0.39}_{-0.41}$   \\
		$a_{\sigma}$           & $0.086$  & $^{+0.025}_{-0.022}$   & $^{+0.037}_{-0.034}$   & $0.0855$ & $^{+0.024}_{-0.022}$   & $^{+0.037}_{-0.034}$   & $0.0856$ & $^{+0.024}_{-0.022}$   & $^{+0.036}_{-0.034}$   & $0.067$  & $^{+0.017}_{-0.014}$   & $^{+0.027}_{-0.021}$ \\
		$a_{\omega}$           & $0.0558$ & $^{+0.046}_{-0.038}$   & $^{+0.07}_{-0.05}$     & $0.0561$ & $^{+0.045}_{-0.038}$   & $^{+0.068}_{-0.051}$   & $0.0567$ & $^{+0.044}_{-0.039}$   & $^{+0.067}_{-0.051}$   & $0.0341$ & $^{+0.033}_{-0.024}$   & $^{+0.052}_{-0.031}$ \\
		$a_{\rho}$             & $0.673$  & $^{+0.43}_{-0.46}$     & $^{+0.56}_{-0.61}$     & $0.563$  & $^{+0.13}_{-0.16}$     & $^{+0.21}_{-0.29}$     & $0.572$  & $^{+0.075}_{-0.085}$   & $^{+0.12}_{-0.15}$     & $0.571$  & $^{+0.071}_{-0.081}$   & $^{+0.11}_{-0.14}$   \\
		$n_\mathrm{sat}$       & $0.152$  & $^{+0.0049}_{-0.0049}$ & $^{+0.0082}_{-0.0081}$ & $0.153$  & $^{+0.0049}_{-0.0049}$ & $^{+0.0082}_{-0.0082}$ & $0.153$  & $^{+0.0049}_{-0.0049}$ & $^{+0.0082}_{-0.0082}$ & $0.152$  & $^{+0.0048}_{-0.0049}$ & $^{+0.008}_{-0.008}$ \\
		$E_\mathrm{sat}$       & $-16.1$  & $^{+0.2}_{-0.2}$       & $^{+0.33}_{-0.33}$     & $-16.1$  & $^{+0.2}_{-0.2}$       & $^{+0.33}_{-0.33}$     & $-16.1$  & $^{+0.2}_{-0.2}$       & $^{+0.33}_{-0.33}$     & $-16.1$  & $^{+0.2}_{-0.2}$       & $^{+0.33}_{-0.33}$   \\
		$K_\mathrm{sat}$       & $245$    & $^{+34}_{-29}$         & $^{+57}_{-42}$         & $247$    & $^{+33}_{-28}$         & $^{+56}_{-41}$         & $247$    & $^{+31}_{-27}$         & $^{+53}_{-41}$         & $262$    & $^{+27}_{-22}$         & $^{+47}_{-33}$       \\
		$Q_\mathrm{sat}$       & $-49.5$  & $^{+160}_{-130}$       & $^{+290}_{-200}$       & $-39.9$  & $^{+160}_{-130}$       & $^{+290}_{-190}$       & $-40.2$  & $^{+150}_{-120}$       & $^{+270}_{-190}$       & $63.3$   & $^{+140}_{-120}$       & $^{+260}_{-170}$     \\
		$Z_\mathrm{sat}$       & $1390$   & $^{+420}_{-840}$       & $^{+600}_{-1700}$      & $1360$   & $^{+410}_{-830}$       & $^{+600}_{-1700}$      & $1360$   & $^{+400}_{-790}$       & $^{+590}_{-1600}$      & $1350$   & $^{+300}_{-580}$       & $^{+430}_{-1200}$    \\
		$J_\mathrm{sym}$       & $32.4$   & $^{+1.8}_{-1.8}$       & $^{+3}_{-3}$           & $32.1$   & $^{+1.8}_{-1.8}$       & $^{+2.9}_{-2.9}$       & $32$     & $^{+1.7}_{-1.7}$       & $^{+2.8}_{-2.7}$       & $32.1$   & $^{+1.7}_{-1.6}$       & $^{+2.8}_{-2.7}$     \\
		$L_\mathrm{sym}$       & $32.1$   & $^{+42}_{-40}$         & $^{+57}_{-57}$         & $42.3$   & $^{+15}_{-13}$         & $^{+26}_{-21}$         & $41.4$   & $^{+8.2}_{-7.5}$       & $^{+14}_{-12}$         & $42.1$   & $^{+8}_{-7.4}$         & $^{+14}_{-12}$       \\
		$K_\mathrm{sym}$       & $-54.7$  & $^{+140}_{-57}$        & $^{+240}_{-79}$        & $-105$   & $^{+27}_{-24}$         & $^{+50}_{-39}$         & $-109$   & $^{+19}_{-20}$         & $^{+32}_{-32}$         & $-104$   & $^{+18}_{-18}$         & $^{+31}_{-30}$       \\
		$Q_\mathrm{sym}$       & $1110$   & $^{+520}_{-950}$       & $^{+750}_{-1100}$      & $932$    & $^{+360}_{-420}$       & $^{+560}_{-700}$       & $954$    & $^{+240}_{-260}$       & $^{+380}_{-430}$       & $972$    & $^{+230}_{-250}$       & $^{+380}_{-430}$     \\
		$Z_\mathrm{sym}$       & $-9450$  & $^{+7700}_{-16000}$    & $^{+8300}_{-22000}$    & $-6440$  & $^{+3100}_{-3800}$     & $^{+4500}_{-6400}$     & $-6610$  & $^{+2100}_{-2200}$     & $^{+3200}_{-3700}$     & $-6870$  & $^{+2000}_{-2200}$     & $^{+3100}_{-3600}$   \\
		$m_\mathrm{eff}$       & $0.659$  & $^{+0.043}_{-0.046}$   & $^{+0.067}_{-0.075}$   & $0.657$  & $^{+0.041}_{-0.045}$   & $^{+0.065}_{-0.073}$   & $0.657$  & $^{+0.04}_{-0.043}$    & $^{+0.064}_{-0.07}$    & $0.657$  & $^{+0.029}_{-0.033}$   & $^{+0.043}_{-0.056}$ \\
		$(E/A)_1$              & $12.5$   & $^{+7.9}_{-5.4}$       & $^{+12}_{-7.1}$        & $10.8$   & $^{+2.7}_{-2.7}$       & $^{+4.4}_{-4.4}$       & $10.8$   & $^{+2.1}_{-2.1}$       & $^{+3.5}_{-3.4}$       & $11$     & $^{+2.1}_{-2}$         & $^{+3.4}_{-3.4}$     \\
		$(E/A)_2$              & $15$     & $^{+3.7}_{-3.1}$       & $^{+6}_{-4.7}$         & $14$     & $^{+2.2}_{-2.2}$       & $^{+3.6}_{-3.7}$       & $14$     & $^{+1.9}_{-1.9}$       & $^{+3.2}_{-3.2}$       & $14.2$   & $^{+1.9}_{-1.9}$       & $^{+3.2}_{-3.1}$     \\
		$(E/A)_3$              & $17.9$   & $^{+2}_{-1.9}$         & $^{+3.4}_{-3.1}$       & $17.8$   & $^{+1.8}_{-1.8}$       & $^{+3}_{-3}$           & $17.6$   & $^{+1.7}_{-1.7}$       & $^{+2.9}_{-2.8}$       & $17.8$   & $^{+1.7}_{-1.7}$       & $^{+2.8}_{-2.8}$     \\
		$P_1$                  & $0.449$  & $^{+0.26}_{-0.65}$     & $^{+0.36}_{-0.99}$     & $0.537$  & $^{+0.1}_{-0.12}$      & $^{+0.16}_{-0.2}$      & $0.529$  & $^{+0.064}_{-0.066}$   & $^{+0.1}_{-0.11}$      & $0.517$  & $^{+0.062}_{-0.064}$   & $^{+0.1}_{-0.11}$    \\
		$P_2$                  & $0.871$  & $^{+1.2}_{-1.6}$       & $^{+1.5}_{-2.3}$       & $1.2$    & $^{+0.39}_{-0.39}$     & $^{+0.65}_{-0.64}$     & $1.17$   & $^{+0.2}_{-0.2}$       & $^{+0.33}_{-0.33}$     & $1.16$   & $^{+0.2}_{-0.2}$       & $^{+0.33}_{-0.33}$   \\
		$P_3$                  & $2.11$   & $^{+2.4}_{-2.1}$       & $^{+3.3}_{-3}$         & $2.67$   & $^{+0.82}_{-0.72}$     & $^{+1.4}_{-1.2}$       & $2.61$   & $^{+0.44}_{-0.42}$     & $^{+0.74}_{-0.69}$     & $2.7$    & $^{+0.42}_{-0.41}$     & $^{+0.72}_{-0.66}$   \\
		$M_\mathrm{G}^*$       & $2.16$   & $^{+0.15}_{-0.11}$     & $^{+0.24}_{-0.14}$     & $2.16$   & $^{+0.15}_{-0.11}$     & $^{+0.24}_{-0.14}$     & $2.16$   & $^{+0.14}_{-0.11}$     & $^{+0.23}_{-0.14}$     & $2.28$   & $^{+0.094}_{-0.06}$    & $^{+0.16}_{-0.076}$  \\
		$M_\mathrm{B}^*$       & $2.57$   & $^{+0.2}_{-0.15}$      & $^{+0.33}_{-0.2}$      & $2.57$   & $^{+0.2}_{-0.15}$      & $^{+0.32}_{-0.19}$     & $2.57$   & $^{+0.19}_{-0.15}$     & $^{+0.31}_{-0.19}$     & $2.74$   & $^{+0.13}_{-0.08}$     & $^{+0.22}_{-0.1}$    \\
		$n_\mathrm{c}^*$       & $0.971$  & $^{+0.082}_{-0.083}$   & $^{+0.13}_{-0.13}$     & $0.98$   & $^{+0.079}_{-0.083}$   & $^{+0.13}_{-0.13}$     & $0.981$  & $^{+0.075}_{-0.081}$   & $^{+0.12}_{-0.13}$     & $0.912$  & $^{+0.044}_{-0.054}$   & $^{+0.065}_{-0.09}$  \\
		$\rho_\mathrm{c,15}^*$ & $2.18$   & $^{+0.19}_{-0.18}$     & $^{+0.31}_{-0.28}$     & $2.21$   & $^{+0.19}_{-0.18}$     & $^{+0.3}_{-0.28}$      & $2.21$   & $^{+0.18}_{-0.17}$     & $^{+0.29}_{-0.27}$     & $2.07$   & $^{+0.11}_{-0.12}$     & $^{+0.16}_{-0.2}$    \\
		$P_\mathrm{c,36}^*$    & $0.782$  & $^{+0.092}_{-0.1}$     & $^{+0.15}_{-0.17}$     & $0.798$  & $^{+0.091}_{-0.097}$   & $^{+0.15}_{-0.16}$     & $0.8$    & $^{+0.088}_{-0.095}$   & $^{+0.14}_{-0.16}$     & $0.805$  & $^{+0.059}_{-0.067}$   & $^{+0.09}_{-0.11}$   \\
		$R_{1.4}$              & $12.6$   & $^{+0.73}_{-0.51}$     & $^{+1.2}_{-0.79}$      & $12.6$   & $^{+0.42}_{-0.42}$     & $^{+0.71}_{-0.68}$     & $12.6$   & $^{+0.34}_{-0.36}$     & $^{+0.55}_{-0.59}$     & $12.8$   & $^{+0.27}_{-0.26}$     & $^{+0.45}_{-0.41}$   \\
		$\Lambda_{1.4}$        & $534$    & $^{+150}_{-110}$       & $^{+280}_{-170}$       & $519$    & $^{+120}_{-100}$       & $^{+200}_{-150}$       & $510$    & $^{+100}_{-92}$        & $^{+180}_{-140}$       & $585$    & $^{+86}_{-72}$         & $^{+150}_{-110}$     \\
		$R_{2.0}$              & $12.2$   & $^{+0.67}_{-0.68}$     & $^{+1.1}_{-1.1}$       & $12.2$   & $^{+0.62}_{-0.64}$     & $^{+0.98}_{-1}$        & $12.2$   & $^{+0.58}_{-0.61}$     & $^{+0.91}_{-0.97}$     & $12.7$   & $^{+0.36}_{-0.32}$     & $^{+0.61}_{-0.48}$   \\
		\bottomrule
		\bottomrule
	\end{tabular}
	\label{tab:Posteriors_0}
\end{table*}

\begin{table*}
	\caption{Parameters of marginalized posterior distributions. Data correspond to runs~4, 5, and 6. Medians (Med.), 68\% and 90\% CI are shown. The median values and the values of the CI are rounded to 3 and 2 significant digits, respectively; \emph{all} trailing zeros after the decimal point are removed. For units, see Table~\ref{tab:Posteriors_0}.}
	\begin{tabular}{lccccccccc}
		\toprule
		\toprule
		\multirow{2}{*}{Par.}  & \multicolumn{3}{c}{run 4}                                & \multicolumn{3}{c}{run 5}                                & \multicolumn{3}{c}{run 6}                               \\
		\cmidrule(lr){2-4}
		\cmidrule(lr){5-7}
		\cmidrule(lr){8-10}
		& Med.     & 68\% CI                & 90\% CI                & Med.     & 68\% CI                & 90\% CI                & Med.     & 68\% CI                & 90\% CI               \\
		\midrule
		$\Gamma_{\sigma,0}$    & $9.33$   & $^{+0.66}_{-0.63}$     & $^{+1.1}_{-0.99}$      & $9.25$   & $^{+0.6}_{-0.58}$      & $^{+0.98}_{-0.93}$     & $9.97$   & $^{+0.46}_{-0.48}$     & $^{+0.75}_{-0.8}$     \\
		$\Gamma_{\omega,0}$    & $11.2$   & $^{+0.99}_{-0.97}$     & $^{+1.6}_{-1.5}$       & $11.1$   & $^{+0.9}_{-0.9}$       & $^{+1.5}_{-1.5}$       & $12.2$   & $^{+0.68}_{-0.72}$     & $^{+1.1}_{-1.2}$      \\
		$\Gamma_{\rho,0}$      & $3.81$   & $^{+0.19}_{-0.21}$     & $^{+0.31}_{-0.35}$     & $3.76$   & $^{+0.15}_{-0.16}$     & $^{+0.24}_{-0.27}$     & $3.8$    & $^{+0.25}_{-0.25}$     & $^{+0.41}_{-0.43}$    \\
		$a_{\sigma}$           & $0.0852$ & $^{+0.024}_{-0.022}$   & $^{+0.037}_{-0.034}$   & $0.0847$ & $^{+0.023}_{-0.021}$   & $^{+0.035}_{-0.032}$   & $0.106$  & $^{+0.017}_{-0.028}$   & $^{+0.025}_{-0.048}$  \\
		$a_{\omega}$           & $0.0557$ & $^{+0.045}_{-0.038}$   & $^{+0.068}_{-0.05}$    & $0.0539$ & $^{+0.043}_{-0.037}$   & $^{+0.065}_{-0.049}$   & $0.0985$ & $^{+0.028}_{-0.041}$   & $^{+0.041}_{-0.071}$  \\
		$a_{\rho}$             & $0.498$  & $^{+0.1}_{-0.095}$     & $^{+0.17}_{-0.16}$     & $0.536$  & $^{+0.06}_{-0.058}$    & $^{+0.1}_{-0.095}$     & $0.6$    & $^{+0.072}_{-0.084}$   & $^{+0.12}_{-0.15}$    \\
		$n_\mathrm{sat}$       & $0.153$  & $^{+0.0048}_{-0.0048}$ & $^{+0.0079}_{-0.0079}$ & $0.154$  & $^{+0.0047}_{-0.0048}$ & $^{+0.0078}_{-0.0078}$ & $0.152$  & $^{+0.0049}_{-0.0048}$ & $^{+0.0081}_{-0.008}$ \\
		$E_\mathrm{sat}$       & $-16.1$  & $^{+0.2}_{-0.2}$       & $^{+0.33}_{-0.33}$     & $-16.1$  & $^{+0.2}_{-0.2}$       & $^{+0.33}_{-0.33}$     & $-16.1$  & $^{+0.2}_{-0.2}$       & $^{+0.33}_{-0.33}$    \\
		$K_\mathrm{sat}$       & $248$    & $^{+34}_{-28}$         & $^{+58}_{-42}$         & $244$    & $^{+30}_{-26}$         & $^{+51}_{-39}$         & $272$    & $^{+27}_{-25}$         & $^{+46}_{-39}$        \\
		$Q_\mathrm{sat}$       & $-36.9$  & $^{+160}_{-130}$       & $^{+290}_{-200}$       & $-52.2$  & $^{+150}_{-120}$       & $^{+260}_{-180}$       & $43.2$   & $^{+140}_{-100}$       & $^{+260}_{-160}$      \\
		$Z_\mathrm{sat}$       & $1350$   & $^{+420}_{-860}$       & $^{+600}_{-1700}$      & $1400$   & $^{+380}_{-730}$       & $^{+560}_{-1500}$      & $517$    & $^{+670}_{-940}$       & $^{+960}_{-1700}$     \\
		$J_\mathrm{sym}$       & $31.4$   & $^{+1.2}_{-1.2}$       & $^{+2}_{-1.9}$         & $31$     & $^{+0.85}_{-0.82}$     & $^{+1.4}_{-1.3}$       & $32.3$   & $^{+1.7}_{-1.7}$       & $^{+2.8}_{-2.7}$      \\
		$L_\mathrm{sym}$       & $48.4$   & $^{+9.4}_{-8.8}$       & $^{+16}_{-14}$         & $44.6$   & $^{+6}_{-5.7}$         & $^{+10}_{-9.2}$        & $43.6$   & $^{+7.8}_{-7.2}$       & $^{+13}_{-12}$        \\
		$K_\mathrm{sym}$       & $-106$   & $^{+19}_{-17}$         & $^{+34}_{-27}$         & $-105$   & $^{+16}_{-16}$         & $^{+26}_{-26}$         & $-98.8$  & $^{+17}_{-17}$         & $^{+28}_{-28}$        \\
		$Q_\mathrm{sym}$       & $737$    & $^{+180}_{-200}$       & $^{+290}_{-340}$       & $821$    & $^{+120}_{-130}$       & $^{+190}_{-230}$       & $891$    & $^{+240}_{-250}$       & $^{+380}_{-430}$      \\
		$Z_\mathrm{sym}$       & $-4850$  & $^{+1400}_{-1800}$     & $^{+2200}_{-3200}$     & $-5470$  & $^{+1000}_{-1100}$     & $^{+1600}_{-1900}$     & $-7300$  & $^{+2100}_{-2200}$     & $^{+3300}_{-3700}$    \\
		$m_\mathrm{eff}$       & $0.657$  & $^{+0.042}_{-0.045}$   & $^{+0.066}_{-0.074}$   & $0.662$  & $^{+0.039}_{-0.042}$   & $^{+0.062}_{-0.068}$   & $0.612$  & $^{+0.033}_{-0.031}$   & $^{+0.054}_{-0.05}$   \\
		$(E/A)_1$              & $9.36$   & $^{+0.52}_{-0.51}$     & $^{+0.85}_{-0.84}$     & $9.45$   & $^{+0.48}_{-0.47}$     & $^{+0.79}_{-0.77}$     & $11.1$   & $^{+2.1}_{-2.1}$       & $^{+3.4}_{-3.4}$      \\
		$(E/A)_2$              & $12.9$   & $^{+0.65}_{-0.65}$     & $^{+1.1}_{-1.1}$       & $12.7$   & $^{+0.52}_{-0.52}$     & $^{+0.87}_{-0.86}$     & $14.3$   & $^{+1.9}_{-1.9}$       & $^{+3.2}_{-3.2}$      \\
		$(E/A)_3$              & $17.1$   & $^{+1.3}_{-1.3}$       & $^{+2.2}_{-2.1}$       & $16.6$   & $^{+0.79}_{-0.78}$     & $^{+1.3}_{-1.3}$       & $18$     & $^{+1.7}_{-1.7}$       & $^{+2.9}_{-2.9}$      \\
		$P_1$                  & $0.56$   & $^{+0.09}_{-0.099}$    & $^{+0.15}_{-0.17}$     & $0.527$  & $^{+0.056}_{-0.059}$   & $^{+0.092}_{-0.1}$     & $0.506$  & $^{+0.064}_{-0.066}$   & $^{+0.1}_{-0.11}$     \\
		$P_2$                  & $1.35$   & $^{+0.27}_{-0.27}$     & $^{+0.45}_{-0.44}$     & $1.24$   & $^{+0.15}_{-0.15}$     & $^{+0.25}_{-0.25}$     & $1.17$   & $^{+0.2}_{-0.19}$      & $^{+0.33}_{-0.32}$    \\
		$P_3$                  & $2.99$   & $^{+0.56}_{-0.51}$     & $^{+0.94}_{-0.83}$     & $2.75$   & $^{+0.35}_{-0.33}$     & $^{+0.58}_{-0.53}$     & $2.79$   & $^{+0.42}_{-0.4}$      & $^{+0.7}_{-0.66}$     \\
		$M_\mathrm{G}^*$       & $2.16$   & $^{+0.15}_{-0.11}$     & $^{+0.24}_{-0.14}$     & $2.15$   & $^{+0.14}_{-0.1}$      & $^{+0.22}_{-0.14}$     & $2.13$   & $^{+0.16}_{-0.097}$    & $^{+0.26}_{-0.12}$    \\
		$M_\mathrm{B}^*$       & $2.56$   & $^{+0.2}_{-0.15}$      & $^{+0.32}_{-0.19}$     & $2.56$   & $^{+0.18}_{-0.14}$     & $^{+0.3}_{-0.19}$      & $2.53$   & $^{+0.21}_{-0.13}$     & $^{+0.36}_{-0.17}$    \\
		$n_\mathrm{c}^*$       & $0.982$  & $^{+0.08}_{-0.084}$    & $^{+0.13}_{-0.14}$     & $0.99$   & $^{+0.074}_{-0.079}$   & $^{+0.12}_{-0.13}$     & $0.96$   & $^{+0.055}_{-0.071}$   & $^{+0.086}_{-0.12}$   \\
		$\rho_\mathrm{c,15}^*$ & $2.21$   & $^{+0.19}_{-0.18}$     & $^{+0.31}_{-0.29}$     & $2.23$   & $^{+0.17}_{-0.17}$     & $^{+0.29}_{-0.27}$     & $2.14$   & $^{+0.12}_{-0.14}$     & $^{+0.2}_{-0.23}$     \\
		$P_\mathrm{c,36}^*$    & $0.801$  & $^{+0.093}_{-0.1}$     & $^{+0.15}_{-0.16}$     & $0.81$   & $^{+0.087}_{-0.092}$   & $^{+0.14}_{-0.15}$     & $0.705$  & $^{+0.074}_{-0.071}$   & $^{+0.12}_{-0.11}$    \\
		$R_{1.4}$              & $12.7$   & $^{+0.39}_{-0.4}$      & $^{+0.63}_{-0.64}$     & $12.6$   & $^{+0.34}_{-0.36}$     & $^{+0.55}_{-0.58}$     & $12.8$   & $^{+0.28}_{-0.27}$     & $^{+0.47}_{-0.45}$    \\
		$\Lambda_{1.4}$        & $527$    & $^{+120}_{-100}$       & $^{+200}_{-160}$       & $505$    & $^{+100}_{-91}$        & $^{+170}_{-140}$       & $563$    & $^{+92}_{-75}$         & $^{+160}_{-120}$      \\
		$R_{2.0}$              & $12.2$   & $^{+0.63}_{-0.65}$     & $^{+0.99}_{-1}$        & $12.1$   & $^{+0.58}_{-0.61}$     & $^{+0.92}_{-0.97}$     & $12.3$   & $^{+0.55}_{-0.53}$     & $^{+0.87}_{-0.81}$    \\
		\bottomrule
		\bottomrule
	\end{tabular}
	\label{tab:Posteriors_1}
\end{table*}

\renewcommand{\arraystretch}{1.0}
\setlength{\tabcolsep}{2.0pt}

A number of features are worth noticing from Tables~\ref{tab:Posteriors_0} and \ref{tab:Posteriors_1}. They are as follows:
\begin{itemize}[nosep]
	\item Run~1 \textit{vs} run~2: Doubling the SDs of the target $P_i$ entails the doubling of the SDs of $a_{\rho}$, $\sym{L}$ and the posterior $P_i$ with $i=1,...,3$; the SDs of $\sym{K}$, $\sym{Q}$, $\sym{Z}$, $E_1$ and $E_2$ increase as well but with smaller factors; modifications of other quantities, including $\Gamma_{\rho,0}$ and $\sym{J}$, are negligible;
	\item Run~3 \textit{vs} run~2: A higher limit for the lower bound of the maximum NS mass reduces the SDs of the input isoscalar parameters; the median values of $a_\sigma$ and $a_\omega$ are also reduced; median values of $\sat{K}$, $\sat{Q}$, related to EOS stiffness, increase; as a consequence the median values of $M_\mathrm{G}^*$ and $M_\mathrm{B}^*$ increase, while those of $n_{c}^*$ and $\rho_c^*$ decrease; EOS stiffening also leads to slightly higher values of $R_{1.4}$ and $\Lambda_{1.4}$; SDs of all NSs properties are decreased;
	\item Run~4 \textit{vs} run~2: Replacing the constraints on PNM pressure with constraints on PNM energy per particle strongly impacts the posteriors of the input isovector parameters and PNM energy per particle and pressure; the small SDs of the target distributions of $(E/A)_i$ are very efficient in reducing the SDs of the posterior of the output distributions of the same quantities; in contrast, the SDs of $P_i$ increase; the latter results in the increase of both median values and variances of $R_{1.4}$ and $\Lambda_{1.4}$; for the median value of $R_{1.4}$ the increase is of 0.8\% while for $\Lambda_{1.4}$ it is of 3\%;
	\item Run~5 \textit{vs} run~2: The most significant consequence of constraining both PNM energy per particle and pressure is the reduction of the SDs of the posterior distributions of  $\left(E/A\right)_i$ and $P_i$; a small modification of the posteriors of the input parameters as well as their medians and variances is also observed; no sizable effect on NS properties is present;
	\item Run~6 \textit{vs} run~2: With the exception of $\sat{n}$, $\sat{E}$, PNM pressure and energies per particle, all considered quantities are modified upon the incorporation of the constraint on $\eff{m}$; our model nevertheless fails in reproducing the target distribution of this quantity, as already commented in relation to Fig.~\ref{Fig:DerPar_All_Hist}.
\end{itemize}

\subsection{Correlations between properties of NS and parameters of NM}
\label{ssec:correl}

%
\begin{figure*}
	\includegraphics[]{"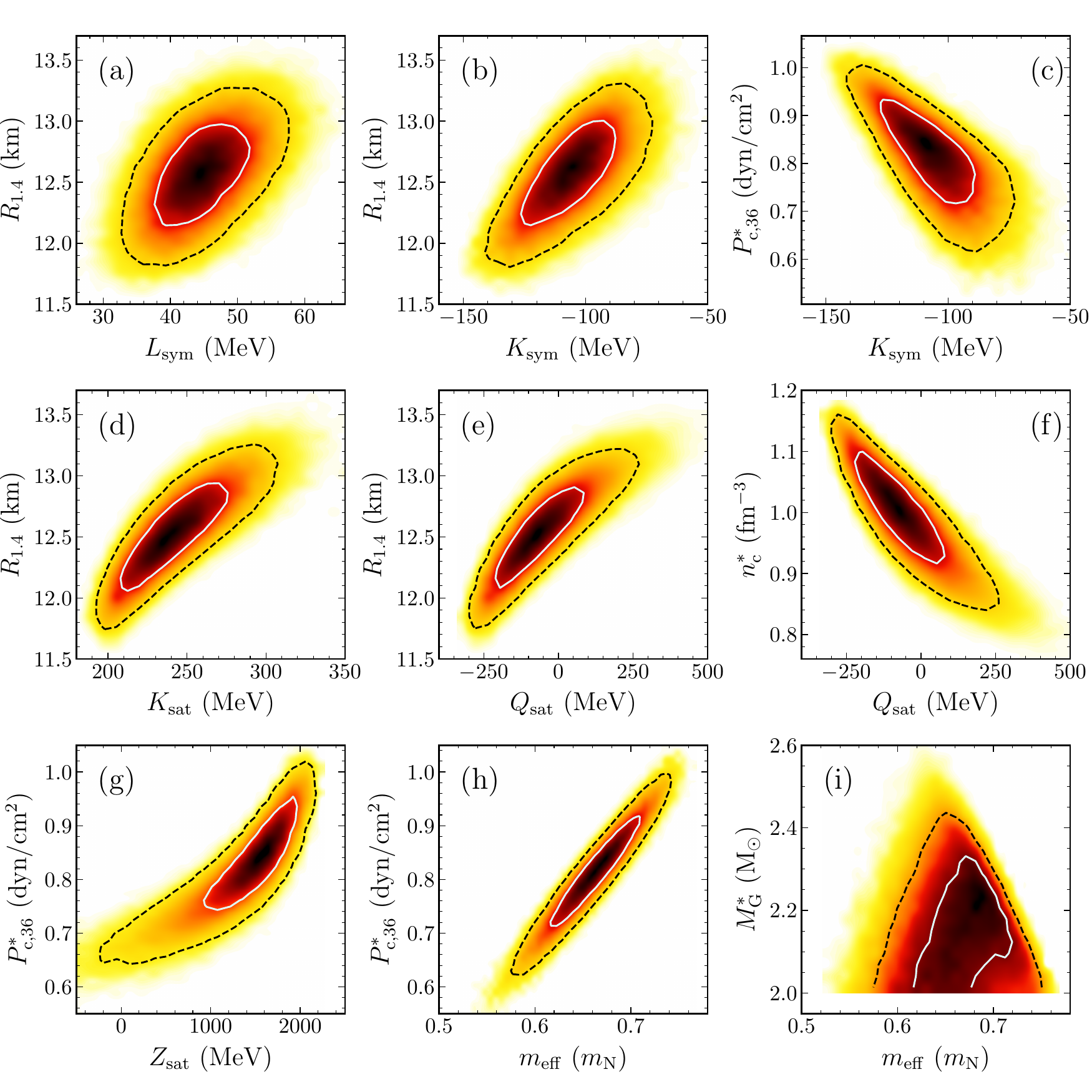"}
	\caption{2D marginalized posterior distributions (correlation plots) for selected NM and NS parameters that were not included in the corner plots on Figs.~\ref{Fig:BasePar_Cor_1xP_1xE}, \ref{Fig:DerPar_sat_Cor_1xP_1xE}, \ref{Fig:DerPar_sym_Cor_1xP_1xE} and \ref{Fig:DerPar_NS_Cor_1xP_1xE}. The color map shows the probability density. The light cyan solid and the black dashed contours  demonstrate 50\% and 90\% CR, respectively. Results correspond to run~5 in Table~\ref{tab:runs}, our fiducial model. See text for details.}
	\label{Fig:Hist2D}
\end{figure*}

Corner plots in Figs.~\ref{Fig:BasePar_Cor_1xP_1xE}, \ref{Fig:DerPar_sat_Cor_1xP_1xE}, \ref{Fig:DerPar_sym_Cor_1xP_1xE} and \ref{Fig:DerPar_NS_Cor_1xP_1xE} demonstrate correlations between various input and output model parameters. A number of correlations among properties of NSs and parameters of NM has been signaled in the literature, too. Examples in this sense are offered by correlations between $\sym{L}$, $\sym{K}$, $\sat{K}$, $\left(12\sat{K}+\sat{Q}\right)$ or a linear combination of them with the radii and tidal deformabilities of NS with $1 \leq M_{\mathrm{G}}/\Msun \leq 1.8$ \cite{Fortin_PRC_2016,Alam_PRC_2016,Margueron_PRC_2018_II,Malik_PRC_2018,Malik_PRC_2020};
correlations between $M_{\mathrm{G}}^*$ and $\sat{K}$, $\sat{Q}$ \cite{Margueron_PRC_2018_II}; correlations between $M_{\mathrm{G}}^*$ and $\eff{m}$ \cite{Weissenborn_NPA_2012,Hornick_2018,Hornick_2021}.

Correlations among key properties of NSs and parameters of NM have been carefully sought for. Fig.~\ref{Fig:Hist2D} illustrates some of the most discussed cases in the literature, e.g., $R_{1.4}-\sym{L}$ (panel (a)), $R_{1.4}-\sym{K}$ (panel (b)), $R_{1.4}-\sat{K}$ (panel (d)), $R_{1.4}-\sat{Q}$ (panel (e)) and $M_{\mathrm{G}}^*-\eff{m}$  (panel (i)) along with a number of other correlations. Considered run is run~5. The color map shows the 2D posterior probability density;  the light cyan solid and the black dashed contours demonstrate 50\% and 90\% CR, respectively. One may notice that, unlike the corner plots (Figs.~\ref{Fig:BasePar_Cor_1xP_1xE}, \ref{Fig:DerPar_sat_Cor_1xP_1xE}, \ref{Fig:DerPar_sym_Cor_1xP_1xE} and \ref{Fig:DerPar_NS_Cor_1xP_1xE}), here the probability density distribution goes outside the 90\% CR contour. This is expected, as the remaining 10\% of the samples indeed lie outside the 90\% CR contour. The fact that we cannot see it on the corner plots is due to some peculiarities in the implementation of the probability density visualization in the \texttt{corner} package\,\footnote{The documentation is available at \url{https://corner.readthedocs.io/} and the github page is \url{https://github.com/dfm/corner.py}} \cite{corner} that we employ to make corner plots. 

It comes out that, contrary to the results in Ref.~\cite{Fortin_PRC_2016}, in our model only a loose correlation exists among $R_{1.4}$ and $\sym{L}$. This situation is attributable to the narrow domain of values allowed for $\sym{L}$. Indeed, the values in Table~\ref{tab:Posteriors_1} indicate that $\sym{L}$ spans a 11.7 MeV (19.2 MeV) range to 68\% (90\%) CI. In contrast, Fig.~9 of Ref.~\cite{Fortin_PRC_2016} accounts for $40 \lesssim \sym{L} \lesssim 130$ MeV. Moreover, the net correlation holds for $\sym{L}\gtrsim 70$ MeV, which is in tension with experimental nuclear data~\cite{Tsang_PRC_2012,Lattimer_EPJA_2014}. In compensation, stronger correlations are obtained between $R_{1.4}$ on the one hand and $\sat{K}$, $\sat{Q}$ and $\sym{K}$ on the other hand. The correlations with the second and third (second) order coefficients in the isoscalar (isovector) channel are in accord (conflict) with the predictions of Ref. \cite{Margueron_PRC_2018_II}, where a non-relativistic framework and an agnostic density functional have been used. This suggests that the first two correlations might be universal while the latter is model-dependent. No correlation among $\eff{m}$ and $M_{\mathrm{G}}^*$ appears in Fig.~\ref{Fig:Hist2D}, panel I. As such, correlations obtained in Refs.~\cite{Weissenborn_NPA_2012,Hornick_2018,Hornick_2021} should be considered a peculiarity of the class of RMF models used in those works rather than a generic behavior of RMF models. As a matter of fact, attentive investigation of Fig.~7 in Ref.~\cite{Weissenborn_NPA_2012} shows that even TM1~\cite{TM1} deviates from the correlation shown by GM1~\cite{GM1-3}, GM3~\cite{GM1-3}, NL3~\cite{NL3}, etc. The difference between these RMF models with non-linear couplings consists in that TM1~\cite{TM1} accounts for an extra coupling, proportional to $\omega^4$.

\begin{turnpage}
	\begin{figure*}
		\includegraphics[]{"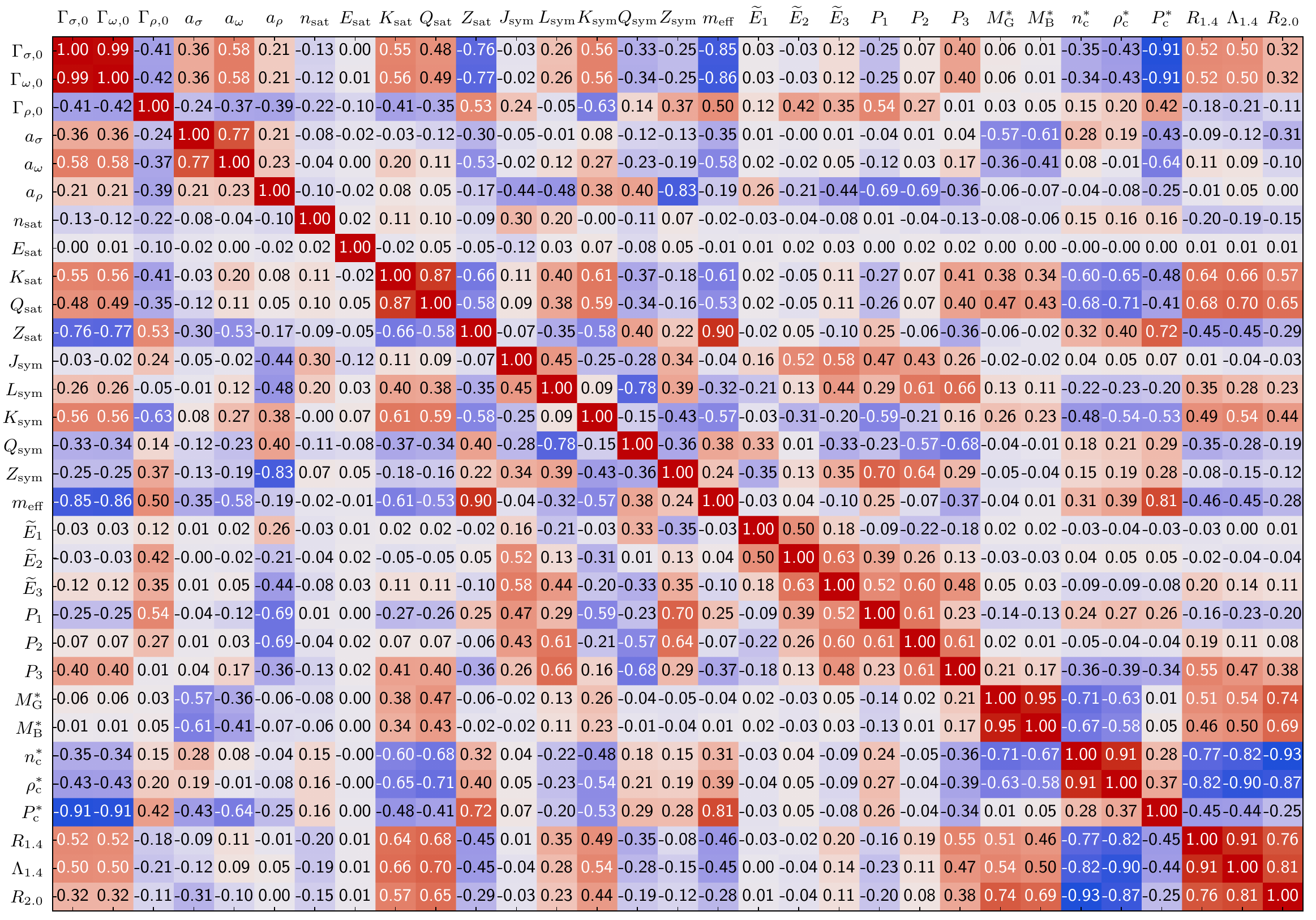"}
		\caption{Matrix of Kendall rank correlation coefficients for our fiducial case (run~5 in Table \ref{tab:runs}). Correlation matrix is, of course, square and symmetric, but to better utilize the rectangular shape of the page, the aspect ratio of the figure was adjusted. $\widetilde{E}_i$ stands for $(E/A)_i$.}
		\label{Fig:CorrPlot_1xP_1xE}
	\end{figure*}
\end{turnpage}

Other correlations plotted in Fig.~\ref{Fig:Hist2D} are the following: $\sym{K} - P_\mathrm{c}^{*}$ [panel (c)], $\sat{Q} - n_\mathrm{c}^*$ [panel (f)], $\sat{Z} - P_\mathrm{c}^{*}$ [panel (g)] and $\eff{m} - P_\mathrm{c}^*$ [panel (h)]. The correlations $\sym{K}-R_{1.4}$ [panel (b)] and $\sym{K}-P_\mathrm{c}^{*}$ [panel (c)] show that, over a significant density range, the stiffness of the symmetry energy gets reflected into the stiffness of the NS EOS. Indeed, models with large values of $\sym{K}$ provide large values of $R_{1.4}$ and low values of $P_\mathrm{c}^{*}$, both signaling low compressibility of the NS matter. Panels (d), (e), (f) show that $\sat{K}$ and $\sat{Q}$ are positively (negatively) correlated with $R_{1.4}$ ($n_\mathrm{c}^*$). As before, this means that they are effective in regulating the stiffness of the NS EOS over a wide density range. According to Fig.~\ref{Fig:CorrPlot_1xP_1xE}, the correlation $R_{2.0} - \sat{Q}$ is as strong as the correlation $R_{1.4}-\sat{Q}$; a weaker correlation holds between $R_{2.0}$ and $\sat{K}$.

A general survey of correlations among the posterior distributions of the input and output parameters is provided in Fig.~\ref{Fig:CorrPlot_1xP_1xE}, where the matrix of Kendall rank correlation coefficients \cite{Kendall_1938} is represented. Kendall coefficients are preferred here to more commonly used Pearson coefficients due to the non-linearity of the most of our correlations. As before, the fiducial run (5) is considered. 

Choosing somewhat arbitrarily the threshold for the absolute value of correlation coefficients to be 0.70, the following additional correlations are worth noticing:
i) $\Gamma_{\sigma,0}$ and $\Gamma_{\omega,0}$ are correlated with $\sat{Z}$, $\eff{m}$ and $P_\mathrm{c}^{*}$; these correlations explain the correlations among $\sat{Z}$, $\eff{m}$ and $P_\mathrm{c}^{*}$ noticed before;
ii) $a_{\rho}$ is correlated with $\sym{Z}$, which in its turn is correlated with $P_1$; a slightly weaker correlation occurs among $a_{\rho}$ and $P_1$.

We note that the presence or the absence of correlations should be considered with great caution. Correlations are \emph{model} and \emph{setup} dependent. For example, in all our runs $\sat{n}$ and $\sat{E}$ appear to be not correlated with any other parameter. Yet, as discussed in Sects.~\ref{ssec:Model} and \ref{ssec:MCMC} $\sat{n}$ is \emph{functionally} dependent on the input parameters. On the other hand, when we had post-filtered the posterior of run~5 to include only a narrow range of values of $\sym{L}$ $(44.5 \leq \sym{L} \leq 45.5~\mathrm{MeV})$, $\sat{n}$ showed a strong (anti-)correlation with $P_3$ and $\sat{E}$ was still uncorrelated with other parameters. Further insight into this question is provided in the Appendix.

\section{Conclusion}
\label{sec:Conclusion}

A modified version of the DD CDF model recently proposed in Ref. \cite{Malik_ApJ_2022} was employed in a statistical Bayes analysis to study the impact of various constraints from nuclear physics and astrophysics on the EOS of dense matter; the effectiveness of progressive incorporation of these constraints; correlations among input parameters of the model, parameters of NM and selected key properties of NSs. 

In addition to constraints on saturation density of SNM; energy per particle of symmetric saturated matter and its compression modulus; symmetry energy at saturation, we alternatively pose constraints on PNM energy per particle and/or pressure as calculated by $\chi$EFT microscopic models; on the Dirac effective mass of the nucleon at saturation; on the lower limit of the maximum NS mass.

Constraints from PNM impact only the isovector behavior of the NM EOS and the result is not the same when conditions are imposed on pressure, energy per nucleon or both quantities. Constraints on PNM pressure (energy per nucleon) lead to higher (lower) values of $\sym{J}$ and $\sym{Q}$ and lower (higher) values of $\sym{Z}$ and $\sym{L}$. The small variance of the energy per particle at low particle density is so effective in constraining the behavior far from saturation that SDs of $\sym{Q}$ and $\sym{Z}$ are by a factor of 0.75 lower for run~4 than for run~2 and the target distribution of $\sym{J}$ is missed. Joint constraints on $P$ and $E/A$ lead to even narrower distributions of the above coefficients in the Taylor expansion and increased tension with the target distribution of the symmetry energy. Properties of NSs sensitive to $\sym{E}(n_\mathrm{B})$, e.g., $R_{1.4}$ and $\Lambda_{1.4}$ are almost identical for runs 2, 4, 5 but the situation might be different for low mass NS.

The momentousness of constraints from PNM becomes even more apparent when posteriors of runs accounting for one or another set of conditions on pressure or energy per nucleon are confronted with posteriors obtained when this information is set aside. The latter is the case of our run~0, which allows for a such a diverse behavior in the isovector channel that even models with negative values of $\sym{L}$ or PNM pressure are allowed.

The posterior distribution of $\eff{m}$ is shifted with respect to the values of standard DD CDF models. Constraints on $\eff{m}$ strongly modify the posterior distributions of $\sat{X}^{(i)}$ with $i\geq 2$ as well as the posterior distributions of all considered NS properties.

A higher value of the lower bound of the maximum NS mass reduces the SDs of the input isoscalar parameters; increases the median values of $\sat{K}$, $\sat{Q}$, $M_{\mathrm {G}}^*$, $R_{2.0}$; decreases the median values of $n_{\mathrm c}^*$, $\rho_{\mathrm c}^*$.

The strongest correlations we found among the parameters of NM and properties of NSs are $\sym{K}-R_{1.4}$, $\sym{K}-P_{\mathrm c}^*$, $\sat{K}-R_{1.4}$, $\sat{Q}-R_{1.4}$, $\sat{Q}-R_{2.0}$, $\sat{Q}-n_{\mathrm c}^*$, $\sat{Z}-P_{\mathrm c}^*$ and $\eff{m}-P_{\mathrm c}^*$. Strong correlations are obtained also between $R_{2.0}$ and $n_{\mathrm c}^*$ and, to a lesser extent, between $R_{1.4}$ and $n_{\mathrm c}^*$.
Out of them $\sat{K}-R_{1.4}$, $\sat{Q}-R_{1.4}$, $R_{2.0}-n_{\mathrm c}^*$ might be universal, while $\sym{K}-R_{1.4}$ is definitely an artifact of the present model. Only a weak correlation was found between $\sym{L}-R_{1.4}$; we explain this result by the narrow domain spanned by $\sym{L}$, due to the constraints from PNM. No correlation was found between $\eff{m}$ and $M_{\mathrm G}^*$; discrepancy with results of Ref.~\cite{Weissenborn_NPA_2012,Hornick_2018,Hornick_2021} is explained by the different classes of models used in these works.

Last but not least, technicalities related to $\sat{n}$, which governs the density dependence of all mesonic fields, are reflected in the posterior distributions of all input and output parameters, although to a different extent. When the saturation density is determined self-consistently for each set of input parameters, the median value of $\sat{K}$ is by $\approx 15$ MeV higher than the one obtained by treating $\sat{n}$ as a free parameter and supplementing the likelihood with a condition that accounts for $P(\sat{n})=0$. Increased values of $\sat{K}$ in our work with respect to Ref.~\cite{Malik_ApJ_2022} also explain the high values of the maximum NS mass obtained here, $\approx 2.7\,\Msun$. 

Discrepancies between our posterior PDFs and those of Ref.~\cite{Malik_ApJ_2022} are illustrative of the model dependence of the results of this kind of studies. More insight into this issue can be gained by comparing the PDFs in Ref.~\cite{Malik_ApJ_2022} and \cite{Malik_RMF_NL_2023}, where the same set of constraints have been imposed to EOSs generated by CDF models with density dependent and non-linear couplings, respectively. The role played by the model is best illustrated by the disagreement in the direct Urca threshold, forbidden in the first class of models and allowed to operate starting from a relatively low density in the second class of models. Other examples are offered by Ref.~\cite{Papakonstantinou_2023}, which confronts posterior distributions corresponding to non-relativistic and relativistic mean-field models of nuclear matter.


\begin{acknowledgments}
  The authors gratefully acknowledge many illuminating discussions with Tuhin Malik and Constan\c{c}a Provid\^encia. 
  Support from a grant of the Ministry of Research, Innovation and Digitization, CNCS/CCCDI – UEFISCDI, Project No. PN-III-P4-ID-PCE-2020-0293 is also acknowledged.
  
  M.V.B. and A.R.R. contributed equally to this work.
\end{acknowledgments}


\appendix*
\section{Complementary information on correlations}
\label{App:Corr}

\begin{turnpage}
	\begin{figure*}
		\includegraphics[]{"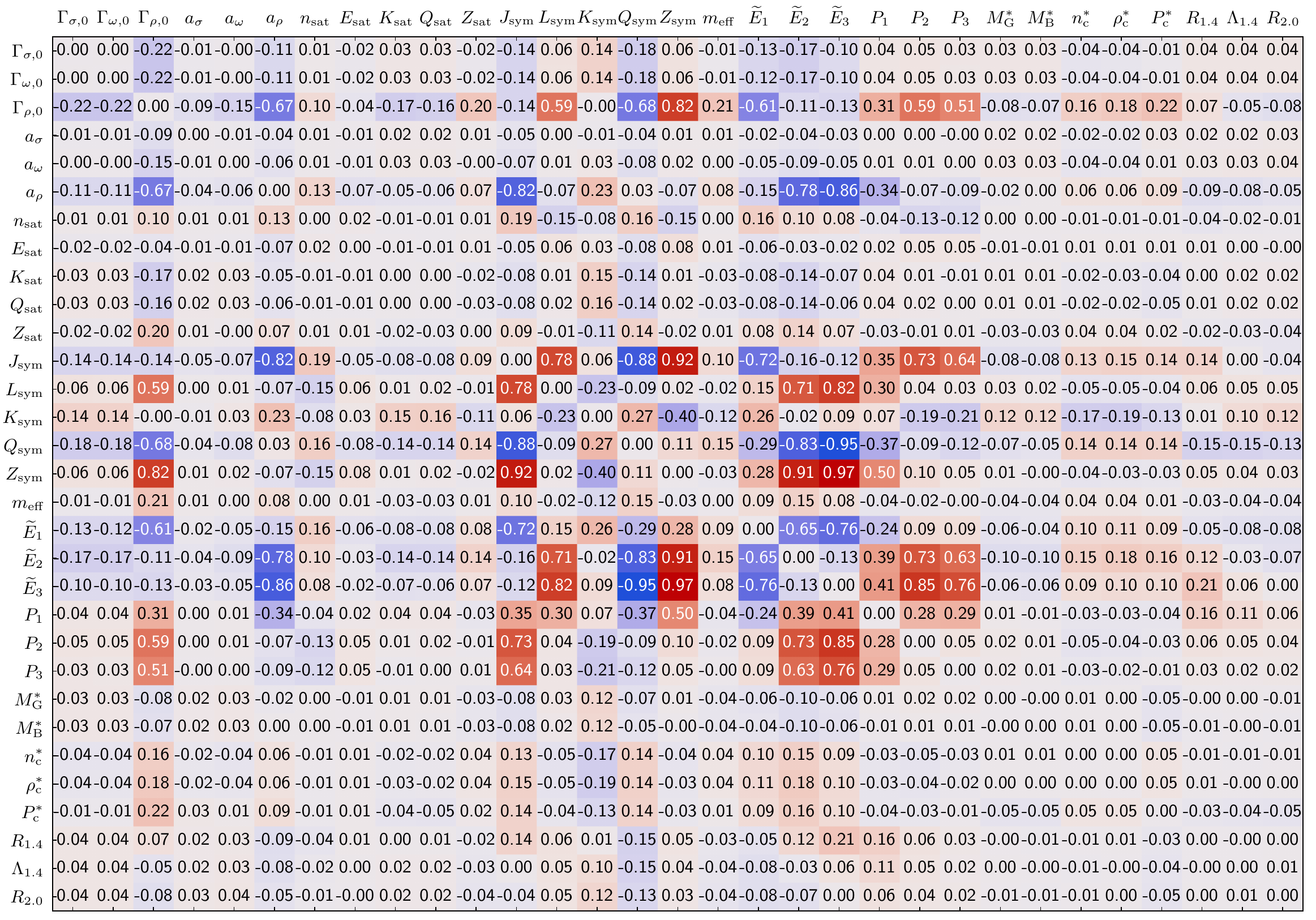"}
		\caption{Difference between matrices of Kendall rank correlation coefficients corresponding to runs~4 and 2. This difference matrix is, of course, square and symmetric. $\widetilde{E}_i$ stands for $(E/A)_i$.}
		\label{Fig:DiffCorr_1xE_to_1xP}
	\end{figure*}
\end{turnpage}

\begin{turnpage}
	\begin{figure*}
		\includegraphics[]{"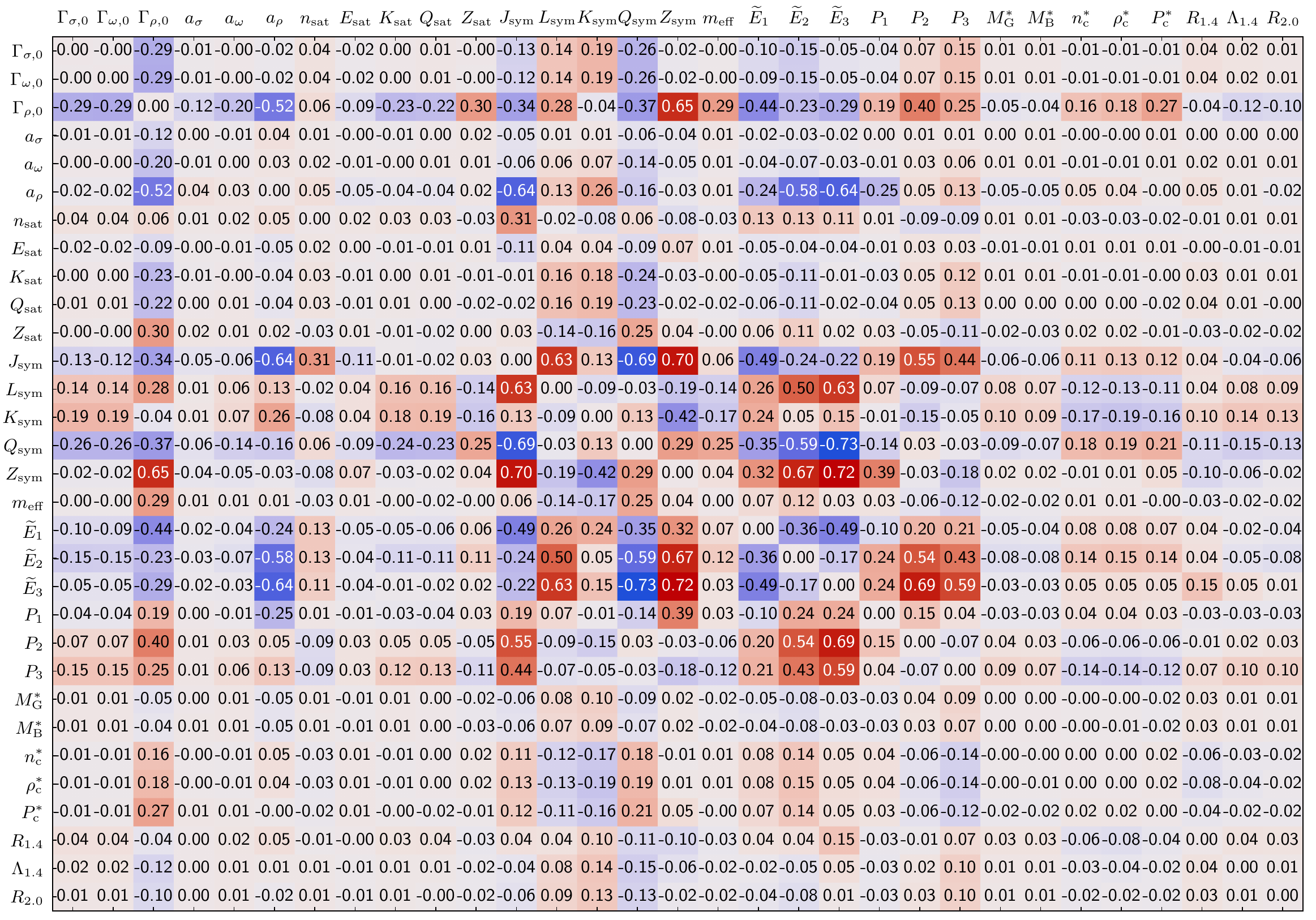"}
		\caption{Difference between matrices of Kendall rank correlation coefficients corresponding to runs~5 and 2. This difference matrix is, of course, square and symmetric. $\widetilde{E}_i$ stands for $(E/A)_i$.}
		\label{Fig:DiffCorr_1xP_1xE_to_1xP}
	\end{figure*}
\end{turnpage}

\begin{turnpage}
	\begin{figure*}
		\includegraphics[]{"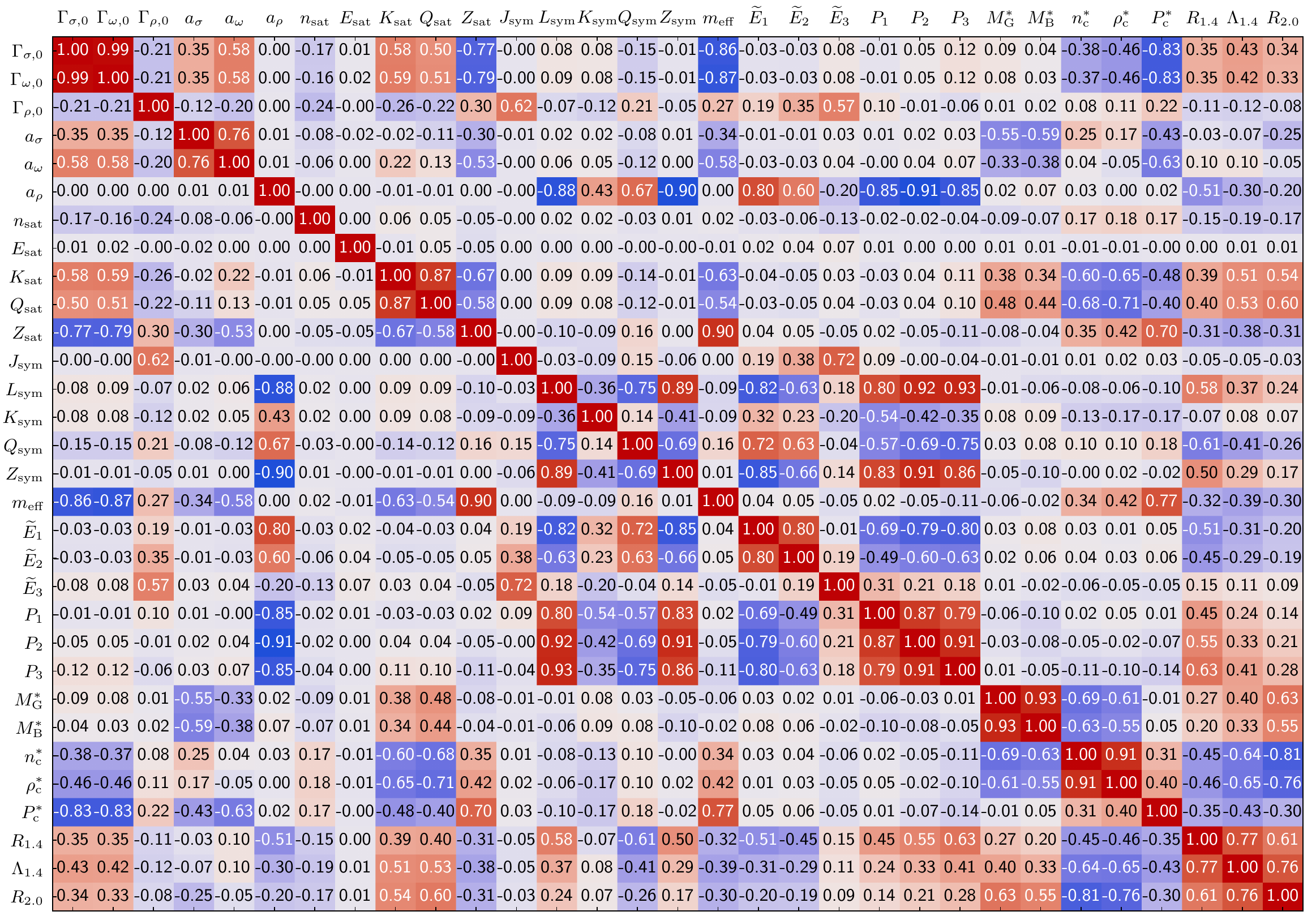"}
		\caption{Matrix of Kendall rank correlation coefficients for run~0 in Table \ref{tab:runs}. Correlation matrix is, of course, square and symmetric, but to better utilize rectangular shape of the page, the aspect ratio of the figure was adjusted. $\widetilde{E}_i$ stands for $(E/A)_i$.}
		\label{Fig:CorrPlot_NONE}
	\end{figure*}
\end{turnpage}

The different roles played by constraints from PNM pressure and energy per particle are investigated in Figs.~\ref{Fig:DiffCorr_1xE_to_1xP}, \ref{Fig:DiffCorr_1xP_1xE_to_1xP} and \ref{Fig:CorrPlot_NONE}. The first two figures report the differences between Kendall rank correlation coefficients computed for run~4 and 2 and run~5 and 2, respectively. The last figure shows the equivalent of Fig.~\ref{Fig:CorrPlot_1xP_1xE} for run~0. In Figs.~\ref{Fig:DiffCorr_1xE_to_1xP} and \ref{Fig:DiffCorr_1xP_1xE_to_1xP} large and positive values correspond to cases where the sign has changed from ``$+$'' in run~4 (5) to ``$-$'' in run~2. Some examples are: 
(i)   $\sym{J}-\sym{L}$, 
(ii)  $\sym{J}-\sym{Z}$,
(iii) $\sym{L}-\widetilde{E}_3$,
(iv)  $\sym{Z}-\widetilde{E}_2$,
(v)   $\sym{Z}-\widetilde{E}_3$,
(vi)  $\widetilde{E}_3-P_2$. 
Here, as in Figs.~\ref{Fig:CorrPlot_1xP_1xE}, \ref{Fig:DiffCorr_1xE_to_1xP} and \ref{Fig:DiffCorr_1xP_1xE_to_1xP}, $\widetilde{E}_i$ stands for $(E/A)_i$. Some of this cases, e.g., (vi) in Fig.~\ref{Fig:DiffCorr_1xE_to_1xP}, correspond to situations where a correlation is present in run~4 and not present in run~2. 

In Figs. ~\ref{Fig:DiffCorr_1xE_to_1xP} and \ref{Fig:DiffCorr_1xP_1xE_to_1xP} large and negative values correspond either to cases where the sign changed from ``$-$'' run~4 (5) to ``$+$'' in run~2, as it is the case of $\sym{Q}-\widetilde{E}_3$, or to cases where a correlation not apparent in run~4 appeared in run~2, as it is the case of $\widetilde{E}_1-\widetilde{E}_2$. 

Suppression of constraints from PNM reveals a number of strong correlations that were not manifesting themselves in run~5. Examples in this sense are offered by: $a_\rho-\sym{L}$, $\widetilde{E}_1-\sym{L}$, $\widetilde{E}_1-\sym{Z}$, $\sym{L}-\sym{Z}$, $\sym{Z}-P_3$, $P_i-P_j$ with $i \neq j$.

These figures once more emphasize that some correlations are strongly model dependent.
%

\bibliographystyle{apsrev4-2}
\bibliography{EOS.bib}

\begin{thebibliography}{79}%
\makeatletter
\providecommand \@ifxundefined [1]{%
 \@ifx{#1\undefined}
}%
\providecommand \@ifnum [1]{%
 \ifnum #1\expandafter \@firstoftwo
 \else \expandafter \@secondoftwo
 \fi
}%
\providecommand \@ifx [1]{%
 \ifx #1\expandafter \@firstoftwo
 \else \expandafter \@secondoftwo
 \fi
}%
\providecommand \natexlab [1]{#1}%
\providecommand \enquote  [1]{``#1''}%
\providecommand \bibnamefont  [1]{#1}%
\providecommand \bibfnamefont [1]{#1}%
\providecommand \citenamefont [1]{#1}%
\providecommand \href@noop [0]{\@secondoftwo}%
\providecommand \href [0]{\begingroup \@sanitize@url \@href}%
\providecommand \@href[1]{\@@startlink{#1}\@@href}%
\providecommand \@@href[1]{\endgroup#1\@@endlink}%
\providecommand \@sanitize@url [0]{\catcode `\\12\catcode `\$12\catcode
  `\&12\catcode `\#12\catcode `\^12\catcode `\_12\catcode `\%12\relax}%
\providecommand \@@startlink[1]{}%
\providecommand \@@endlink[0]{}%
\providecommand \url  [0]{\begingroup\@sanitize@url \@url }%
\providecommand \@url [1]{\endgroup\@href {#1}{\urlprefix }}%
\providecommand \urlprefix  [0]{URL }%
\providecommand \Eprint [0]{\href }%
\providecommand \doibase [0]{https://doi.org/}%
\providecommand \selectlanguage [0]{\@gobble}%
\providecommand \bibinfo  [0]{\@secondoftwo}%
\providecommand \bibfield  [0]{\@secondoftwo}%
\providecommand \translation [1]{[#1]}%
\providecommand \BibitemOpen [0]{}%
\providecommand \bibitemStop [0]{}%
\providecommand \bibitemNoStop [0]{.\EOS\space}%
\providecommand \EOS [0]{\spacefactor3000\relax}%
\providecommand \BibitemShut  [1]{\csname bibitem#1\endcsname}%
\let\auto@bib@innerbib\@empty
\bibitem [{\citenamefont {Oertel}\ \emph {et~al.}(2017)\citenamefont {Oertel},
  \citenamefont {Hempel}, \citenamefont {Kl\"ahn},\ and\ \citenamefont
  {Typel}}]{Oertel_RMP_2017}%
  \BibitemOpen
  \bibfield  {author} {\bibinfo {author} {\bibfnamefont {M.}~\bibnamefont
  {Oertel}}, \bibinfo {author} {\bibfnamefont {M.}~\bibnamefont {Hempel}},
  \bibinfo {author} {\bibfnamefont {T.}~\bibnamefont {Kl\"ahn}},\ and\ \bibinfo
  {author} {\bibfnamefont {S.}~\bibnamefont {Typel}},\ }\href
  {https://doi.org/10.1103/RevModPhys.89.015007} {\bibfield  {journal}
  {\bibinfo  {journal} {Rev. Mod. Phys.}\ }\textbf {\bibinfo {volume} {89}},\
  \bibinfo {pages} {015007} (\bibinfo {year} {2017})},\ \Eprint
  {https://arxiv.org/abs/1610.03361} {arXiv:1610.03361 [astro-ph.HE]}
  \BibitemShut {NoStop}%
\bibitem [{\citenamefont {Burgio}\ \emph {et~al.}(2021)\citenamefont {Burgio},
  \citenamefont {Schulze}, \citenamefont {Vidana},\ and\ \citenamefont
  {Wei}}]{Burgio_PPNP_2021}%
  \BibitemOpen
  \bibfield  {author} {\bibinfo {author} {\bibfnamefont {G.~F.}\ \bibnamefont
  {Burgio}}, \bibinfo {author} {\bibfnamefont {H.~J.}\ \bibnamefont {Schulze}},
  \bibinfo {author} {\bibfnamefont {I.}~\bibnamefont {Vidana}},\ and\ \bibinfo
  {author} {\bibfnamefont {J.~B.}\ \bibnamefont {Wei}},\ }\href
  {https://doi.org/10.1016/j.ppnp.2021.103879} {\bibfield  {journal} {\bibinfo
  {journal} {Prog. Part. Nucl. Phys.}\ }\textbf {\bibinfo {volume} {120}},\
  \bibinfo {pages} {103879} (\bibinfo {year} {2021})},\ \Eprint
  {https://arxiv.org/abs/2105.03747} {arXiv:2105.03747 [nucl-th]} \BibitemShut
  {NoStop}%
\bibitem [{\citenamefont {Lattimer}\ and\ \citenamefont
  {Prakash}(2007)}]{Lattimer_PhysRep_2007}%
  \BibitemOpen
  \bibfield  {author} {\bibinfo {author} {\bibfnamefont {J.~M.}\ \bibnamefont
  {Lattimer}}\ and\ \bibinfo {author} {\bibfnamefont {M.}~\bibnamefont
  {Prakash}},\ }\href {https://doi.org/10.1016/j.physrep.2007.02.003}
  {\bibfield  {journal} {\bibinfo  {journal} {Phys. Rept.}\ }\textbf {\bibinfo
  {volume} {442}},\ \bibinfo {pages} {109} (\bibinfo {year} {2007})},\ \Eprint
  {https://arxiv.org/abs/astro-ph/0612440} {arXiv:astro-ph/0612440}
  \BibitemShut {NoStop}%
\bibitem [{\citenamefont {Fortin}\ \emph {et~al.}(2016)\citenamefont {Fortin},
  \citenamefont {Provid{\^e}ncia}, \citenamefont {Raduta}, \citenamefont
  {Gulminelli}, \citenamefont {Zdunik}, \citenamefont {Haensel},\ and\
  \citenamefont {Bejger}}]{Fortin_PRC_2016}%
  \BibitemOpen
  \bibfield  {author} {\bibinfo {author} {\bibfnamefont {M.}~\bibnamefont
  {Fortin}}, \bibinfo {author} {\bibfnamefont {C.}~\bibnamefont
  {Provid{\^e}ncia}}, \bibinfo {author} {\bibfnamefont {A.~R.}\ \bibnamefont
  {Raduta}}, \bibinfo {author} {\bibfnamefont {F.}~\bibnamefont {Gulminelli}},
  \bibinfo {author} {\bibfnamefont {J.~L.}\ \bibnamefont {Zdunik}}, \bibinfo
  {author} {\bibfnamefont {P.}~\bibnamefont {Haensel}},\ and\ \bibinfo {author}
  {\bibfnamefont {M.}~\bibnamefont {Bejger}},\ }\href
  {https://doi.org/10.1103/PhysRevC.94.035804} {\bibfield  {journal} {\bibinfo
  {journal} {Phys. Rev. C}\ }\textbf {\bibinfo {volume} {94}},\ \bibinfo
  {pages} {035804} (\bibinfo {year} {2016})}\BibitemShut {NoStop}%
\bibitem [{\citenamefont {Demorest}\ \emph {et~al.}(2010)\citenamefont
  {Demorest}, \citenamefont {Pennucci}, \citenamefont {Ransom}, \citenamefont
  {Roberts},\ and\ \citenamefont {Hessels}}]{Demorest_Nature_2010}%
  \BibitemOpen
  \bibfield  {author} {\bibinfo {author} {\bibfnamefont {P.~B.}\ \bibnamefont
  {Demorest}}, \bibinfo {author} {\bibfnamefont {T.}~\bibnamefont {Pennucci}},
  \bibinfo {author} {\bibfnamefont {S.~M.}\ \bibnamefont {Ransom}}, \bibinfo
  {author} {\bibfnamefont {M.~S.~E.}\ \bibnamefont {Roberts}},\ and\ \bibinfo
  {author} {\bibfnamefont {J.~W.~T.}\ \bibnamefont {Hessels}},\ }\href
  {https://doi.org/10.1038/nature09466} {\bibfield  {journal} {\bibinfo
  {journal} {Nature}\ }\textbf {\bibinfo {volume} {467}},\ \bibinfo {pages}
  {1081} (\bibinfo {year} {2010})},\ \Eprint {https://arxiv.org/abs/1010.5788}
  {arXiv:1010.5788 [astro-ph.HE]} \BibitemShut {NoStop}%
\bibitem [{\citenamefont {{Antoniadis}}\ \emph {et~al.}(2013)\citenamefont
  {{Antoniadis}}, \citenamefont {{Freire}}, \citenamefont {{Wex}},
  \citenamefont {{Tauris}}, \citenamefont {{Lynch}}, \citenamefont {{van
  Kerkwijk}}, \citenamefont {{Kramer}}, \citenamefont {{Bassa}}, \citenamefont
  {{Dhillon}}, \citenamefont {{Driebe}}, \citenamefont {{Hessels}},
  \citenamefont {{Kaspi}}, \citenamefont {{Kondratiev}}, \citenamefont
  {{Langer}}, \citenamefont {{Marsh}}, \citenamefont {{McLaughlin}},
  \citenamefont {{Pennucci}}, \citenamefont {{Ransom}}, \citenamefont
  {{Stairs}}, \citenamefont {{van Leeuwen}}, \citenamefont {{Verbiest}},\ and\
  \citenamefont {{Whelan}}}]{Antoniadis2013}%
  \BibitemOpen
  \bibfield  {author} {\bibinfo {author} {\bibfnamefont {J.}~\bibnamefont
  {{Antoniadis}}}, \bibinfo {author} {\bibfnamefont {P.~C.~C.}\ \bibnamefont
  {{Freire}}}, \bibinfo {author} {\bibfnamefont {N.}~\bibnamefont {{Wex}}},
  \bibinfo {author} {\bibfnamefont {T.~M.}\ \bibnamefont {{Tauris}}}, \bibinfo
  {author} {\bibfnamefont {R.~S.}\ \bibnamefont {{Lynch}}}, \bibinfo {author}
  {\bibfnamefont {M.~H.}\ \bibnamefont {{van Kerkwijk}}}, \bibinfo {author}
  {\bibfnamefont {M.}~\bibnamefont {{Kramer}}}, \bibinfo {author}
  {\bibfnamefont {C.}~\bibnamefont {{Bassa}}}, \bibinfo {author} {\bibfnamefont
  {V.~S.}\ \bibnamefont {{Dhillon}}}, \bibinfo {author} {\bibfnamefont
  {T.}~\bibnamefont {{Driebe}}}, \bibinfo {author} {\bibfnamefont {J.~W.~T.}\
  \bibnamefont {{Hessels}}}, \bibinfo {author} {\bibfnamefont {V.~M.}\
  \bibnamefont {{Kaspi}}}, \bibinfo {author} {\bibfnamefont {V.~I.}\
  \bibnamefont {{Kondratiev}}}, \bibinfo {author} {\bibfnamefont
  {N.}~\bibnamefont {{Langer}}}, \bibinfo {author} {\bibfnamefont {T.~R.}\
  \bibnamefont {{Marsh}}}, \bibinfo {author} {\bibfnamefont {M.~A.}\
  \bibnamefont {{McLaughlin}}}, \bibinfo {author} {\bibfnamefont {T.~T.}\
  \bibnamefont {{Pennucci}}}, \bibinfo {author} {\bibfnamefont {S.~M.}\
  \bibnamefont {{Ransom}}}, \bibinfo {author} {\bibfnamefont {I.~H.}\
  \bibnamefont {{Stairs}}}, \bibinfo {author} {\bibfnamefont {J.}~\bibnamefont
  {{van Leeuwen}}}, \bibinfo {author} {\bibfnamefont {J.~P.~W.}\ \bibnamefont
  {{Verbiest}}},\ and\ \bibinfo {author} {\bibfnamefont {D.~G.}\ \bibnamefont
  {{Whelan}}},\ }\href {https://doi.org/10.1126/science.1233232} {\bibfield
  {journal} {\bibinfo  {journal} {Science}\ }\textbf {\bibinfo {volume}
  {340}},\ \bibinfo {pages} {448} (\bibinfo {year} {2013})}\BibitemShut
  {NoStop}%
\bibitem [{\citenamefont {Arzoumanian}\ \emph {et~al.}(2018)\citenamefont
  {Arzoumanian} \emph {et~al.}}]{Arzoumanian_ApJSS_2018}%
  \BibitemOpen
  \bibfield  {author} {\bibinfo {author} {\bibfnamefont {Z.}~\bibnamefont
  {Arzoumanian}} \emph {et~al.} (\bibinfo {collaboration} {NANOGrav}),\ }\href
  {https://doi.org/10.3847/1538-4365/aab5b0} {\bibfield  {journal} {\bibinfo
  {journal} {Astrophys. J. Suppl.}\ }\textbf {\bibinfo {volume} {235}},\
  \bibinfo {pages} {37} (\bibinfo {year} {2018})},\ \Eprint
  {https://arxiv.org/abs/1801.01837} {arXiv:1801.01837 [astro-ph.HE]}
  \BibitemShut {NoStop}%
\bibitem [{\citenamefont {Cromartie}\ \emph {et~al.}(2020)\citenamefont
  {Cromartie} \emph {et~al.}}]{Cromartie2020}%
  \BibitemOpen
  \bibfield  {author} {\bibinfo {author} {\bibfnamefont {H.~T.}\ \bibnamefont
  {Cromartie}} \emph {et~al.} (\bibinfo {collaboration} {NANOGrav}),\ }\href
  {https://doi.org/10.1038/s41550-019-0880-2} {\bibfield  {journal} {\bibinfo
  {journal} {Nature Astron.}\ }\textbf {\bibinfo {volume} {4}},\ \bibinfo
  {pages} {72} (\bibinfo {year} {2020})},\ \Eprint
  {https://arxiv.org/abs/1904.06759} {arXiv:1904.06759 [astro-ph.HE]}
  \BibitemShut {NoStop}%
\bibitem [{\citenamefont {Fonseca}\ \emph {et~al.}(2021)\citenamefont
  {Fonseca}, \citenamefont {Cromartie}, \citenamefont {Pennucci}, \citenamefont
  {Ray}, \citenamefont {Kirichenko}, \citenamefont {Ransom}, \citenamefont
  {Demorest}, \citenamefont {Stairs}, \citenamefont {Arzoumanian},
  \citenamefont {Guillemot}, \citenamefont {Parthasarathy}, \citenamefont
  {Kerr}, \citenamefont {Cognard}, \citenamefont {Baker}, \citenamefont
  {Blumer}, \citenamefont {Brook}, \citenamefont {DeCesar}, \citenamefont
  {Dolch}, \citenamefont {Dong}, \citenamefont {Ferrara}, \citenamefont
  {Fiore}, \citenamefont {Garver-Daniels}, \citenamefont {Good}, \citenamefont
  {Jennings}, \citenamefont {Jones}, \citenamefont {Kaspi}, \citenamefont
  {Lam}, \citenamefont {Lorimer}, \citenamefont {Luo}, \citenamefont {McEwen},
  \citenamefont {McKee}, \citenamefont {McLaughlin}, \citenamefont {McMann},
  \citenamefont {Meyers}, \citenamefont {Naidu}, \citenamefont {Ng},
  \citenamefont {Nice}, \citenamefont {Pol}, \citenamefont {Radovan},
  \citenamefont {Shapiro-Albert}, \citenamefont {Tan}, \citenamefont
  {Tendulkar}, \citenamefont {Swiggum}, \citenamefont {Wahl},\ and\
  \citenamefont {Zhu}}]{Fonseca_2021}%
  \BibitemOpen
  \bibfield  {author} {\bibinfo {author} {\bibfnamefont {E.}~\bibnamefont
  {Fonseca}}, \bibinfo {author} {\bibfnamefont {H.~T.}\ \bibnamefont
  {Cromartie}}, \bibinfo {author} {\bibfnamefont {T.~T.}\ \bibnamefont
  {Pennucci}}, \bibinfo {author} {\bibfnamefont {P.~S.}\ \bibnamefont {Ray}},
  \bibinfo {author} {\bibfnamefont {A.~Y.}\ \bibnamefont {Kirichenko}},
  \bibinfo {author} {\bibfnamefont {S.~M.}\ \bibnamefont {Ransom}}, \bibinfo
  {author} {\bibfnamefont {P.~B.}\ \bibnamefont {Demorest}}, \bibinfo {author}
  {\bibfnamefont {I.~H.}\ \bibnamefont {Stairs}}, \bibinfo {author}
  {\bibfnamefont {Z.}~\bibnamefont {Arzoumanian}}, \bibinfo {author}
  {\bibfnamefont {L.}~\bibnamefont {Guillemot}}, \bibinfo {author}
  {\bibfnamefont {A.}~\bibnamefont {Parthasarathy}}, \bibinfo {author}
  {\bibfnamefont {M.}~\bibnamefont {Kerr}}, \bibinfo {author} {\bibfnamefont
  {I.}~\bibnamefont {Cognard}}, \bibinfo {author} {\bibfnamefont {P.~T.}\
  \bibnamefont {Baker}}, \bibinfo {author} {\bibfnamefont {H.}~\bibnamefont
  {Blumer}}, \bibinfo {author} {\bibfnamefont {P.~R.}\ \bibnamefont {Brook}},
  \bibinfo {author} {\bibfnamefont {M.}~\bibnamefont {DeCesar}}, \bibinfo
  {author} {\bibfnamefont {T.}~\bibnamefont {Dolch}}, \bibinfo {author}
  {\bibfnamefont {F.~A.}\ \bibnamefont {Dong}}, \bibinfo {author}
  {\bibfnamefont {E.~C.}\ \bibnamefont {Ferrara}}, \bibinfo {author}
  {\bibfnamefont {W.}~\bibnamefont {Fiore}}, \bibinfo {author} {\bibfnamefont
  {N.}~\bibnamefont {Garver-Daniels}}, \bibinfo {author} {\bibfnamefont
  {D.~C.}\ \bibnamefont {Good}}, \bibinfo {author} {\bibfnamefont
  {R.}~\bibnamefont {Jennings}}, \bibinfo {author} {\bibfnamefont {M.~L.}\
  \bibnamefont {Jones}}, \bibinfo {author} {\bibfnamefont {V.~M.}\ \bibnamefont
  {Kaspi}}, \bibinfo {author} {\bibfnamefont {M.~T.}\ \bibnamefont {Lam}},
  \bibinfo {author} {\bibfnamefont {D.~R.}\ \bibnamefont {Lorimer}}, \bibinfo
  {author} {\bibfnamefont {J.}~\bibnamefont {Luo}}, \bibinfo {author}
  {\bibfnamefont {A.}~\bibnamefont {McEwen}}, \bibinfo {author} {\bibfnamefont
  {J.~W.}\ \bibnamefont {McKee}}, \bibinfo {author} {\bibfnamefont {M.~A.}\
  \bibnamefont {McLaughlin}}, \bibinfo {author} {\bibfnamefont
  {N.}~\bibnamefont {McMann}}, \bibinfo {author} {\bibfnamefont {B.~W.}\
  \bibnamefont {Meyers}}, \bibinfo {author} {\bibfnamefont {A.}~\bibnamefont
  {Naidu}}, \bibinfo {author} {\bibfnamefont {C.}~\bibnamefont {Ng}}, \bibinfo
  {author} {\bibfnamefont {D.~J.}\ \bibnamefont {Nice}}, \bibinfo {author}
  {\bibfnamefont {N.}~\bibnamefont {Pol}}, \bibinfo {author} {\bibfnamefont
  {H.~A.}\ \bibnamefont {Radovan}}, \bibinfo {author} {\bibfnamefont
  {B.}~\bibnamefont {Shapiro-Albert}}, \bibinfo {author} {\bibfnamefont
  {C.~M.}\ \bibnamefont {Tan}}, \bibinfo {author} {\bibfnamefont {S.~P.}\
  \bibnamefont {Tendulkar}}, \bibinfo {author} {\bibfnamefont {J.~K.}\
  \bibnamefont {Swiggum}}, \bibinfo {author} {\bibfnamefont {H.~M.}\
  \bibnamefont {Wahl}},\ and\ \bibinfo {author} {\bibfnamefont {W.~W.}\
  \bibnamefont {Zhu}},\ }\href {https://doi.org/10.3847/2041-8213/ac03b8}
  {\bibfield  {journal} {\bibinfo  {journal} {Astrophys. J. Lett.}\ }\textbf
  {\bibinfo {volume} {915}},\ \bibinfo {pages} {L12} (\bibinfo {year}
  {2021})}\BibitemShut {NoStop}%
\bibitem [{\citenamefont {{Sedrakian}}\ \emph {et~al.}(2022)\citenamefont
  {{Sedrakian}}, \citenamefont {{Li}},\ and\ \citenamefont
  {{Weber}}}]{Sedrakian_PPNP_2022}%
  \BibitemOpen
  \bibfield  {author} {\bibinfo {author} {\bibfnamefont {A.}~\bibnamefont
  {{Sedrakian}}}, \bibinfo {author} {\bibfnamefont {J.-J.}\ \bibnamefont
  {{Li}}},\ and\ \bibinfo {author} {\bibfnamefont {F.}~\bibnamefont
  {{Weber}}},\ }\href@noop {} {\bibfield  {journal} {\bibinfo  {journal} {arXiv
  e-prints}\ ,\ \bibinfo {eid} {arXiv:2212.01086}} (\bibinfo {year} {2022})},\
  \Eprint {https://arxiv.org/abs/2212.01086} {arXiv:2212.01086 [nucl-th]}
  \BibitemShut {NoStop}%
\bibitem [{\citenamefont {Abbott}\ and\ \citenamefont
  {et~al.}(2017)}]{Abbott_PRL119_161101}%
  \BibitemOpen
  \bibfield  {author} {\bibinfo {author} {\bibfnamefont {B.~P.}\ \bibnamefont
  {Abbott}}\ and\ \bibinfo {author} {\bibnamefont {et~al.}} (\bibinfo
  {collaboration} {LIGO Scientific Collaboration and Virgo Collaboration}),\
  }\href {https://doi.org/10.1103/PhysRevLett.119.161101} {\bibfield  {journal}
  {\bibinfo  {journal} {Phys. Rev. Lett.}\ }\textbf {\bibinfo {volume} {119}},\
  \bibinfo {pages} {161101} (\bibinfo {year} {2017})}\BibitemShut {NoStop}%
\bibitem [{\citenamefont {{Abbott}}\ and\ \citenamefont
  {et~al.}(2017)}]{Abbott_ApJ2017ApJ_L12}%
  \BibitemOpen
  \bibfield  {author} {\bibinfo {author} {\bibfnamefont {B.~P.}\ \bibnamefont
  {{Abbott}}}\ and\ \bibinfo {author} {\bibnamefont {et~al.}},\ }\href
  {https://doi.org/10.3847/2041-8213/aa91c9} {\bibfield  {journal} {\bibinfo
  {journal} {Astrophys. J. Lett.}\ }\textbf {\bibinfo {volume} {848}},\
  \bibinfo {eid} {L12} (\bibinfo {year} {2017})},\ \Eprint
  {https://arxiv.org/abs/1710.05833} {arXiv:1710.05833 [astro-ph.HE]}
  \BibitemShut {NoStop}%
\bibitem [{\citenamefont {Miller}\ \emph {et~al.}(2019)\citenamefont {Miller}
  \emph {et~al.}}]{Miller_2019}%
  \BibitemOpen
  \bibfield  {author} {\bibinfo {author} {\bibfnamefont {M.~C.}\ \bibnamefont
  {Miller}} \emph {et~al.},\ }\href {https://doi.org/10.3847/2041-8213/ab50c5}
  {\bibfield  {journal} {\bibinfo  {journal} {Astrophys. J. Lett.}\ }\textbf
  {\bibinfo {volume} {887}},\ \bibinfo {pages} {L24} (\bibinfo {year}
  {2019})},\ \Eprint {https://arxiv.org/abs/1912.05705} {arXiv:1912.05705
  [astro-ph.HE]} \BibitemShut {NoStop}%
\bibitem [{\citenamefont {Riley}\ \emph {et~al.}(2019)\citenamefont {Riley}
  \emph {et~al.}}]{Riley_2019}%
  \BibitemOpen
  \bibfield  {author} {\bibinfo {author} {\bibfnamefont {T.~E.}\ \bibnamefont
  {Riley}} \emph {et~al.},\ }\href {https://doi.org/10.3847/2041-8213/ab481c}
  {\bibfield  {journal} {\bibinfo  {journal} {Astrophys. J. Lett.}\ }\textbf
  {\bibinfo {volume} {887}},\ \bibinfo {pages} {L21} (\bibinfo {year}
  {2019})},\ \Eprint {https://arxiv.org/abs/1912.05702} {arXiv:1912.05702
  [astro-ph.HE]} \BibitemShut {NoStop}%
\bibitem [{\citenamefont {Miller}\ \emph {et~al.}(2021)\citenamefont {Miller},
  \citenamefont {Lamb}, \citenamefont {Dittmann}, \citenamefont {Bogdanov},
  \citenamefont {Arzoumanian}, \citenamefont {Gendreau}, \citenamefont
  {Guillot}, \citenamefont {Ho}, \citenamefont {Lattimer}, \citenamefont
  {Loewenstein}, \citenamefont {Morsink}, \citenamefont {Ray}, \citenamefont
  {Wolff}, \citenamefont {Baker}, \citenamefont {Cazeau}, \citenamefont
  {Manthripragada}, \citenamefont {Markwardt}, \citenamefont {Okajima},
  \citenamefont {Pollard}, \citenamefont {Cognard}, \citenamefont {Cromartie},
  \citenamefont {Fonseca}, \citenamefont {Guillemot}, \citenamefont {Kerr},
  \citenamefont {Parthasarathy}, \citenamefont {Pennucci}, \citenamefont
  {Ransom},\ and\ \citenamefont {Stairs}}]{Miller_may2021}%
  \BibitemOpen
  \bibfield  {author} {\bibinfo {author} {\bibfnamefont {M.~C.}\ \bibnamefont
  {Miller}}, \bibinfo {author} {\bibfnamefont {F.~K.}\ \bibnamefont {Lamb}},
  \bibinfo {author} {\bibfnamefont {A.~J.}\ \bibnamefont {Dittmann}}, \bibinfo
  {author} {\bibfnamefont {S.}~\bibnamefont {Bogdanov}}, \bibinfo {author}
  {\bibfnamefont {Z.}~\bibnamefont {Arzoumanian}}, \bibinfo {author}
  {\bibfnamefont {K.~C.}\ \bibnamefont {Gendreau}}, \bibinfo {author}
  {\bibfnamefont {S.}~\bibnamefont {Guillot}}, \bibinfo {author} {\bibfnamefont
  {W.~C.~G.}\ \bibnamefont {Ho}}, \bibinfo {author} {\bibfnamefont {J.~M.}\
  \bibnamefont {Lattimer}}, \bibinfo {author} {\bibfnamefont {M.}~\bibnamefont
  {Loewenstein}}, \bibinfo {author} {\bibfnamefont {S.~M.}\ \bibnamefont
  {Morsink}}, \bibinfo {author} {\bibfnamefont {P.~S.}\ \bibnamefont {Ray}},
  \bibinfo {author} {\bibfnamefont {M.~T.}\ \bibnamefont {Wolff}}, \bibinfo
  {author} {\bibfnamefont {C.~L.}\ \bibnamefont {Baker}}, \bibinfo {author}
  {\bibfnamefont {T.}~\bibnamefont {Cazeau}}, \bibinfo {author} {\bibfnamefont
  {S.}~\bibnamefont {Manthripragada}}, \bibinfo {author} {\bibfnamefont
  {C.~B.}\ \bibnamefont {Markwardt}}, \bibinfo {author} {\bibfnamefont
  {T.}~\bibnamefont {Okajima}}, \bibinfo {author} {\bibfnamefont
  {S.}~\bibnamefont {Pollard}}, \bibinfo {author} {\bibfnamefont
  {I.}~\bibnamefont {Cognard}}, \bibinfo {author} {\bibfnamefont {H.~T.}\
  \bibnamefont {Cromartie}}, \bibinfo {author} {\bibfnamefont {E.}~\bibnamefont
  {Fonseca}}, \bibinfo {author} {\bibfnamefont {L.}~\bibnamefont {Guillemot}},
  \bibinfo {author} {\bibfnamefont {M.}~\bibnamefont {Kerr}}, \bibinfo {author}
  {\bibfnamefont {A.}~\bibnamefont {Parthasarathy}}, \bibinfo {author}
  {\bibfnamefont {T.~T.}\ \bibnamefont {Pennucci}}, \bibinfo {author}
  {\bibfnamefont {S.}~\bibnamefont {Ransom}},\ and\ \bibinfo {author}
  {\bibfnamefont {I.}~\bibnamefont {Stairs}},\ }\href
  {https://doi.org/10.3847/2041-8213/ac089b} {\bibfield  {journal} {\bibinfo
  {journal} {Astrophys. J. Lett.}\ }\textbf {\bibinfo {volume} {918}},\
  \bibinfo {pages} {L28} (\bibinfo {year} {2021})},\ \Eprint
  {https://arxiv.org/abs/2105.06979} {arXiv:2105.06979} \BibitemShut {NoStop}%
\bibitem [{\citenamefont {Riley}\ \emph {et~al.}(2021)\citenamefont {Riley},
  \citenamefont {Watts}, \citenamefont {Ray}, \citenamefont {Bogdanov},
  \citenamefont {Guillot}, \citenamefont {Morsink}, \citenamefont {Bilous},
  \citenamefont {Arzoumanian}, \citenamefont {Choudhury}, \citenamefont
  {Deneva}, \citenamefont {Gendreau}, \citenamefont {Harding}, \citenamefont
  {Ho}, \citenamefont {Lattimer}, \citenamefont {Loewenstein}, \citenamefont
  {Ludlam}, \citenamefont {Markwardt}, \citenamefont {Okajima}, \citenamefont
  {Prescod-Weinstein}, \citenamefont {Remillard}, \citenamefont {Wolff},
  \citenamefont {Fonseca}, \citenamefont {Cromartie}, \citenamefont {Kerr},
  \citenamefont {Pennucci}, \citenamefont {Parthasarathy}, \citenamefont
  {Ransom}, \citenamefont {Stairs}, \citenamefont {Guillemot},\ and\
  \citenamefont {Cognard}}]{Riley_may2021}%
  \BibitemOpen
  \bibfield  {author} {\bibinfo {author} {\bibfnamefont {T.~E.}\ \bibnamefont
  {Riley}}, \bibinfo {author} {\bibfnamefont {A.~L.}\ \bibnamefont {Watts}},
  \bibinfo {author} {\bibfnamefont {P.~S.}\ \bibnamefont {Ray}}, \bibinfo
  {author} {\bibfnamefont {S.}~\bibnamefont {Bogdanov}}, \bibinfo {author}
  {\bibfnamefont {S.}~\bibnamefont {Guillot}}, \bibinfo {author} {\bibfnamefont
  {S.~M.}\ \bibnamefont {Morsink}}, \bibinfo {author} {\bibfnamefont {A.~V.}\
  \bibnamefont {Bilous}}, \bibinfo {author} {\bibfnamefont {Z.}~\bibnamefont
  {Arzoumanian}}, \bibinfo {author} {\bibfnamefont {D.}~\bibnamefont
  {Choudhury}}, \bibinfo {author} {\bibfnamefont {J.~S.}\ \bibnamefont
  {Deneva}}, \bibinfo {author} {\bibfnamefont {K.~C.}\ \bibnamefont
  {Gendreau}}, \bibinfo {author} {\bibfnamefont {A.~K.}\ \bibnamefont
  {Harding}}, \bibinfo {author} {\bibfnamefont {W.~C.~G.}\ \bibnamefont {Ho}},
  \bibinfo {author} {\bibfnamefont {J.~M.}\ \bibnamefont {Lattimer}}, \bibinfo
  {author} {\bibfnamefont {M.}~\bibnamefont {Loewenstein}}, \bibinfo {author}
  {\bibfnamefont {R.~M.}\ \bibnamefont {Ludlam}}, \bibinfo {author}
  {\bibfnamefont {C.~B.}\ \bibnamefont {Markwardt}}, \bibinfo {author}
  {\bibfnamefont {T.}~\bibnamefont {Okajima}}, \bibinfo {author} {\bibfnamefont
  {C.}~\bibnamefont {Prescod-Weinstein}}, \bibinfo {author} {\bibfnamefont
  {R.~A.}\ \bibnamefont {Remillard}}, \bibinfo {author} {\bibfnamefont {M.~T.}\
  \bibnamefont {Wolff}}, \bibinfo {author} {\bibfnamefont {E.}~\bibnamefont
  {Fonseca}}, \bibinfo {author} {\bibfnamefont {H.~T.}\ \bibnamefont
  {Cromartie}}, \bibinfo {author} {\bibfnamefont {M.}~\bibnamefont {Kerr}},
  \bibinfo {author} {\bibfnamefont {T.~T.}\ \bibnamefont {Pennucci}}, \bibinfo
  {author} {\bibfnamefont {A.}~\bibnamefont {Parthasarathy}}, \bibinfo {author}
  {\bibfnamefont {S.}~\bibnamefont {Ransom}}, \bibinfo {author} {\bibfnamefont
  {I.}~\bibnamefont {Stairs}}, \bibinfo {author} {\bibfnamefont
  {L.}~\bibnamefont {Guillemot}},\ and\ \bibinfo {author} {\bibfnamefont
  {I.}~\bibnamefont {Cognard}},\ }\href
  {https://doi.org/10.3847/2041-8213/ac0a81} {\bibfield  {journal} {\bibinfo
  {journal} {Astrophys. J. Lett.}\ }\textbf {\bibinfo {volume} {918}},\
  \bibinfo {pages} {L27} (\bibinfo {year} {2021})},\ \Eprint
  {https://arxiv.org/abs/2105.06980} {arXiv:2105.06980 [astro-ph.HE]}
  \BibitemShut {NoStop}%
\bibitem [{\citenamefont {Raaijmakers}\ \emph {et~al.}(2021)\citenamefont
  {Raaijmakers}, \citenamefont {Greif}, \citenamefont {Hebeler}, \citenamefont
  {Hinderer}, \citenamefont {Nissanke}, \citenamefont {Schwenk}, \citenamefont
  {Riley}, \citenamefont {Watts}, \citenamefont {Lattimer},\ and\ \citenamefont
  {Ho}}]{Raaijmakers_may2021}%
  \BibitemOpen
  \bibfield  {author} {\bibinfo {author} {\bibfnamefont {G.}~\bibnamefont
  {Raaijmakers}}, \bibinfo {author} {\bibfnamefont {S.~K.}\ \bibnamefont
  {Greif}}, \bibinfo {author} {\bibfnamefont {K.}~\bibnamefont {Hebeler}},
  \bibinfo {author} {\bibfnamefont {T.}~\bibnamefont {Hinderer}}, \bibinfo
  {author} {\bibfnamefont {S.}~\bibnamefont {Nissanke}}, \bibinfo {author}
  {\bibfnamefont {A.}~\bibnamefont {Schwenk}}, \bibinfo {author} {\bibfnamefont
  {T.~E.}\ \bibnamefont {Riley}}, \bibinfo {author} {\bibfnamefont {A.~L.}\
  \bibnamefont {Watts}}, \bibinfo {author} {\bibfnamefont {J.~M.}\ \bibnamefont
  {Lattimer}},\ and\ \bibinfo {author} {\bibfnamefont {W.~C.~G.}\ \bibnamefont
  {Ho}},\ }\href {https://doi.org/10.3847/2041-8213/ac089a} {\bibfield
  {journal} {\bibinfo  {journal} {Astrophys. J. Lett.}\ }\textbf {\bibinfo
  {volume} {918}},\ \bibinfo {pages} {L29} (\bibinfo {year} {2021})},\ \Eprint
  {https://arxiv.org/abs/2105.06981} {arXiv:2105.06981 [astro-ph.HE]}
  \BibitemShut {NoStop}%
\bibitem [{\citenamefont {Abbott}\ \emph {et~al.}(2020)\citenamefont {Abbott},
  \citenamefont {Abbott}, \citenamefont {Abbott}, \citenamefont {Abraham},
  \citenamefont {Acernese}, \citenamefont {Ackley}, \citenamefont {Adams},
  \citenamefont {Adhikari}, \citenamefont {Adya}, \citenamefont {Affeldt},
  \citenamefont {Agathos}, \citenamefont {Agatsuma}, \citenamefont {Aggarwal},
  \citenamefont {Aguiar},\ and\ \citenamefont {et~al.}}]{Abbott_ApJL_2020}%
  \BibitemOpen
  \bibfield  {author} {\bibinfo {author} {\bibfnamefont {B.~P.}\ \bibnamefont
  {Abbott}}, \bibinfo {author} {\bibfnamefont {R.}~\bibnamefont {Abbott}},
  \bibinfo {author} {\bibfnamefont {T.~D.}\ \bibnamefont {Abbott}}, \bibinfo
  {author} {\bibfnamefont {S.}~\bibnamefont {Abraham}}, \bibinfo {author}
  {\bibfnamefont {F.}~\bibnamefont {Acernese}}, \bibinfo {author}
  {\bibfnamefont {K.}~\bibnamefont {Ackley}}, \bibinfo {author} {\bibfnamefont
  {C.}~\bibnamefont {Adams}}, \bibinfo {author} {\bibfnamefont {R.~X.}\
  \bibnamefont {Adhikari}}, \bibinfo {author} {\bibfnamefont {V.~B.}\
  \bibnamefont {Adya}}, \bibinfo {author} {\bibfnamefont {C.}~\bibnamefont
  {Affeldt}}, \bibinfo {author} {\bibfnamefont {M.}~\bibnamefont {Agathos}},
  \bibinfo {author} {\bibfnamefont {K.}~\bibnamefont {Agatsuma}}, \bibinfo
  {author} {\bibfnamefont {N.}~\bibnamefont {Aggarwal}}, \bibinfo {author}
  {\bibfnamefont {O.~D.}\ \bibnamefont {Aguiar}},\ and\ \bibinfo {author}
  {\bibnamefont {et~al.}},\ }\href {https://doi.org/10.3847/2041-8213/ab75f5}
  {\bibfield  {journal} {\bibinfo  {journal} {Astrophys. J. Lett.}\ }\textbf
  {\bibinfo {volume} {892}},\ \bibinfo {pages} {L3} (\bibinfo {year}
  {2020})}\BibitemShut {NoStop}%
\bibitem [{\citenamefont {Lim}\ and\ \citenamefont
  {Holt}(2019)}]{Lim_EPJA_2019}%
  \BibitemOpen
  \bibfield  {author} {\bibinfo {author} {\bibfnamefont {Y.}~\bibnamefont
  {Lim}}\ and\ \bibinfo {author} {\bibfnamefont {J.~W.}\ \bibnamefont {Holt}},\
  }\href {https://doi.org/10.1140/epja/i2019-12917-9} {\bibfield  {journal}
  {\bibinfo  {journal} {Eur. Phys. J. A}\ }\textbf {\bibinfo {volume} {55}},\
  \bibinfo {pages} {209} (\bibinfo {year} {2019})},\ \Eprint
  {https://arxiv.org/abs/1902.05502} {arXiv:1902.05502 [nucl-th]} \BibitemShut
  {NoStop}%
\bibitem [{\citenamefont {Zhang}\ and\ \citenamefont
  {Li}(2019)}]{Zhang_ApJ_2019}%
  \BibitemOpen
  \bibfield  {author} {\bibinfo {author} {\bibfnamefont {N.-B.}\ \bibnamefont
  {Zhang}}\ and\ \bibinfo {author} {\bibfnamefont {B.-A.}\ \bibnamefont {Li}},\
  }\href {https://doi.org/10.3847/1538-4357/ab24cb} {\bibfield  {journal}
  {\bibinfo  {journal} {Astrophys. J.}\ }\textbf {\bibinfo {volume} {879}},\
  \bibinfo {pages} {99} (\bibinfo {year} {2019})}\BibitemShut {NoStop}%
\bibitem [{\citenamefont {Ferreira}\ \emph {et~al.}(2020)\citenamefont
  {Ferreira}, \citenamefont {Fortin}, \citenamefont {Malik}, \citenamefont
  {Agrawal},\ and\ \citenamefont {Provid\^encia}}]{Ferreira_PRD_2020}%
  \BibitemOpen
  \bibfield  {author} {\bibinfo {author} {\bibfnamefont {M.}~\bibnamefont
  {Ferreira}}, \bibinfo {author} {\bibfnamefont {M.}~\bibnamefont {Fortin}},
  \bibinfo {author} {\bibfnamefont {T.}~\bibnamefont {Malik}}, \bibinfo
  {author} {\bibfnamefont {B.~K.}\ \bibnamefont {Agrawal}},\ and\ \bibinfo
  {author} {\bibfnamefont {C.}~\bibnamefont {Provid\^encia}},\ }\href
  {https://doi.org/10.1103/PhysRevD.101.043021} {\bibfield  {journal} {\bibinfo
   {journal} {Phys. Rev. D}\ }\textbf {\bibinfo {volume} {101}},\ \bibinfo
  {pages} {043021} (\bibinfo {year} {2020})}\BibitemShut {NoStop}%
\bibitem [{\citenamefont {G\"uven}\ \emph {et~al.}(2020)\citenamefont
  {G\"uven}, \citenamefont {Bozkurt}, \citenamefont {Khan},\ and\ \citenamefont
  {Margueron}}]{Guven_PRC_2020}%
  \BibitemOpen
  \bibfield  {author} {\bibinfo {author} {\bibfnamefont {H.}~\bibnamefont
  {G\"uven}}, \bibinfo {author} {\bibfnamefont {K.}~\bibnamefont {Bozkurt}},
  \bibinfo {author} {\bibfnamefont {E.}~\bibnamefont {Khan}},\ and\ \bibinfo
  {author} {\bibfnamefont {J.}~\bibnamefont {Margueron}},\ }\href
  {https://doi.org/10.1103/PhysRevC.102.015805} {\bibfield  {journal} {\bibinfo
   {journal} {Phys. Rev. C}\ }\textbf {\bibinfo {volume} {102}},\ \bibinfo
  {pages} {015805} (\bibinfo {year} {2020})}\BibitemShut {NoStop}%
\bibitem [{\citenamefont {Ferreira}\ and\ \citenamefont
  {Provid\^encia}(2021{\natexlab{a}})}]{Ferreira_PRD_2021}%
  \BibitemOpen
  \bibfield  {author} {\bibinfo {author} {\bibfnamefont {M.}~\bibnamefont
  {Ferreira}}\ and\ \bibinfo {author} {\bibfnamefont {C.}~\bibnamefont
  {Provid\^encia}},\ }\href {https://doi.org/10.1103/PhysRevD.104.063006}
  {\bibfield  {journal} {\bibinfo  {journal} {Phys. Rev. D}\ }\textbf {\bibinfo
  {volume} {104}},\ \bibinfo {pages} {063006} (\bibinfo {year}
  {2021}{\natexlab{a}})}\BibitemShut {NoStop}%
\bibitem [{\citenamefont {Ferreira}\ and\ \citenamefont
  {Provid\^encia}(2021{\natexlab{b}})}]{Ferreira_JCAP_2021}%
  \BibitemOpen
  \bibfield  {author} {\bibinfo {author} {\bibfnamefont {M.}~\bibnamefont
  {Ferreira}}\ and\ \bibinfo {author} {\bibfnamefont {C.}~\bibnamefont
  {Provid\^encia}},\ }\href {https://doi.org/10.1088/1475-7516/2021/07/011}
  {\bibfield  {journal} {\bibinfo  {journal} {JCAP}\ }\textbf {\bibinfo
  {volume} {07}},\ \bibinfo {pages} {011}},\ \Eprint
  {https://arxiv.org/abs/1910.05554} {arXiv:1910.05554 [nucl-th]} \BibitemShut
  {NoStop}%
\bibitem [{\citenamefont {Ghosh}\ \emph
  {et~al.}(2022{\natexlab{a}})\citenamefont {Ghosh}, \citenamefont
  {Chatterjee},\ and\ \citenamefont {Schaffner-Bielich}}]{Ghosh_EPJA_2022}%
  \BibitemOpen
  \bibfield  {author} {\bibinfo {author} {\bibfnamefont {S.}~\bibnamefont
  {Ghosh}}, \bibinfo {author} {\bibfnamefont {D.}~\bibnamefont {Chatterjee}},\
  and\ \bibinfo {author} {\bibfnamefont {J.}~\bibnamefont
  {Schaffner-Bielich}},\ }\href
  {https://doi.org/10.1140/epja/s10050-022-00679-w} {\bibfield  {journal}
  {\bibinfo  {journal} {Eur. Phys. J. A}\ }\textbf {\bibinfo {volume} {58}},\
  \bibinfo {pages} {37} (\bibinfo {year} {2022}{\natexlab{a}})},\ \Eprint
  {https://arxiv.org/abs/2107.09371} {arXiv:2107.09371 [astro-ph.HE]}
  \BibitemShut {NoStop}%
\bibitem [{\citenamefont {Malik}\ \emph {et~al.}(2022)\citenamefont {Malik},
  \citenamefont {Ferreira}, \citenamefont {Agrawal},\ and\ \citenamefont
  {Provid\^encia}}]{Malik_ApJ_2022}%
  \BibitemOpen
  \bibfield  {author} {\bibinfo {author} {\bibfnamefont {T.}~\bibnamefont
  {Malik}}, \bibinfo {author} {\bibfnamefont {M.}~\bibnamefont {Ferreira}},
  \bibinfo {author} {\bibfnamefont {B.~K.}\ \bibnamefont {Agrawal}},\ and\
  \bibinfo {author} {\bibfnamefont {C.}~\bibnamefont {Provid\^encia}},\ }\href
  {https://doi.org/10.3847/1538-4357/ac5d3c} {\bibfield  {journal} {\bibinfo
  {journal} {Astrophys. J.}\ }\textbf {\bibinfo {volume} {930}},\ \bibinfo
  {pages} {17} (\bibinfo {year} {2022})},\ \Eprint
  {https://arxiv.org/abs/2201.12552} {arXiv:2201.12552 [nucl-th]} \BibitemShut
  {NoStop}%
\bibitem [{\citenamefont {{Patra}}\ \emph {et~al.}(2022)\citenamefont
  {{Patra}}, \citenamefont {{Imam}}, \citenamefont {{Agrawal}}, \citenamefont
  {{Mukherjee}},\ and\ \citenamefont {{Malik}}}]{Patra_PRD_2022}%
  \BibitemOpen
  \bibfield  {author} {\bibinfo {author} {\bibfnamefont {N.~K.}\ \bibnamefont
  {{Patra}}}, \bibinfo {author} {\bibfnamefont {S.~M.~A.}\ \bibnamefont
  {{Imam}}}, \bibinfo {author} {\bibfnamefont {B.~K.}\ \bibnamefont
  {{Agrawal}}}, \bibinfo {author} {\bibfnamefont {A.}~\bibnamefont
  {{Mukherjee}}},\ and\ \bibinfo {author} {\bibfnamefont {T.}~\bibnamefont
  {{Malik}}},\ }\href {https://doi.org/10.1103/PhysRevD.106.043024} {\bibfield
  {journal} {\bibinfo  {journal} {\prd}\ }\textbf {\bibinfo {volume} {106}},\
  \bibinfo {eid} {043024} (\bibinfo {year} {2022})},\ \Eprint
  {https://arxiv.org/abs/2203.08521} {arXiv:2203.08521 [nucl-th]} \BibitemShut
  {NoStop}%
\bibitem [{\citenamefont {{Malik}}\ \emph {et~al.}(2023)\citenamefont
  {{Malik}}, \citenamefont {{Ferreira}},\ and\ \citenamefont
  {{Provid{\^e}ncia}}}]{Malik_RMF_NL_2023}%
  \BibitemOpen
  \bibfield  {author} {\bibinfo {author} {\bibfnamefont {T.}~\bibnamefont
  {{Malik}}}, \bibinfo {author} {\bibfnamefont {M.}~\bibnamefont
  {{Ferreira}}},\ and\ \bibinfo {author} {\bibfnamefont {C.}~\bibnamefont
  {{Provid{\^e}ncia}}},\ }\href {https://doi.org/10.48550/arXiv.2301.08169}
  {\bibfield  {journal} {\bibinfo  {journal} {arXiv e-prints}\ ,\ \bibinfo
  {eid} {arXiv:2301.08169}} (\bibinfo {year} {2023})},\ \Eprint
  {https://arxiv.org/abs/2301.08169} {arXiv:2301.08169 [nucl-th]} \BibitemShut
  {NoStop}%
\bibitem [{\citenamefont {{Zhou}}\ \emph {et~al.}(2023)\citenamefont {{Zhou}},
  \citenamefont {{Xu}},\ and\ \citenamefont
  {{Papakonstantinou}}}]{Papakonstantinou_2023}%
  \BibitemOpen
  \bibfield  {author} {\bibinfo {author} {\bibfnamefont {J.}~\bibnamefont
  {{Zhou}}}, \bibinfo {author} {\bibfnamefont {J.}~\bibnamefont {{Xu}}},\ and\
  \bibinfo {author} {\bibfnamefont {P.}~\bibnamefont {{Papakonstantinou}}},\
  }\href {https://doi.org/10.48550/arXiv.2301.07904} {\bibfield  {journal}
  {\bibinfo  {journal} {arXiv e-prints}\ ,\ \bibinfo {eid} {arXiv:2301.07904}}
  (\bibinfo {year} {2023})},\ \Eprint {https://arxiv.org/abs/2301.07904}
  {arXiv:2301.07904 [nucl-th]} \BibitemShut {NoStop}%
\bibitem [{\citenamefont {Malik}\ and\ \citenamefont
  {Provid\^encia}(2022)}]{Malik_PRD_2022}%
  \BibitemOpen
  \bibfield  {author} {\bibinfo {author} {\bibfnamefont {T.}~\bibnamefont
  {Malik}}\ and\ \bibinfo {author} {\bibfnamefont {C.}~\bibnamefont
  {Provid\^encia}},\ }\href {https://doi.org/10.1103/PhysRevD.106.063024}
  {\bibfield  {journal} {\bibinfo  {journal} {Phys. Rev. D}\ }\textbf {\bibinfo
  {volume} {106}},\ \bibinfo {pages} {063024} (\bibinfo {year}
  {2022})}\BibitemShut {NoStop}%
\bibitem [{\citenamefont {Ghosh}\ \emph
  {et~al.}(2022{\natexlab{b}})\citenamefont {Ghosh}, \citenamefont {Pradhan},
  \citenamefont {Chatterjee},\ and\ \citenamefont
  {Schaffner-Bielich}}]{Ghosh_FASS_2022}%
  \BibitemOpen
  \bibfield  {author} {\bibinfo {author} {\bibfnamefont {S.}~\bibnamefont
  {Ghosh}}, \bibinfo {author} {\bibfnamefont {B.~K.}\ \bibnamefont {Pradhan}},
  \bibinfo {author} {\bibfnamefont {D.}~\bibnamefont {Chatterjee}},\ and\
  \bibinfo {author} {\bibfnamefont {J.}~\bibnamefont {Schaffner-Bielich}},\
  }\href {https://doi.org/10.3389/fspas.2022.864294} {\bibfield  {journal}
  {\bibinfo  {journal} {Front. Astron. Space Sci.}\ }\textbf {\bibinfo {volume}
  {9}},\ \bibinfo {pages} {864294} (\bibinfo {year} {2022}{\natexlab{b}})},\
  \Eprint {https://arxiv.org/abs/2203.03156} {arXiv:2203.03156 [astro-ph.HE]}
  \BibitemShut {NoStop}%
\bibitem [{\citenamefont {Annala}\ \emph {et~al.}(2018)\citenamefont {Annala},
  \citenamefont {Gorda}, \citenamefont {Kurkela},\ and\ \citenamefont
  {Vuorinen}}]{Annala_PRL_2018}%
  \BibitemOpen
  \bibfield  {author} {\bibinfo {author} {\bibfnamefont {E.}~\bibnamefont
  {Annala}}, \bibinfo {author} {\bibfnamefont {T.}~\bibnamefont {Gorda}},
  \bibinfo {author} {\bibfnamefont {A.}~\bibnamefont {Kurkela}},\ and\ \bibinfo
  {author} {\bibfnamefont {A.}~\bibnamefont {Vuorinen}},\ }\href
  {https://doi.org/10.1103/PhysRevLett.120.172703} {\bibfield  {journal}
  {\bibinfo  {journal} {Phys. Rev. Lett.}\ }\textbf {\bibinfo {volume} {120}},\
  \bibinfo {pages} {172703} (\bibinfo {year} {2018})}\BibitemShut {NoStop}%
\bibitem [{\citenamefont {Annala}\ \emph {et~al.}(2020)\citenamefont {Annala},
  \citenamefont {Gorda}, \citenamefont {Kurkela}, \citenamefont {N\"attil\"a},\
  and\ \citenamefont {Vuorinen}}]{Annala_Nature_2020}%
  \BibitemOpen
  \bibfield  {author} {\bibinfo {author} {\bibfnamefont {E.}~\bibnamefont
  {Annala}}, \bibinfo {author} {\bibfnamefont {T.}~\bibnamefont {Gorda}},
  \bibinfo {author} {\bibfnamefont {A.}~\bibnamefont {Kurkela}}, \bibinfo
  {author} {\bibfnamefont {J.}~\bibnamefont {N\"attil\"a}},\ and\ \bibinfo
  {author} {\bibfnamefont {A.}~\bibnamefont {Vuorinen}},\ }\href
  {https://doi.org/10.1038/s41567-020-0914-9} {\bibfield  {journal} {\bibinfo
  {journal} {Nature Phys.}\ }\textbf {\bibinfo {volume} {16}},\ \bibinfo
  {pages} {907} (\bibinfo {year} {2020})},\ \Eprint
  {https://arxiv.org/abs/1903.09121} {arXiv:1903.09121 [astro-ph.HE]}
  \BibitemShut {NoStop}%
\bibitem [{\citenamefont {Ayriyan}\ \emph {et~al.}(2021)\citenamefont
  {Ayriyan}, \citenamefont {Blaschke}, \citenamefont {Grunfeld}, \citenamefont
  {Alvarez-Castillo}, \citenamefont {Grigorian},\ and\ \citenamefont
  {Abgaryan}}]{Blaschke_EPJA_2021}%
  \BibitemOpen
  \bibfield  {author} {\bibinfo {author} {\bibfnamefont {A.}~\bibnamefont
  {Ayriyan}}, \bibinfo {author} {\bibfnamefont {D.}~\bibnamefont {Blaschke}},
  \bibinfo {author} {\bibfnamefont {A.~G.}\ \bibnamefont {Grunfeld}}, \bibinfo
  {author} {\bibfnamefont {D.}~\bibnamefont {Alvarez-Castillo}}, \bibinfo
  {author} {\bibfnamefont {H.}~\bibnamefont {Grigorian}},\ and\ \bibinfo
  {author} {\bibfnamefont {V.}~\bibnamefont {Abgaryan}},\ }\href
  {https://doi.org/10.1140/epja/s10050-021-00619-0} {\bibfield  {journal}
  {\bibinfo  {journal} {Eur. Phys. J. A}\ }\textbf {\bibinfo {volume} {57}},\
  \bibinfo {pages} {318} (\bibinfo {year} {2021})},\ \Eprint
  {https://arxiv.org/abs/2102.13485} {arXiv:2102.13485 [astro-ph.HE]}
  \BibitemShut {NoStop}%
\bibitem [{\citenamefont {Huth}\ \emph {et~al.}(2021)\citenamefont {Huth},
  \citenamefont {Wellenhofer},\ and\ \citenamefont {Schwenk}}]{Huth_PRC_2021}%
  \BibitemOpen
  \bibfield  {author} {\bibinfo {author} {\bibfnamefont {S.}~\bibnamefont
  {Huth}}, \bibinfo {author} {\bibfnamefont {C.}~\bibnamefont {Wellenhofer}},\
  and\ \bibinfo {author} {\bibfnamefont {A.}~\bibnamefont {Schwenk}},\ }\href
  {https://doi.org/10.1103/PhysRevC.103.025803} {\bibfield  {journal} {\bibinfo
   {journal} {Phys. Rev. C}\ }\textbf {\bibinfo {volume} {103}},\ \bibinfo
  {pages} {025803} (\bibinfo {year} {2021})}\BibitemShut {NoStop}%
\bibitem [{\citenamefont {{Li}}\ \emph {et~al.}(2021)\citenamefont {{Li}},
  \citenamefont {{Miao}}, \citenamefont {{Han}},\ and\ \citenamefont
  {{Zhang}}}]{Li_ApJ_2021}%
  \BibitemOpen
  \bibfield  {author} {\bibinfo {author} {\bibfnamefont {A.}~\bibnamefont
  {{Li}}}, \bibinfo {author} {\bibfnamefont {Z.}~\bibnamefont {{Miao}}},
  \bibinfo {author} {\bibfnamefont {S.}~\bibnamefont {{Han}}},\ and\ \bibinfo
  {author} {\bibfnamefont {B.}~\bibnamefont {{Zhang}}},\ }\href
  {https://doi.org/10.3847/1538-4357/abf355} {\bibfield  {journal} {\bibinfo
  {journal} {\apj}\ }\textbf {\bibinfo {volume} {913}},\ \bibinfo {eid} {27}
  (\bibinfo {year} {2021})},\ \Eprint {https://arxiv.org/abs/2103.15119}
  {arXiv:2103.15119 [astro-ph.HE]} \BibitemShut {NoStop}%
\bibitem [{\citenamefont {Dutra}\ \emph {et~al.}(2014)\citenamefont {Dutra},
  \citenamefont {Lourenço}, \citenamefont {Avancini}, \citenamefont {Carlson},
  \citenamefont {Delfino}, \citenamefont {Menezes}, \citenamefont
  {Provid\^encia}, \citenamefont {Typel},\ and\ \citenamefont
  {Stone}}]{Dutra_RMF_PRC_2014}%
  \BibitemOpen
  \bibfield  {author} {\bibinfo {author} {\bibfnamefont {M.}~\bibnamefont
  {Dutra}}, \bibinfo {author} {\bibfnamefont {O.}~\bibnamefont {Lourenço}},
  \bibinfo {author} {\bibfnamefont {S.~S.}\ \bibnamefont {Avancini}}, \bibinfo
  {author} {\bibfnamefont {B.~V.}\ \bibnamefont {Carlson}}, \bibinfo {author}
  {\bibfnamefont {A.}~\bibnamefont {Delfino}}, \bibinfo {author} {\bibfnamefont
  {D.~P.}\ \bibnamefont {Menezes}}, \bibinfo {author} {\bibfnamefont
  {C.}~\bibnamefont {Provid\^encia}}, \bibinfo {author} {\bibfnamefont
  {S.}~\bibnamefont {Typel}},\ and\ \bibinfo {author} {\bibfnamefont {J.~R.}\
  \bibnamefont {Stone}},\ }\href {https://doi.org/10.1103/PhysRevC.90.055203}
  {\bibfield  {journal} {\bibinfo  {journal} {Phys. Rev.}\ }\textbf {\bibinfo
  {volume} {C90}},\ \bibinfo {pages} {055203} (\bibinfo {year} {2014})},\
  \Eprint {https://arxiv.org/abs/1405.3633} {arXiv:1405.3633 [nucl-th]}
  \BibitemShut {NoStop}%
\bibitem [{\citenamefont {Typel}\ \emph {et~al.}(2010)\citenamefont {Typel},
  \citenamefont {R\"opke}, \citenamefont {Kl\"ahn}, \citenamefont {Blaschke},\
  and\ \citenamefont {Wolter}}]{DD2}%
  \BibitemOpen
  \bibfield  {author} {\bibinfo {author} {\bibfnamefont {S.}~\bibnamefont
  {Typel}}, \bibinfo {author} {\bibfnamefont {G.}~\bibnamefont {R\"opke}},
  \bibinfo {author} {\bibfnamefont {T.}~\bibnamefont {Kl\"ahn}}, \bibinfo
  {author} {\bibfnamefont {D.}~\bibnamefont {Blaschke}},\ and\ \bibinfo
  {author} {\bibfnamefont {H.~H.}\ \bibnamefont {Wolter}},\ }\href
  {https://doi.org/10.1103/PhysRevC.81.015803} {\bibfield  {journal} {\bibinfo
  {journal} {Phys. Rev. C}\ }\textbf {\bibinfo {volume} {81}},\ \bibinfo
  {pages} {015803} (\bibinfo {year} {2010})}\BibitemShut {NoStop}%
\bibitem [{\citenamefont {Lalazissis}\ \emph {et~al.}(2005)\citenamefont
  {Lalazissis}, \citenamefont {Nik\ifmmode \check{s}\else
  \v{s}\fi{}i\ifmmode~\acute{c}\else \'{c}\fi{}}, \citenamefont {Vretenar},\
  and\ \citenamefont {Ring}}]{Lalazissis_PRC_2005}%
  \BibitemOpen
  \bibfield  {author} {\bibinfo {author} {\bibfnamefont {G.~A.}\ \bibnamefont
  {Lalazissis}}, \bibinfo {author} {\bibfnamefont {T.}~\bibnamefont
  {Nik\ifmmode \check{s}\else \v{s}\fi{}i\ifmmode~\acute{c}\else \'{c}\fi{}}},
  \bibinfo {author} {\bibfnamefont {D.}~\bibnamefont {Vretenar}},\ and\
  \bibinfo {author} {\bibfnamefont {P.}~\bibnamefont {Ring}},\ }\href
  {https://doi.org/10.1103/PhysRevC.71.024312} {\bibfield  {journal} {\bibinfo
  {journal} {Phys. Rev. C}\ }\textbf {\bibinfo {volume} {71}},\ \bibinfo
  {pages} {024312} (\bibinfo {year} {2005})}\BibitemShut {NoStop}%
\bibitem [{\citenamefont {Hebeler}\ \emph {et~al.}(2013)\citenamefont
  {Hebeler}, \citenamefont {Lattimer}, \citenamefont {Pethick},\ and\
  \citenamefont {Schwenk}}]{Hebeler_ApJ_2013}%
  \BibitemOpen
  \bibfield  {author} {\bibinfo {author} {\bibfnamefont {K.}~\bibnamefont
  {Hebeler}}, \bibinfo {author} {\bibfnamefont {J.~M.}\ \bibnamefont
  {Lattimer}}, \bibinfo {author} {\bibfnamefont {C.~J.}\ \bibnamefont
  {Pethick}},\ and\ \bibinfo {author} {\bibfnamefont {A.}~\bibnamefont
  {Schwenk}},\ }\href {https://doi.org/10.1088/0004-637X/773/1/11} {\bibfield
  {journal} {\bibinfo  {journal} {Astrophys. J.}\ }\textbf {\bibinfo {volume}
  {773}},\ \bibinfo {pages} {11} (\bibinfo {year} {2013})},\ \Eprint
  {https://arxiv.org/abs/1303.4662} {arXiv:1303.4662 [astro-ph.SR]}
  \BibitemShut {NoStop}%
\bibitem [{\citenamefont {{Abbott}}\ and\ \citenamefont
  {et~al.}(2018)}]{Abbott_PRL_121}%
  \BibitemOpen
  \bibfield  {author} {\bibinfo {author} {\bibfnamefont {B.~P.}\ \bibnamefont
  {{Abbott}}}\ and\ \bibinfo {author} {\bibnamefont {et~al.}} (\bibinfo
  {collaboration} {The LIGO Scientific Collaboration and the Virgo
  Collaboration}),\ }\href {https://doi.org/10.1103/PhysRevLett.121.161101}
  {\bibfield  {journal} {\bibinfo  {journal} {Phys. Rev. Lett.}\ }\textbf
  {\bibinfo {volume} {121}},\ \bibinfo {pages} {161101} (\bibinfo {year}
  {2018})}\BibitemShut {NoStop}%
\bibitem [{\citenamefont {{Haensel}}\ \emph {et~al.}(1989)\citenamefont
  {{Haensel}}, \citenamefont {{Zdunik}},\ and\ \citenamefont
  {{Dobaczewski}}}]{HDZ_1989}%
  \BibitemOpen
  \bibfield  {author} {\bibinfo {author} {\bibfnamefont {P.}~\bibnamefont
  {{Haensel}}}, \bibinfo {author} {\bibfnamefont {J.~L.}\ \bibnamefont
  {{Zdunik}}},\ and\ \bibinfo {author} {\bibfnamefont {J.}~\bibnamefont
  {{Dobaczewski}}},\ }\href@noop {} {\bibfield  {journal} {\bibinfo  {journal}
  {Astronomy and Astrophysics}\ }\textbf {\bibinfo {volume} {222}},\ \bibinfo
  {pages} {353} (\bibinfo {year} {1989})}\BibitemShut {NoStop}%
\bibitem [{\citenamefont {Negele}\ and\ \citenamefont
  {Vautherin}(1973)}]{NV_1973}%
  \BibitemOpen
  \bibfield  {author} {\bibinfo {author} {\bibfnamefont {J.}~\bibnamefont
  {Negele}}\ and\ \bibinfo {author} {\bibfnamefont {D.}~\bibnamefont
  {Vautherin}},\ }\href
  {https://doi.org/https://doi.org/10.1016/0375-9474(73)90349-7} {\bibfield
  {journal} {\bibinfo  {journal} {Nuclear Physics A}\ }\textbf {\bibinfo
  {volume} {207}},\ \bibinfo {pages} {298} (\bibinfo {year}
  {1973})}\BibitemShut {NoStop}%
\bibitem [{\citenamefont {Typel}\ and\ \citenamefont
  {Wolter}(1999)}]{Typel_NPA_1999}%
  \BibitemOpen
  \bibfield  {author} {\bibinfo {author} {\bibfnamefont {S.}~\bibnamefont
  {Typel}}\ and\ \bibinfo {author} {\bibfnamefont {H.}~\bibnamefont {Wolter}},\
  }\href {https://doi.org/https://doi.org/10.1016/S0375-9474(99)00310-3}
  {\bibfield  {journal} {\bibinfo  {journal} {Nuclear Physics A}\ }\textbf
  {\bibinfo {volume} {656}},\ \bibinfo {pages} {331} (\bibinfo {year}
  {1999})}\BibitemShut {NoStop}%
\bibitem [{\citenamefont {Todd-Rutel}\ and\ \citenamefont
  {Piekarewicz}(2005)}]{Todd_PRL_2005}%
  \BibitemOpen
  \bibfield  {author} {\bibinfo {author} {\bibfnamefont {B.~G.}\ \bibnamefont
  {Todd-Rutel}}\ and\ \bibinfo {author} {\bibfnamefont {J.}~\bibnamefont
  {Piekarewicz}},\ }\href {https://doi.org/10.1103/PhysRevLett.95.122501}
  {\bibfield  {journal} {\bibinfo  {journal} {Phys. Rev. Lett.}\ }\textbf
  {\bibinfo {volume} {95}},\ \bibinfo {pages} {122501} (\bibinfo {year}
  {2005})}\BibitemShut {NoStop}%
\bibitem [{\citenamefont {{Shlomo, S.}}\ \emph {et~al.}(2006)\citenamefont
  {{Shlomo, S.}}, \citenamefont {{Kolomietz, V. M.}},\ and\ \citenamefont
  {{Col\`o, G.}}}]{Shlomo_EPJA_2006}%
  \BibitemOpen
  \bibfield  {author} {\bibinfo {author} {\bibnamefont {{Shlomo, S.}}},
  \bibinfo {author} {\bibnamefont {{Kolomietz, V. M.}}},\ and\ \bibinfo
  {author} {\bibnamefont {{Col\`o, G.}}},\ }\href
  {https://doi.org/10.1140/epja/i2006-10100-3} {\bibfield  {journal} {\bibinfo
  {journal} {Eur. Phys. J. A}\ }\textbf {\bibinfo {volume} {30}},\ \bibinfo
  {pages} {23} (\bibinfo {year} {2006})}\BibitemShut {NoStop}%
\bibitem [{\citenamefont {Essick}\ \emph {et~al.}(2021)\citenamefont {Essick},
  \citenamefont {Landry}, \citenamefont {Schwenk},\ and\ \citenamefont
  {Tews}}]{Essick_PRC_2021}%
  \BibitemOpen
  \bibfield  {author} {\bibinfo {author} {\bibfnamefont {R.}~\bibnamefont
  {Essick}}, \bibinfo {author} {\bibfnamefont {P.}~\bibnamefont {Landry}},
  \bibinfo {author} {\bibfnamefont {A.}~\bibnamefont {Schwenk}},\ and\ \bibinfo
  {author} {\bibfnamefont {I.}~\bibnamefont {Tews}},\ }\href
  {https://doi.org/10.1103/PhysRevC.104.065804} {\bibfield  {journal} {\bibinfo
   {journal} {Phys. Rev. C}\ }\textbf {\bibinfo {volume} {104}},\ \bibinfo
  {pages} {065804} (\bibinfo {year} {2021})}\BibitemShut {NoStop}%
\bibitem [{\citenamefont {Typel}(2005)}]{Typel_PRC_2005}%
  \BibitemOpen
  \bibfield  {author} {\bibinfo {author} {\bibfnamefont {S.}~\bibnamefont
  {Typel}},\ }\href {https://doi.org/10.1103/PhysRevC.71.064301} {\bibfield
  {journal} {\bibinfo  {journal} {Phys. Rev. C}\ }\textbf {\bibinfo {volume}
  {71}},\ \bibinfo {pages} {064301} (\bibinfo {year} {2005})}\BibitemShut
  {NoStop}%
\bibitem [{\citenamefont {Margueron}\ \emph
  {et~al.}(2018{\natexlab{a}})\citenamefont {Margueron}, \citenamefont
  {Hoffmann~Casali},\ and\ \citenamefont {Gulminelli}}]{Margueron_PRC_2018}%
  \BibitemOpen
  \bibfield  {author} {\bibinfo {author} {\bibfnamefont {J.}~\bibnamefont
  {Margueron}}, \bibinfo {author} {\bibfnamefont {R.}~\bibnamefont
  {Hoffmann~Casali}},\ and\ \bibinfo {author} {\bibfnamefont {F.}~\bibnamefont
  {Gulminelli}},\ }\href {https://doi.org/10.1103/PhysRevC.97.025805}
  {\bibfield  {journal} {\bibinfo  {journal} {Phys. Rev. C}\ }\textbf {\bibinfo
  {volume} {97}},\ \bibinfo {pages} {025805} (\bibinfo {year}
  {2018}{\natexlab{a}})}\BibitemShut {NoStop}%
\bibitem [{\citenamefont {Blaizot}\ and\ \citenamefont
  {Friman}(1981)}]{Blaizot_NPA_1981}%
  \BibitemOpen
  \bibfield  {author} {\bibinfo {author} {\bibfnamefont {J.}~\bibnamefont
  {Blaizot}}\ and\ \bibinfo {author} {\bibfnamefont {B.}~\bibnamefont
  {Friman}},\ }\href
  {https://doi.org/https://doi.org/10.1016/0375-9474(81)90087-7} {\bibfield
  {journal} {\bibinfo  {journal} {Nuclear Physics A}\ }\textbf {\bibinfo
  {volume} {372}},\ \bibinfo {pages} {69} (\bibinfo {year} {1981})}\BibitemShut
  {NoStop}%
\bibitem [{\citenamefont {Bonasera}\ \emph {et~al.}(2018)\citenamefont
  {Bonasera}, \citenamefont {Anders},\ and\ \citenamefont
  {Shlomo}}]{Bonasera_PRC_2018}%
  \BibitemOpen
  \bibfield  {author} {\bibinfo {author} {\bibfnamefont {G.}~\bibnamefont
  {Bonasera}}, \bibinfo {author} {\bibfnamefont {M.~R.}\ \bibnamefont
  {Anders}},\ and\ \bibinfo {author} {\bibfnamefont {S.}~\bibnamefont
  {Shlomo}},\ }\href {https://doi.org/10.1103/PhysRevC.98.054316} {\bibfield
  {journal} {\bibinfo  {journal} {Phys. Rev. C}\ }\textbf {\bibinfo {volume}
  {98}},\ \bibinfo {pages} {054316} (\bibinfo {year} {2018})}\BibitemShut
  {NoStop}%
\bibitem [{\citenamefont {{Typel, Stefan}}\ and\ \citenamefont {{Alvear
  Terrero, Diana}}(2020)}]{Typel_EPJA_2020}%
  \BibitemOpen
  \bibfield  {author} {\bibinfo {author} {\bibnamefont {{Typel, Stefan}}}\ and\
  \bibinfo {author} {\bibnamefont {{Alvear Terrero, Diana}}},\ }\href
  {https://doi.org/10.1140/epja/s10050-020-00172-2} {\bibfield  {journal}
  {\bibinfo  {journal} {Eur. Phys. J. A}\ }\textbf {\bibinfo {volume} {56}},\
  \bibinfo {pages} {160} (\bibinfo {year} {2020})}\BibitemShut {NoStop}%
\bibitem [{\citenamefont {{Somasundaram}}\ \emph {et~al.}(2021)\citenamefont
  {{Somasundaram}}, \citenamefont {{Drischler}}, \citenamefont {{Tews}},\ and\
  \citenamefont {{Margueron}}}]{Somasundaram_2021}%
  \BibitemOpen
  \bibfield  {author} {\bibinfo {author} {\bibfnamefont {R.}~\bibnamefont
  {{Somasundaram}}}, \bibinfo {author} {\bibfnamefont {C.}~\bibnamefont
  {{Drischler}}}, \bibinfo {author} {\bibfnamefont {I.}~\bibnamefont
  {{Tews}}},\ and\ \bibinfo {author} {\bibfnamefont {J.}~\bibnamefont
  {{Margueron}}},\ }\href {https://doi.org/10.1103/PhysRevC.103.045803}
  {\bibfield  {journal} {\bibinfo  {journal} {Phys. Rev. C}\ }\textbf {\bibinfo
  {volume} {103}},\ \bibinfo {eid} {045803} (\bibinfo {year} {2021})},\ \Eprint
  {https://arxiv.org/abs/2009.04737} {arXiv:2009.04737 [nucl-th]} \BibitemShut
  {NoStop}%
\bibitem [{\citenamefont {{Foreman-Mackey}}\ \emph {et~al.}(2013)\citenamefont
  {{Foreman-Mackey}}, \citenamefont {{Hogg}}, \citenamefont {{Lang}},\ and\
  \citenamefont {{Goodman}}}]{emcee}%
  \BibitemOpen
  \bibfield  {author} {\bibinfo {author} {\bibfnamefont {D.}~\bibnamefont
  {{Foreman-Mackey}}}, \bibinfo {author} {\bibfnamefont {D.~W.}\ \bibnamefont
  {{Hogg}}}, \bibinfo {author} {\bibfnamefont {D.}~\bibnamefont {{Lang}}},\
  and\ \bibinfo {author} {\bibfnamefont {J.}~\bibnamefont {{Goodman}}},\ }\href
  {https://doi.org/10.1086/670067} {\bibfield  {journal} {\bibinfo  {journal}
  {PASP}\ }\textbf {\bibinfo {volume} {125}},\ \bibinfo {pages} {306} (\bibinfo
  {year} {2013})},\ \Eprint {https://arxiv.org/abs/1202.3665} {arXiv:1202.3665
  [astro-ph.IM]} \BibitemShut {NoStop}%
\bibitem [{\citenamefont {{Foreman-Mackey}}\ \emph {et~al.}(2019)\citenamefont
  {{Foreman-Mackey}}, \citenamefont {{Farr}}, \citenamefont {{Sinha}},
  \citenamefont {{Archibald}}, \citenamefont {{Hogg}}, \citenamefont
  {{Sanders}}, \citenamefont {{Zuntz}}, \citenamefont {{Williams}},
  \citenamefont {{Nelson}}, \citenamefont {{de Val-Borro}}, \citenamefont
  {{Erhardt}}, \citenamefont {{Pashchenko}},\ and\ \citenamefont
  {{Pla}}}]{emcee3}%
  \BibitemOpen
  \bibfield  {author} {\bibinfo {author} {\bibfnamefont {D.}~\bibnamefont
  {{Foreman-Mackey}}}, \bibinfo {author} {\bibfnamefont {W.}~\bibnamefont
  {{Farr}}}, \bibinfo {author} {\bibfnamefont {M.}~\bibnamefont {{Sinha}}},
  \bibinfo {author} {\bibfnamefont {A.}~\bibnamefont {{Archibald}}}, \bibinfo
  {author} {\bibfnamefont {D.}~\bibnamefont {{Hogg}}}, \bibinfo {author}
  {\bibfnamefont {J.}~\bibnamefont {{Sanders}}}, \bibinfo {author}
  {\bibfnamefont {J.}~\bibnamefont {{Zuntz}}}, \bibinfo {author} {\bibfnamefont
  {P.}~\bibnamefont {{Williams}}}, \bibinfo {author} {\bibfnamefont
  {A.}~\bibnamefont {{Nelson}}}, \bibinfo {author} {\bibfnamefont
  {M.}~\bibnamefont {{de Val-Borro}}}, \bibinfo {author} {\bibfnamefont
  {T.}~\bibnamefont {{Erhardt}}}, \bibinfo {author} {\bibfnamefont
  {I.}~\bibnamefont {{Pashchenko}}},\ and\ \bibinfo {author} {\bibfnamefont
  {O.}~\bibnamefont {{Pla}}},\ }\href {https://doi.org/10.21105/joss.01864}
  {\bibfield  {journal} {\bibinfo  {journal} {JOSS}\ }\textbf {\bibinfo
  {volume} {4}},\ \bibinfo {eid} {1864} (\bibinfo {year} {2019})},\ \Eprint
  {https://arxiv.org/abs/1911.07688} {arXiv:1911.07688 [astro-ph.IM]}
  \BibitemShut {NoStop}%
\bibitem [{\citenamefont {{Goodman}}\ and\ \citenamefont
  {{Weare}}(2010)}]{Goodman_2010}%
  \BibitemOpen
  \bibfield  {author} {\bibinfo {author} {\bibfnamefont {J.}~\bibnamefont
  {{Goodman}}}\ and\ \bibinfo {author} {\bibfnamefont {J.}~\bibnamefont
  {{Weare}}},\ }\href {https://doi.org/10.2140/camcos.2010.5.65} {\bibfield
  {journal} {\bibinfo  {journal} {Comm. App. Math. and Comp. Sci.}\ }\textbf
  {\bibinfo {volume} {5}},\ \bibinfo {pages} {65} (\bibinfo {year}
  {2010})}\BibitemShut {NoStop}%
\bibitem [{\citenamefont {{Gelman}}\ and\ \citenamefont
  {{Rubin}}(1992)}]{Gelman_1992}%
  \BibitemOpen
  \bibfield  {author} {\bibinfo {author} {\bibfnamefont {A.}~\bibnamefont
  {{Gelman}}}\ and\ \bibinfo {author} {\bibfnamefont {D.~B.}\ \bibnamefont
  {{Rubin}}},\ }\href {https://doi.org/10.1214/ss/1177011136} {\bibfield
  {journal} {\bibinfo  {journal} {Stat. Sci.}\ }\textbf {\bibinfo {volume}
  {7}},\ \bibinfo {pages} {457} (\bibinfo {year} {1992})}\BibitemShut {NoStop}%
\bibitem [{\citenamefont {{Skilling}}(2004)}]{Skilling_2004}%
  \BibitemOpen
  \bibfield  {author} {\bibinfo {author} {\bibfnamefont {J.}~\bibnamefont
  {{Skilling}}},\ }in\ \href {https://doi.org/10.1063/1.1835238} {\emph
  {\bibinfo {booktitle} {Bayesian Inference and Maximum Entropy Methods in
  Science and Engineering: 24th International Workshop on Bayesian Inference
  and Maximum Entropy Methods in Science and Engineering}}},\ \bibinfo {series}
  {American Institute of Physics Conference Series}, Vol.\ \bibinfo {volume}
  {735},\ \bibinfo {editor} {edited by\ \bibinfo {editor} {\bibfnamefont
  {R.}~\bibnamefont {{Fischer}}}, \bibinfo {editor} {\bibfnamefont
  {R.}~\bibnamefont {{Preuss}}},\ and\ \bibinfo {editor} {\bibfnamefont
  {U.~V.}\ \bibnamefont {{Toussaint}}}}\ (\bibinfo {year} {2004})\ pp.\
  \bibinfo {pages} {395--405}\BibitemShut {NoStop}%
\bibitem [{\citenamefont {{Skilling}}(2006)}]{Skilling_2006}%
  \BibitemOpen
  \bibfield  {author} {\bibinfo {author} {\bibfnamefont {J.}~\bibnamefont
  {{Skilling}}},\ }\href {https://doi.org/10.1214/06-BA127} {\bibfield
  {journal} {\bibinfo  {journal} {Bayesian Analysis}\ }\textbf {\bibinfo
  {volume} {1}},\ \bibinfo {pages} {833 } (\bibinfo {year} {2006})}\BibitemShut
  {NoStop}%
\bibitem [{\citenamefont {{Higson}}\ \emph {et~al.}(2019)\citenamefont
  {{Higson}}, \citenamefont {{Handley}}, \citenamefont {{Hobson}},\ and\
  \citenamefont {{Lasenby}}}]{Higson_2019}%
  \BibitemOpen
  \bibfield  {author} {\bibinfo {author} {\bibfnamefont {E.}~\bibnamefont
  {{Higson}}}, \bibinfo {author} {\bibfnamefont {W.}~\bibnamefont {{Handley}}},
  \bibinfo {author} {\bibfnamefont {M.}~\bibnamefont {{Hobson}}},\ and\
  \bibinfo {author} {\bibfnamefont {A.}~\bibnamefont {{Lasenby}}},\ }\href
  {https://doi.org/10.1007/s11222-018-9844-0} {\bibfield  {journal} {\bibinfo
  {journal} {Stat. and Comput.}\ }\textbf {\bibinfo {volume} {29}},\ \bibinfo
  {pages} {891} (\bibinfo {year} {2019})},\ \Eprint
  {https://arxiv.org/abs/1704.03459} {arXiv:1704.03459 [stat.CO]} \BibitemShut
  {NoStop}%
\bibitem [{\citenamefont {{Speagle}}(2020)}]{dynesty}%
  \BibitemOpen
  \bibfield  {author} {\bibinfo {author} {\bibfnamefont {J.~S.}\ \bibnamefont
  {{Speagle}}},\ }\href {https://doi.org/10.1093/mnras/staa278} {\bibfield
  {journal} {\bibinfo  {journal} {MNRAS}\ }\textbf {\bibinfo {volume} {493}},\
  \bibinfo {pages} {3132} (\bibinfo {year} {2020})},\ \Eprint
  {https://arxiv.org/abs/1904.02180} {arXiv:1904.02180 [astro-ph.IM]}
  \BibitemShut {NoStop}%
\bibitem [{\citenamefont {{Koposov}}\ \emph {et~al.}(2022)\citenamefont
  {{Koposov}}, \citenamefont {{Speagle}}, \citenamefont {{Barbary}},
  \citenamefont {{Ashton}}, \citenamefont {{Buchner}}, \citenamefont
  {{Scheffler}}, \citenamefont {{Cook}}, \citenamefont {{Talbot}},
  \citenamefont {{Guillochon}}, \citenamefont {{Cubillos}}, \citenamefont
  {{Asensio Ramos}}, \citenamefont {{Johnson}}, \citenamefont {{Lang}},
  \citenamefont {{Ilya}}, \citenamefont {{Dartiailh}}, \citenamefont {{Nitz}},
  \citenamefont {{McCluskey}}, \citenamefont {{Archibald}}, \citenamefont
  {{Deil}}, \citenamefont {{Foreman-Mackey}}, \citenamefont {{Goldstein}},
  \citenamefont {{Tollerud}}, \citenamefont {{Leja}}, \citenamefont {{Kirk}},
  \citenamefont {{Pitkin}}, \citenamefont {{Sheehan}}, \citenamefont
  {{Cargile}}, \citenamefont {{Ruskin23}}, \citenamefont {{Angus}},\ and\
  \citenamefont {{Daylan}}}]{dynesty2}%
  \BibitemOpen
  \bibfield  {author} {\bibinfo {author} {\bibfnamefont {S.}~\bibnamefont
  {{Koposov}}}, \bibinfo {author} {\bibfnamefont {J.}~\bibnamefont
  {{Speagle}}}, \bibinfo {author} {\bibfnamefont {K.}~\bibnamefont
  {{Barbary}}}, \bibinfo {author} {\bibfnamefont {G.}~\bibnamefont {{Ashton}}},
  \bibinfo {author} {\bibfnamefont {J.}~\bibnamefont {{Buchner}}}, \bibinfo
  {author} {\bibfnamefont {C.}~\bibnamefont {{Scheffler}}}, \bibinfo {author}
  {\bibfnamefont {B.}~\bibnamefont {{Cook}}}, \bibinfo {author} {\bibfnamefont
  {C.}~\bibnamefont {{Talbot}}}, \bibinfo {author} {\bibfnamefont
  {J.}~\bibnamefont {{Guillochon}}}, \bibinfo {author} {\bibfnamefont
  {P.}~\bibnamefont {{Cubillos}}}, \bibinfo {author} {\bibfnamefont
  {A.}~\bibnamefont {{Asensio Ramos}}}, \bibinfo {author} {\bibfnamefont
  {B.}~\bibnamefont {{Johnson}}}, \bibinfo {author} {\bibfnamefont
  {D.}~\bibnamefont {{Lang}}}, \bibinfo {author} {\bibnamefont {{Ilya}}},
  \bibinfo {author} {\bibfnamefont {M.}~\bibnamefont {{Dartiailh}}}, \bibinfo
  {author} {\bibfnamefont {A.}~\bibnamefont {{Nitz}}}, \bibinfo {author}
  {\bibfnamefont {A.}~\bibnamefont {{McCluskey}}}, \bibinfo {author}
  {\bibfnamefont {A.}~\bibnamefont {{Archibald}}}, \bibinfo {author}
  {\bibfnamefont {C.}~\bibnamefont {{Deil}}}, \bibinfo {author} {\bibfnamefont
  {D.}~\bibnamefont {{Foreman-Mackey}}}, \bibinfo {author} {\bibfnamefont
  {D.}~\bibnamefont {{Goldstein}}}, \bibinfo {author} {\bibfnamefont
  {E.}~\bibnamefont {{Tollerud}}}, \bibinfo {author} {\bibfnamefont
  {J.}~\bibnamefont {{Leja}}}, \bibinfo {author} {\bibfnamefont
  {M.}~\bibnamefont {{Kirk}}}, \bibinfo {author} {\bibfnamefont
  {M.}~\bibnamefont {{Pitkin}}}, \bibinfo {author} {\bibfnamefont
  {P.}~\bibnamefont {{Sheehan}}}, \bibinfo {author} {\bibfnamefont
  {P.}~\bibnamefont {{Cargile}}}, \bibinfo {author} {\bibnamefont
  {{Ruskin23}}}, \bibinfo {author} {\bibfnamefont {R.}~\bibnamefont
  {{Angus}}},\ and\ \bibinfo {author} {\bibfnamefont {T.}~\bibnamefont
  {{Daylan}}},\ }\href {https://doi.org/10.5281/zenodo.6456387} {\bibinfo
  {title} {{joshspeagle/dynesty: v1.2.2}}},\ \bibinfo {howpublished} {Zenodo}
  (\bibinfo {year} {2022})\BibitemShut {NoStop}%
\bibitem [{\citenamefont {Khan}\ and\ \citenamefont
  {Margueron}(2013)}]{Khan_PRC_2013}%
  \BibitemOpen
  \bibfield  {author} {\bibinfo {author} {\bibfnamefont {E.}~\bibnamefont
  {Khan}}\ and\ \bibinfo {author} {\bibfnamefont {J.}~\bibnamefont
  {Margueron}},\ }\href {https://doi.org/10.1103/PhysRevC.88.034319} {\bibfield
   {journal} {\bibinfo  {journal} {Phys. Rev. C}\ }\textbf {\bibinfo {volume}
  {88}},\ \bibinfo {pages} {034319} (\bibinfo {year} {2013})}\BibitemShut
  {NoStop}%
\bibitem [{\citenamefont {{Jiang}}\ \emph {et~al.}(2022)\citenamefont
  {{Jiang}}, \citenamefont {{Ecker}},\ and\ \citenamefont
  {{Rezzolla}}}]{Rezzolla_2022}%
  \BibitemOpen
  \bibfield  {author} {\bibinfo {author} {\bibfnamefont {J.-L.}\ \bibnamefont
  {{Jiang}}}, \bibinfo {author} {\bibfnamefont {C.}~\bibnamefont {{Ecker}}},\
  and\ \bibinfo {author} {\bibfnamefont {L.}~\bibnamefont {{Rezzolla}}},\
  }\href {https://doi.org/10.48550/arXiv.2211.00018} {\bibfield  {journal}
  {\bibinfo  {journal} {arXiv e-prints}\ ,\ \bibinfo {eid} {arXiv:2211.00018}}
  (\bibinfo {year} {2022})},\ \Eprint {https://arxiv.org/abs/2211.00018}
  {arXiv:2211.00018 [gr-qc]} \BibitemShut {NoStop}%
\bibitem [{\citenamefont {{Abbott}}\ and\ \citenamefont
  {et~al.}(2020)}]{GW190814}%
  \BibitemOpen
  \bibfield  {author} {\bibinfo {author} {\bibfnamefont {R.}~\bibnamefont
  {{Abbott}}}\ and\ \bibinfo {author} {\bibnamefont {et~al.}} (\bibinfo
  {collaboration} {LIGO Scientific Collaboration and Virgo Collaboration}),\
  }\href {https://doi.org/10.3847/2041-8213/ab960f} {\bibfield  {journal}
  {\bibinfo  {journal} {Astrophys. J. Lett.}\ }\textbf {\bibinfo {volume}
  {896}},\ \bibinfo {eid} {L44} (\bibinfo {year} {2020})},\ \Eprint
  {https://arxiv.org/abs/2006.12611} {arXiv:2006.12611 [astro-ph.HE]}
  \BibitemShut {NoStop}%
\bibitem [{\citenamefont {Alam}\ \emph {et~al.}(2016)\citenamefont {Alam},
  \citenamefont {Agrawal}, \citenamefont {Fortin}, \citenamefont {Pais},
  \citenamefont {Provid\^encia}, \citenamefont {Raduta},\ and\ \citenamefont
  {Sulaksono}}]{Alam_PRC_2016}%
  \BibitemOpen
  \bibfield  {author} {\bibinfo {author} {\bibfnamefont {N.}~\bibnamefont
  {Alam}}, \bibinfo {author} {\bibfnamefont {B.~K.}\ \bibnamefont {Agrawal}},
  \bibinfo {author} {\bibfnamefont {M.}~\bibnamefont {Fortin}}, \bibinfo
  {author} {\bibfnamefont {H.}~\bibnamefont {Pais}}, \bibinfo {author}
  {\bibfnamefont {C.}~\bibnamefont {Provid\^encia}}, \bibinfo {author}
  {\bibfnamefont {A.~R.}\ \bibnamefont {Raduta}},\ and\ \bibinfo {author}
  {\bibfnamefont {A.}~\bibnamefont {Sulaksono}},\ }\href
  {https://doi.org/10.1103/PhysRevC.94.052801} {\bibfield  {journal} {\bibinfo
  {journal} {Phys. Rev. C}\ }\textbf {\bibinfo {volume} {94}},\ \bibinfo
  {pages} {052801} (\bibinfo {year} {2016})}\BibitemShut {NoStop}%
\bibitem [{\citenamefont {Margueron}\ \emph
  {et~al.}(2018{\natexlab{b}})\citenamefont {Margueron}, \citenamefont
  {Hoffmann~Casali},\ and\ \citenamefont {Gulminelli}}]{Margueron_PRC_2018_II}%
  \BibitemOpen
  \bibfield  {author} {\bibinfo {author} {\bibfnamefont {J.}~\bibnamefont
  {Margueron}}, \bibinfo {author} {\bibfnamefont {R.}~\bibnamefont
  {Hoffmann~Casali}},\ and\ \bibinfo {author} {\bibfnamefont {F.}~\bibnamefont
  {Gulminelli}},\ }\href {https://doi.org/10.1103/PhysRevC.97.025806}
  {\bibfield  {journal} {\bibinfo  {journal} {Phys. Rev. C}\ }\textbf {\bibinfo
  {volume} {97}},\ \bibinfo {pages} {025806} (\bibinfo {year}
  {2018}{\natexlab{b}})}\BibitemShut {NoStop}%
\bibitem [{\citenamefont {Malik}\ \emph {et~al.}(2018)\citenamefont {Malik},
  \citenamefont {Alam}, \citenamefont {Fortin}, \citenamefont {Provid\^encia},
  \citenamefont {Agrawal}, \citenamefont {Jha}, \citenamefont {Kumar},\ and\
  \citenamefont {Patra}}]{Malik_PRC_2018}%
  \BibitemOpen
  \bibfield  {author} {\bibinfo {author} {\bibfnamefont {T.}~\bibnamefont
  {Malik}}, \bibinfo {author} {\bibfnamefont {N.}~\bibnamefont {Alam}},
  \bibinfo {author} {\bibfnamefont {M.}~\bibnamefont {Fortin}}, \bibinfo
  {author} {\bibfnamefont {C.}~\bibnamefont {Provid\^encia}}, \bibinfo {author}
  {\bibfnamefont {B.~K.}\ \bibnamefont {Agrawal}}, \bibinfo {author}
  {\bibfnamefont {T.~K.}\ \bibnamefont {Jha}}, \bibinfo {author} {\bibfnamefont
  {B.}~\bibnamefont {Kumar}},\ and\ \bibinfo {author} {\bibfnamefont {S.~K.}\
  \bibnamefont {Patra}},\ }\href {https://doi.org/10.1103/PhysRevC.98.035804}
  {\bibfield  {journal} {\bibinfo  {journal} {Phys. Rev. C}\ }\textbf {\bibinfo
  {volume} {98}},\ \bibinfo {pages} {035804} (\bibinfo {year}
  {2018})}\BibitemShut {NoStop}%
\bibitem [{\citenamefont {Malik}\ \emph {et~al.}(2020)\citenamefont {Malik},
  \citenamefont {Agrawal}, \citenamefont {Provid\^encia},\ and\ \citenamefont
  {De}}]{Malik_PRC_2020}%
  \BibitemOpen
  \bibfield  {author} {\bibinfo {author} {\bibfnamefont {T.}~\bibnamefont
  {Malik}}, \bibinfo {author} {\bibfnamefont {B.~K.}\ \bibnamefont {Agrawal}},
  \bibinfo {author} {\bibfnamefont {C.}~\bibnamefont {Provid\^encia}},\ and\
  \bibinfo {author} {\bibfnamefont {J.~N.}\ \bibnamefont {De}},\ }\href
  {https://doi.org/10.1103/PhysRevC.102.052801} {\bibfield  {journal} {\bibinfo
   {journal} {Phys. Rev. C}\ }\textbf {\bibinfo {volume} {102}},\ \bibinfo
  {pages} {052801} (\bibinfo {year} {2020})}\BibitemShut {NoStop}%
\bibitem [{\citenamefont {Weissenborn}\ \emph {et~al.}(2012)\citenamefont
  {Weissenborn}, \citenamefont {Chatterjee},\ and\ \citenamefont
  {Schaffner-Bielich}}]{Weissenborn_NPA_2012}%
  \BibitemOpen
  \bibfield  {author} {\bibinfo {author} {\bibfnamefont {S.}~\bibnamefont
  {Weissenborn}}, \bibinfo {author} {\bibfnamefont {D.}~\bibnamefont
  {Chatterjee}},\ and\ \bibinfo {author} {\bibfnamefont {J.}~\bibnamefont
  {Schaffner-Bielich}},\ }\href
  {https://doi.org/https://doi.org/10.1016/j.nuclphysa.2012.02.012} {\bibfield
  {journal} {\bibinfo  {journal} {Nuclear Physics A}\ }\textbf {\bibinfo
  {volume} {881}},\ \bibinfo {pages} {62} (\bibinfo {year} {2012})},\ \bibinfo
  {note} {progress in Strangeness Nuclear Physics}\BibitemShut {NoStop}%
\bibitem [{\citenamefont {{Hornick}}\ \emph {et~al.}(2018)\citenamefont
  {{Hornick}}, \citenamefont {{Tolos}}, \citenamefont {{Zacchi}}, \citenamefont
  {{Christian}},\ and\ \citenamefont {{Schaffner-Bielich}}}]{Hornick_2018}%
  \BibitemOpen
  \bibfield  {author} {\bibinfo {author} {\bibfnamefont {N.}~\bibnamefont
  {{Hornick}}}, \bibinfo {author} {\bibfnamefont {L.}~\bibnamefont {{Tolos}}},
  \bibinfo {author} {\bibfnamefont {A.}~\bibnamefont {{Zacchi}}}, \bibinfo
  {author} {\bibfnamefont {J.-E.}\ \bibnamefont {{Christian}}},\ and\ \bibinfo
  {author} {\bibfnamefont {J.}~\bibnamefont {{Schaffner-Bielich}}},\ }\href
  {https://doi.org/10.1103/PhysRevC.98.065804} {\bibfield  {journal} {\bibinfo
  {journal} {Phys. Rev. C}\ }\textbf {\bibinfo {volume} {98}},\ \bibinfo {eid}
  {065804} (\bibinfo {year} {2018})},\ \Eprint
  {https://arxiv.org/abs/1808.06808} {arXiv:1808.06808 [astro-ph.HE]}
  \BibitemShut {NoStop}%
\bibitem [{\citenamefont {{Hornick}}\ \emph {et~al.}(2021)\citenamefont
  {{Hornick}}, \citenamefont {{Tolos}}, \citenamefont {{Zacchi}}, \citenamefont
  {{Christian}},\ and\ \citenamefont {{Schaffner-Bielich}}}]{Hornick_2021}%
  \BibitemOpen
  \bibfield  {author} {\bibinfo {author} {\bibfnamefont {N.}~\bibnamefont
  {{Hornick}}}, \bibinfo {author} {\bibfnamefont {L.}~\bibnamefont {{Tolos}}},
  \bibinfo {author} {\bibfnamefont {A.}~\bibnamefont {{Zacchi}}}, \bibinfo
  {author} {\bibfnamefont {J.-E.}\ \bibnamefont {{Christian}}},\ and\ \bibinfo
  {author} {\bibfnamefont {J.}~\bibnamefont {{Schaffner-Bielich}}},\ }\href
  {https://doi.org/10.1103/PhysRevC.103.039902} {\bibfield  {journal} {\bibinfo
   {journal} {Phys. Rev. C}\ }\textbf {\bibinfo {volume} {103}},\ \bibinfo
  {eid} {039902} (\bibinfo {year} {2021})}\BibitemShut {NoStop}%
\bibitem [{\citenamefont {{Foreman-Mackey}}(2016)}]{corner}%
  \BibitemOpen
  \bibfield  {author} {\bibinfo {author} {\bibfnamefont {D.}~\bibnamefont
  {{Foreman-Mackey}}},\ }\href {https://doi.org/10.21105/joss.00024} {\bibfield
   {journal} {\bibinfo  {journal} {JOSS}\ }\textbf {\bibinfo {volume} {1}},\
  \bibinfo {pages} {24} (\bibinfo {year} {2016})}\BibitemShut {NoStop}%
\bibitem [{\citenamefont {Tsang}\ \emph {et~al.}(2012)\citenamefont {Tsang},
  \citenamefont {Stone}, \citenamefont {Camera}, \citenamefont {Danielewicz},
  \citenamefont {Gandolfi}, \citenamefont {Hebeler}, \citenamefont {Horowitz},
  \citenamefont {Lee}, \citenamefont {Lynch}, \citenamefont {Kohley},
  \citenamefont {Lemmon}, \citenamefont {M\"oller}, \citenamefont {Murakami},
  \citenamefont {Riordan}, \citenamefont {Roca-Maza}, \citenamefont
  {Sammarruca}, \citenamefont {Steiner}, \citenamefont {Vida\~na},\ and\
  \citenamefont {Yennello}}]{Tsang_PRC_2012}%
  \BibitemOpen
  \bibfield  {author} {\bibinfo {author} {\bibfnamefont {M.~B.}\ \bibnamefont
  {Tsang}}, \bibinfo {author} {\bibfnamefont {J.~R.}\ \bibnamefont {Stone}},
  \bibinfo {author} {\bibfnamefont {F.}~\bibnamefont {Camera}}, \bibinfo
  {author} {\bibfnamefont {P.}~\bibnamefont {Danielewicz}}, \bibinfo {author}
  {\bibfnamefont {S.}~\bibnamefont {Gandolfi}}, \bibinfo {author}
  {\bibfnamefont {K.}~\bibnamefont {Hebeler}}, \bibinfo {author} {\bibfnamefont
  {C.~J.}\ \bibnamefont {Horowitz}}, \bibinfo {author} {\bibfnamefont
  {J.}~\bibnamefont {Lee}}, \bibinfo {author} {\bibfnamefont {W.~G.}\
  \bibnamefont {Lynch}}, \bibinfo {author} {\bibfnamefont {Z.}~\bibnamefont
  {Kohley}}, \bibinfo {author} {\bibfnamefont {R.}~\bibnamefont {Lemmon}},
  \bibinfo {author} {\bibfnamefont {P.}~\bibnamefont {M\"oller}}, \bibinfo
  {author} {\bibfnamefont {T.}~\bibnamefont {Murakami}}, \bibinfo {author}
  {\bibfnamefont {S.}~\bibnamefont {Riordan}}, \bibinfo {author} {\bibfnamefont
  {X.}~\bibnamefont {Roca-Maza}}, \bibinfo {author} {\bibfnamefont
  {F.}~\bibnamefont {Sammarruca}}, \bibinfo {author} {\bibfnamefont {A.~W.}\
  \bibnamefont {Steiner}}, \bibinfo {author} {\bibfnamefont {I.}~\bibnamefont
  {Vida\~na}},\ and\ \bibinfo {author} {\bibfnamefont {S.~J.}\ \bibnamefont
  {Yennello}},\ }\href {https://doi.org/10.1103/PhysRevC.86.015803} {\bibfield
  {journal} {\bibinfo  {journal} {Phys. Rev. C}\ }\textbf {\bibinfo {volume}
  {86}},\ \bibinfo {pages} {015803} (\bibinfo {year} {2012})}\BibitemShut
  {NoStop}%
\bibitem [{\citenamefont {{Lattimer}}\ and\ \citenamefont
  {{Steiner}}(2014)}]{Lattimer_EPJA_2014}%
  \BibitemOpen
  \bibfield  {author} {\bibinfo {author} {\bibfnamefont {J.~M.}\ \bibnamefont
  {{Lattimer}}}\ and\ \bibinfo {author} {\bibfnamefont {A.~W.}\ \bibnamefont
  {{Steiner}}},\ }\href {https://doi.org/10.1140/epja/i2014-14040-y} {\bibfield
   {journal} {\bibinfo  {journal} {European Physical Journal A}\ }\textbf
  {\bibinfo {volume} {50}},\ \bibinfo {eid} {40} (\bibinfo {year} {2014})},\
  \Eprint {https://arxiv.org/abs/1403.1186} {arXiv:1403.1186 [nucl-th]}
  \BibitemShut {NoStop}%
\bibitem [{\citenamefont {Sugahara}\ and\ \citenamefont {Toki}(1994)}]{TM1}%
  \BibitemOpen
  \bibfield  {author} {\bibinfo {author} {\bibfnamefont {Y.}~\bibnamefont
  {Sugahara}}\ and\ \bibinfo {author} {\bibfnamefont {H.}~\bibnamefont
  {Toki}},\ }\href
  {https://doi.org/https://doi.org/10.1016/0375-9474(94)90923-7} {\bibfield
  {journal} {\bibinfo  {journal} {Nuclear Physics A}\ }\textbf {\bibinfo
  {volume} {579}},\ \bibinfo {pages} {557} (\bibinfo {year}
  {1994})}\BibitemShut {NoStop}%
\bibitem [{\citenamefont {Glendenning}\ and\ \citenamefont
  {Moszkowski}(1991)}]{GM1-3}%
  \BibitemOpen
  \bibfield  {author} {\bibinfo {author} {\bibfnamefont {N.~K.}\ \bibnamefont
  {Glendenning}}\ and\ \bibinfo {author} {\bibfnamefont {S.~A.}\ \bibnamefont
  {Moszkowski}},\ }\href {https://doi.org/10.1103/PhysRevLett.67.2414}
  {\bibfield  {journal} {\bibinfo  {journal} {Phys. Rev. Lett.}\ }\textbf
  {\bibinfo {volume} {67}},\ \bibinfo {pages} {2414} (\bibinfo {year}
  {1991})}\BibitemShut {NoStop}%
\bibitem [{\citenamefont {Lalazissis}\ \emph {et~al.}(1997)\citenamefont
  {Lalazissis}, \citenamefont {K\"onig},\ and\ \citenamefont {Ring}}]{NL3}%
  \BibitemOpen
  \bibfield  {author} {\bibinfo {author} {\bibfnamefont {G.~A.}\ \bibnamefont
  {Lalazissis}}, \bibinfo {author} {\bibfnamefont {J.}~\bibnamefont
  {K\"onig}},\ and\ \bibinfo {author} {\bibfnamefont {P.}~\bibnamefont
  {Ring}},\ }\href {https://doi.org/10.1103/PhysRevC.55.540} {\bibfield
  {journal} {\bibinfo  {journal} {Phys. Rev. C}\ }\textbf {\bibinfo {volume}
  {55}},\ \bibinfo {pages} {540} (\bibinfo {year} {1997})}\BibitemShut
  {NoStop}%
\bibitem [{\citenamefont {Kendall}(1938)}]{Kendall_1938}%
  \BibitemOpen
  \bibfield  {author} {\bibinfo {author} {\bibfnamefont {M.~G.}\ \bibnamefont
  {Kendall}},\ }\href {https://doi.org/10.1093/biomet/30.1-2.81} {\bibfield
  {journal} {\bibinfo  {journal} {Biometrika}\ }\textbf {\bibinfo {volume}
  {30}},\ \bibinfo {pages} {81} (\bibinfo {year} {1938})}\BibitemShut {NoStop}%
\end{thebibliography}%

\end{document}